\documentclass[11pt,a4paper]{article}
\pdfoutput=1
\usepackage{jheppub}
\usepackage{cite}
\usepackage{graphicx}
\usepackage{longtable}
\usepackage{wrapfig}
\usepackage{rotating}
\usepackage[normalem]{ulem}
\usepackage{amsmath}
\usepackage{amssymb}
\usepackage{capt-of}
\usepackage{hyperref}
\usepackage{mathrsfs}
\usepackage{bbold}
\usepackage{cancel}
\usepackage{braket}
\usepackage{setspace}
\usepackage{amsmath,amsfonts,epsfig}
\usepackage{tcolorbox}
\usepackage{cite}
\usepackage{subcaption}
\usepackage{xstring}
\usepackage[normalem]{ulem}
\usepackage{braket}
\usepackage{caption}
\usepackage{cancel}
\usepackage{subcaption}
\usepackage{multirow, graphicx,amssymb,url,mathrsfs,amsmath}
\usepackage{wrapfig,boxedminipage,epsfig}
\usepackage{amsxtra,amstext,latexsym,dsfont,amsfonts}
\usepackage{color}
\usepackage{tcolorbox}
\usepackage{float}
\usepackage{slashed}
\usepackage{calligra}
\usepackage{enumerate}
\usepackage{mathtools}
\DeclareFontShape{T1}{calligra}{m}{n}{<->s*[2.2]callig15}{}
\DeclareMathAlphabet{\mathcalligra}{T1}{calligra}{m}{n}

\graphicspath{{./figs/}}

\newcommand\GM[1][]{{\rm GM}^{\IfInteger{#1}{(#1)}{\mathtt{(#1)}}}}
\renewcommand\S[1][]{S^{\IfInteger{#1}{(#1)}{\mathtt{(#1)}}}}
\newcommand\A[1][]{\mathcal{A}^{\IfInteger{#1}{(#1)}{\mathtt{(#1)}}}}

\newcommand{\be}{\begin{equation}}
\newcommand{\ee}{\end{equation}}

\title{More on genuine multi-entropy and holography}

\author[a, b]{Norihiro Iizuka}
 
\author[c]{Simon Lin} 
 
\author[d]{and Mitsuhiro Nishida}
 
\affiliation[a]{\it Department of Physics, National Tsing Hua University, Hsinchu 300044, Taiwan}

\affiliation[b]{\it Yukawa Institute for Theoretical Physics, Kyoto University, Kyoto 606-8502, Japan}

\affiliation[c]{\it New York University Abu Dhabi, P.O. Box 129188, Abu Dhabi, United Arab Emirates}

\affiliation[d]{\it National Institute of Technology, Yuge College, Ehime 794-2593, Japan}
 
\emailAdd{iizuka@phys.nthu.edu.tw}
\emailAdd{simonlin@nyu.edu}
\emailAdd{mnishida124@gmail.com}

\abstract{  
By generalizing the construction of genuine multi-entropy $\GM[q]$ for genuine multi-partite entanglement proposed in the previous paper \cite{Iizuka:2025ioc}, we give a prescription on how to construct $\GM[q]$ systematically for any $\mathtt{q}$. The crucial point is that our construction naturally fits to the partition number $p(\mathtt{a})$ of integer $\mathtt{a}$. For general $\mathtt{q}$, $\GM[q]$ contains $N (\mathtt{q}) = p(\mathtt{q})-p(\mathtt{q}-1)-1$ number of free parameters. Furthermore, these give $N (\mathtt{q})+1$ number of new diagnostics for genuine $\mathtt{q}$-partite entanglement. Especially for $\mathtt{q}=4$ case, this reproduces not only the known diagnostics pointed out by \cite{Balasubramanian:2014hda}, but also a new diagnostics for quadripartite entanglement. We also study these $\GM[q]$ for $\mathtt{q} = 4, 5$ in holography and show that these are of the order of ${\cal{O}}\left(1/G_N \right)$ both analytically and numerically. Our results give evidence that genuine multi-partite entanglement is ubiquitous in holography. We discuss the connection to quantum error correction and the role of genuine multi-partite entanglement in bulk reconstruction.
}

\begin{document}
\maketitle

\section{Introduction}
Quantum entanglement and the geometry of space-time are very closely related concepts. 
In particular, the Ryu-Takayanagi (RT) formula \cite{Ryu:2006bv}, which relates the entanglement entropy to the area of the minimal surface in the bulk geometry, has made many researchers in high-energy physics and gravity interested in quantum entanglement. 
In recent years, it has become increasingly apparent that bipartite entanglement is not enough to account for certain features in holography. One such example is a quantity known as the Markov Gap \cite{Akers:2019gcv, Hayden:2021gno}, which led to the rebuttal of the ``mostly bipartite conjecture'' \cite{Cui:2018dyq} for holographic CFT states. 

It is known that the entanglement entropy is the unique entanglement measure for bipartite entanglement in pure states \cite{Donald2001TheUT}. However, for multi-partite entanglement, there is no unique and unified entanglement measure, and various quantities have been proposed and investigated to detect multi-partite entanglement \cite{Walter:2016lgl}. We are interested in the study of $\mathtt{q}$-partite entanglement of a physical system where we divide the entire system into $\mathtt{q}$ smaller subsystems. To investigate such multi-partite entanglement, one needs to define a quantity that can be used to detect $\mathtt{q}$-partite entanglement systematically for any given $\mathtt{q}$.
One of such candidates is the multi-entropy \cite{Gadde:2022cqi,Penington:2022dhr,Gadde:2023zzj}, which is defined as a symmetric contraction of density matrices on a multi-partite quantum system. The multi-entropy can be thought of as the multi-partite generalization of entanglement entropy, and can be defined for any positive integer $\mathtt{q}$.

However, the multi-entropy itself is not a great measure of $\texttt{q}$-partite entanglement since the $\mathtt{q}$-partite multi-entropy can be nonzero even for pure states that are tensor products of bipartite entangled states. In order to better study multi-partite entanglements, it would be helpful to construct a multi-partite entanglement measure, which takes non-zero values only when the system is a genuine multi-partite entangled state\footnote{Here by \emph{genuine multi-partite entangled states} we mean states that cannot be decomposed into tensor products of smaller-partite entangled states.}. Even better, one can hope for a \emph{$\mathtt{q}$-partite entanglement measure} that goes further such that it vanishes for all genuine $\mathtt{\tilde{q}}$-partite entangled states with $\mathtt{\tilde{q}} < \mathtt{q}$. As shown in \cite{Iizuka:2025ioc}, a class of $\mathtt{q}$-partite entanglement measure, called \emph{genuine multi-entropy}, can be constructed by considering simple linear combinations of multi-entropies and by demanding that it vanishes on lower-partite entangled states. In particular \cite{Iizuka:2025ioc} worked out the explicit examples for $\mathtt{q}=3$ and $4$.

In this paper, we further study aspects of the genuine multi-entropy for generic values of $\mathtt{q}$.
We provide a prescription for systematically constructing the genuine multi-entropy for any $\mathtt{q}$, and as concrete examples, we demonstrate it by explicitly working out the construction for $\mathtt{q}=3,4,5$.
In our systematic construction, the integer partition of $\mathtt{q}$ plays an important role, which represents how to divide $\mathtt{q}$ subsystems for each multi-entropy in the genuine multi-entropy. By using the partition number of $\mathtt{q}$, we derive the number of free parameters in the $\mathtt{q}$-partite genuine multi-entropy. Corresponding to the existence of free parameters, independent diagnostics of genuine multi-partite entanglement can be extracted from the genuine multi-entropy.
We emphasize that our prescription provides a totally generic and systematic way to construct $\mathtt{q}$-partite entanglement measures.

We also establish geometrical arguments of the genuine multi-entropy in holography, assuming the holographic proposal by multiway cuts in the bulk \cite{Gadde:2022cqi,Gadde:2023zzj}. In the AdS$_3$ vacuum geometry, we derive new geometrical inequalities and perform several analytical and numerical computations of the genuine multi-entropy by using the minimal area of multiway cuts. Our results show that the genuine multi-entropy in holography is nonzero in general, which implies the importance of genuine multi-partite entanglement in holography.

The organization of this paper is as follows. We define the genuine multi-entropy and explain its properties in Section \ref{sec:DefMulti}. In Section \ref{sec:GMq}, we describe how to systematically construct the genuine multi-entropy using the integer partition for any $\mathtt{q}$. In Section \ref{sec:holographicGM}, we investigate geometrical aspects of the holographic genuine multi-entropy based on multiway cuts in the bulk. We conclude with discussions on a connection between genuine multi-partite entanglement and holographic quantum error correction for the bulk IR reconstruction in Section \ref{sec:dis}.
Several appendices are for a refined lower bound on the holographic multi-entropy in Appendix \ref{app:lowerbound}, more detailed expressions of $\GM[5]_n$ in Appendix \ref{detailGM5}, analytic formulas of holographic genuine multi-entropy in Appendix \ref{app:analytic}, cancellations of the UV divergence in Appendix \ref{app:UV_cancellation}, and $\GM[4]_n$ of four-qubit states in Appendix \ref{app:fourqubit}.

\section{Definition of genuine multi-entropy}\label{sec:DefMulti}
We define the $n$-th Rényi multi-entropy of a pure state $\ket{\psi}$ on  $\mathtt{q}$-partite subsystems $A_1$, $A_2$, $\dots$, $A_\mathtt{q}$ by \cite{Gadde:2022cqi, Penington:2022dhr, Gadde:2023zzj, Gadde:2023zni}
\begin{align}
\label{thedefinition}
S^{(\mathtt{q})}_n(A_1:A_2:\dots:A_\mathtt{q}) &:= \frac{1}{1-n}\frac{1}{n^{\mathtt{q}-2}}\log \frac{Z^{(\mathtt{q})}_n}{(Z^{(\mathtt{q})}_1)^{n^{\mathtt{q}-1}}},\\
Z^{(\mathtt{q})}_n &:= \bra{\psi}^{\otimes n^{\mathtt{q}-1}} \Sigma_1(g_1)\Sigma_2(g_2)\dots\Sigma_\mathtt{q}(g_\mathtt{q})\ket{\psi}^{\otimes n^{\mathtt{q}-1}},
\end{align}
where $\Sigma_\mathtt{k}(g_\mathtt{k})$ are twist operators for the permutation action of \(g_\mathtt{k}\) on indices of density matrices for $A_\mathtt{k}$. The action of \(g_\mathtt{k}\) can be expressed as
\begin{align}
\label{gkdefinition}
g_{\mathtt{k}} & \cdot (x_1,\dots,x_\mathtt{k},\dots,x_{\mathtt{q}-1}) = (x_1,\dots,x_{\mathtt{k}}+1,\dots,x_{\mathtt{q}-1}), \quad 1\le \mathtt{k} \le \mathtt{q}-1, \\
g_\mathtt{q} &= e ,
\end{align}
where \((x_1,x_2,\dots,x_{\mathtt{q}-1})\) represents an integer lattice point on a $(\mathtt{q}-1)$-dimensional hypercube of length \(n\) with identification of \(x_\mathtt{k}= n + 1 \) and \(x_\mathtt{k}=1\). The $\mathtt{q}$-partite multi-entropy $S^{(\mathtt{q})}(A_1:A_2:\dots:A_\mathtt{q})$ is defined by taking the $n\to1$ limit as  
\begin{align}
\label{multidef}
S^{(\mathtt{q})}(A_1:A_2:\dots:A_\mathtt{q}):=\lim_{n\to1}S^{(\mathtt{q})}_n(A_1:A_2:\dots:A_\mathtt{q}).
\end{align}

The R\'enyi multi-entropy $S^{(\mathtt{q})}_n(A_1:A_2:\dots:A_\mathtt{q})$ has the following nice properties \cite{Penington:2022dhr, Gadde:2023zni}:
\begin{itemize}
\item $S^{(\mathtt{q})}_n(A_1:A_2:\dots:A_\mathtt{q})$ is invariant under local unitary transformations of $|\psi\rangle$ on each subsystem. 
\item $S^{(\mathtt{q})}_n(A_1:A_2:\dots:A_\mathtt{q})$ is symmetric under permutation of the subsystems $A_1,A_2,\dots,A_\mathtt{q}$.
\item $S^{(\mathtt{q})}_n(A_1:A_2:\dots:A_\mathtt{q})$ is additive under tensor products of pure states. 
More specifically, $S^{(\mathtt{q})}_n(A_1 B_1 :A_2 B_2 :\dots:A_\mathtt{q} B_\mathtt{q})$ of  $\ket{\psi}=\ket{\psi_A}\otimes\ket{\psi_B}$ on Hilbert space $H_{A_1B_1 A_2 B_2 \dots A_\mathtt{q} B_\mathtt{q}}^\psi=H_{A_1A_2\dots A_\mathtt{q}}^{\psi_A}\otimes H_{B_1B_2\dots B_\mathtt{q}}^{\psi_B}$is given by the following sum
  \begin{align}
    \begin{split}
      &S^{(\mathtt{q})}_n(A_1B_1:A_2B_2:\dots:A_\mathtt{q}B_\mathtt{q})_{\ket{\psi}}\\
    &=~S^{(\mathtt{q})}_n(A_1:A_2:\dots:A_\mathtt{q})_{\ket{\psi_A}}+S^{(\mathtt{q})}_n(B_1:B_2:\dots:B_\mathtt{q})_{\ket{\psi_B}}.\label{additive}
    \end{split}    
\end{align}  
\end{itemize}

However, $S_n^{(\mathtt{q})}$ itself is not a great measure of $\texttt{q}$-partite entanglement since it is also sensitive to all $\mathtt{\tilde{q}}$-partite entanglements for $\mathtt{\tilde{q}}<\mathtt{q}$.
To study genuine $\mathtt{q}$-partite entanglements, we define the {\it genuine} $\mathtt{q}$-partite R\'enyi multi-entropy $\GM[{\mathtt{q}}]_n(A_1:A_2:\dots:A_\mathtt{q})$  with the following properties:
\begin{tcolorbox}
\begin{itemize}
    \item $\GM[{\mathtt{q}}]_n(A_1:A_2:\dots:A_\mathtt{q})$ includes the $\mathtt{q}$-partite R\'enyi multi-entropy $S^{(\mathtt{q})}_n(A_1:A_2:\dots:A_\mathtt{q})$.
    \item $\GM[{\mathtt{q}}]_n(A_1:A_2:\dots:A_\mathtt{q})$ vanishes for all states that can be factorized as 
    \begin{align}
    \label{genuinecondition}
    \ket{\psi_\mathtt{q}}_{A_1 \cdots A_\mathtt{q}}=\ket{\psi_\mathtt{\tilde{q}}}_{A_1 \cdots A_\mathtt{\tilde{q}}} \otimes \ket{\psi_{\mathtt{q}-\mathtt{\tilde{q}}}}_{A_\mathtt{\tilde{q}+1} \cdots A_\mathtt{\tilde{q}}} 
    \end{align}
    where $\mathtt{\tilde{q}} = 1, 2, \cdots \mathtt{q}-1$. 
\end{itemize}
\end{tcolorbox}
\noindent
In a nutshell, one should think of $\GM[{\mathtt{q}}]_n(A_1:A_2:\dots:A_\mathtt{q})$ as an ``irreducible representation decomposition'' of the different numbers of multi-partite entanglements contained in the multi-entropy. 
We construct $\GM[{\mathtt{q}}]_n(A_1:A_2:\dots:A_\mathtt{q})$ by considering appropriate symmetric linear combinations of R\'enyi multi-entropies. Note that such a construction of $\GM[{\mathtt{q}}]_n(A_1:A_2:\dots:A_\mathtt{q})$ also vanishes for tensor products of eq.~\eqref{genuinecondition} states due to the additive property given by eq.~(\ref{additive}).

Concrete examples of $\GM[{\mathtt{q}}]_n(A_1:A_2:\dots:A_\mathtt{q})$ are worked out for $\mathtt{q}=3$ and $\mathtt{q}=4$ in \cite{Iizuka:2025ioc}.  For the $\mathtt{q}=3$ case, the genuine R\'enyi multi-entropy $\GM[3]_n(A:B:C)$ is defined as\footnote{The importance of the  $\mathtt{q}=3$ combination, given by the r.h.s. of eq.~\eqref{q3genuinemulti},  was first pointed out in \cite{Penington:2022dhr, Harper:2024ker} and studied in \cite{Harper:2024ker, Liu:2024ulq}.}
\begin{align}
\label{q3genuinemulti}
\GM[3]_n(A:B:C) &:=S^{(3)}_n(A:B:C) \nonumber \\
 &\quad -\frac{1}{2}\left(S_n^{(2)}(AB:C)+S_n^{(2)}(AC:B)+S_n^{(2)}(BC:A)\right).
\end{align}
For the $\mathtt{q}=4$ case, the genuine R\'enyi multi-entropy $\GM[4]_n(A:B:C:D)$ is defined as \cite{ Iizuka:2025ioc}
\begin{align}
\begin{split}
& \GM[4]_n(A:B:C:D) :=  S_n^{(4)}(A:B:C:D) \\
& \qquad -\frac{1}{3}\Big(S_n^{(3)}(AB:C:D)+S_n^{(3)}(AC:B:D)
+S_n^{(3)}(AD:B:C)\\
& \qquad\;\;\;\;\;\;\;+S_n^{(3)}(BC:A:D)+S_n^{(3)}(BD:A:C)+S_n^{(3)}(CD:A:B)\Big) \\
& \qquad+a\left(S_n^{(2)}(AB:CD)+S_n^{(2)}(AC:BD)+S_n^{(2)}(AD:BC)\right)  \\
& \qquad+\left(\frac{1}{3}-a\right)\left(S_n^{(2)}(ABC:D)+S_n^{(2)}(ABD:C)+S_n^{(2)}(ACD:B)+S_n^{(2)}(BCD:A)\right),
\end{split}
\end{align}
which includes one real free parameter $a$.

\begin{figure}[t]
    \centering
    \includegraphics[width=15cm]{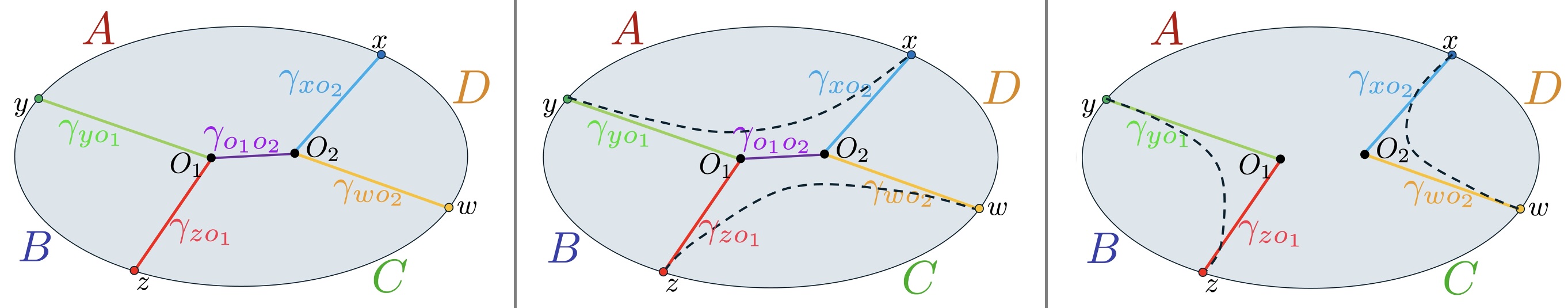}
    \caption{Left: For the case where holographic multi-entropy $S^{(\mathtt{4})}(A:B:C:D)$ is given by the sum of the five surfaces, $\gamma_{xo_2}$, $\gamma_{yo_1}$, $\gamma_{zo_1}$, $\gamma_{wo_2}$ and $\gamma_{o_1o_2}$. Middle and Right: Subtracting the RT surfaces for bipartite entanglement entropy gives positive in both cases. These figures are taken from \cite{Iizuka:2025ioc}. }
    \label{fig:q42type}
\end{figure}

It is known that bipartite entanglement is not enough to account for some features in holography \cite{Akers:2019gcv, Hayden:2021gno}. One can also see this from the following linear combination of $S^{(\mathtt{q})}_n(A_1:A_2:\dots:A_{\mathtt{q}})$ and bipartite R\'enyi entropy $S^{(2)}_n(A_\mathtt{k})$:
\begin{align}
\label{excludebicoombi}
S^{(\mathtt{q})}_n(A_1: A_2: \cdots :A_\mathtt{q})-\frac{1}{2}\left( \sum_{\mathtt{k}=1}^\mathtt{q} S^{(2)}_n(A_\mathtt{k}) \right),   
\end{align}
where $S^{(2)}_n(A_\mathtt{k})$ is bipartite R\'enyi entropy between subregion $A_\mathtt{k}$ and the rest. Although this linear combination is different from $\GM[\mathtt{q}]_n(A_1:A_2:\dots:A_\mathtt{q})$, note that 
\begin{enumerate}
\item[A:] The combination of eq.~\eqref{excludebicoombi} excludes all bipartite entanglement contributions.
\item[B:] For $\mathtt{q}=3$ case, this reduces to the genuine R\'enyi multi-entropy  $\GM[3]_n$. 
\end{enumerate}
By taking the $n\to1$ limit of this combination, it reduces to the linear combination of entanglement entropy and multi-entropy defined by eq.~\eqref{multidef}. Assuming that holographic multi-entropy is given by the area of minimal $\mathtt{q}$-way cut \cite{Gadde:2022cqi, Gadde:2023zzj} as shown in Figure \ref{fig:q42type} for $\mathtt{q}=4$, one can obtain the inequality \cite{Iizuka:2025ioc}
\begin{align}
\label{Sqinequality}
 S^{(\mathtt{q})}(A_1: A_2: \cdots :A_\mathtt{q})-\frac{1}{2}\left( \sum_{\mathtt{k}=1}^\mathtt{q} S^{(2)}(A_\mathtt{k}) \right) > 0 \,, 
\end{align}
if all the boundary subregions $A_1, A_2,...,A_\mathtt{q}$ are connected\footnote{We assume there is only one asymptotic boundary.}\footnote{To the best of our knowledge, the left-hand side of \eqref{Sqinequality} is nonnegative even for generic non-holographic states when \(\mathtt{q}=3\). For general \(\mathtt{q}\), we are not aware of any counterexamples.}. 

In fact, as shown in Appendix \ref{app:lowerbound}, in AdS$_3$/CFT$_2$ it is possible obtain a better bound than \eqref{Sqinequality}:
  \begin{align}
    \label{eq:hpt_bound_new}
    \begin{split}
     \S[q](A_1:\cdots:A_{\mathtt{q}}) - \frac{1}{2}\left(\sum_{\mathtt{k}=1}^{\mathtt{q}} \S[2](A_{\mathtt{k}}) \right) \ge \frac{3}{4G_N}\log\frac{2}{\sqrt{3}} \times (\# \text{ of vertices in minimal \texttt{q}-way cut}).    
    \end{split}
  \end{align}
  Several remarks regarding this inequality:
  \begin{enumerate}
  \item We emphasize that \eqref{eq:hpt_bound_new}  is a new result which to our knowledge has not appeared elsewhere.
  \item \eqref{eq:hpt_bound_new} is more general than \eqref{Sqinequality} in the sense that we do not need to assume if the boundary subregions are connected or disconnected. Furthermore, if the boundary subregions are disconnected, the RHS of \eqref{eq:hpt_bound_new} can be zero because in that case, the number of vertices can also be zero.
  \item \eqref{eq:hpt_bound_new} relates the amount of higher-partite ($\mathtt{q}\ge 3$) entanglement to codimension-3 ``corner'' objects in the bulk. This is akin to the bound on the Markov gap \cite{Hayden:2021gno} in AdS$_3$, an independent measure of tripartite entanglement. Our result is more general in the sense that the quantity on the LHS of \eqref{eq:hpt_bound_new} is sensitive to all $\mathtt{q}\ge 3$ multi-partite entanglements.
  \item The minimal $\mathtt{q}$-way cut can only contain equiangular trivalent vertices \cite{7f59d6a7-717a-33f0-b433-de2c5f4f8dd4}. In the case of connected subregions on a single boundary, a minimal $\mathtt{q}$-way cut always has $\mathtt{q}-2$ trivalent vertices. Therefore, the difference $\S[q](A_1:\cdots:A_\mathtt{q})-\frac{1}{2}\sum_\mathtt{k} \S[2](A_\mathtt{k})$ is larger as one increases $\mathtt{q}$. \eqref{eq:hpt_bound_new} then implies that multi-partite entanglement is increasingly more important when one considers higher-partite divisions of the boundary.
 \item Continuing upon 4., since $\S[q]-\frac{1}{2}\sum_\mathtt{k} \S[2]$ is sensitive to all $\mathtt{q}\ge 3$ multi-partite entanglements, \eqref{eq:hpt_bound_new} cannot tell us whether only some particular or all $\mathtt{q}$ are important. We will answer this question by considering the genuine multi-entropy $\GM[q](A_1:A_2:\dots:A_\mathtt{q})$, which is defined as the $n\to1$ limit of genuine R\'enyi multi-entropy 
\begin{align}
\GM[q](A_1:A_2:\dots:A_\mathtt{q}) &\coloneqq \lim_{n\to1}\GM[{\mathtt{q}}]_n(A_1:A_2:\dots:A_\mathtt{q}) \,,
\end{align} 
 and its holographic dual in Sec.~\ref{sec:holographicGM}.
  \end{enumerate}

\section{Genuine $\mathtt{q}$-partite Rényi multi-entropy $\GM[\mathtt{q}]_n$}\label{sec:GMq}
In this section, we study genuine multi-entropies in more detail. We will begin by outlining a systematic prescription on the construction of genuine multi-entropies for any $\mathtt{q}$, working out explicit examples along the way. We will see that the structure of integer partitions $p(\mathtt{q})$ is crucial in the construction of $\GM[q]_n$, and that it determines the number of free parameters in $\GM[q]_n$. In addition to discussion on general properties of $\GM[\mathtt{q}]_n$, we will also consider particular examples where $\mathtt{q}=4,5$, and see how they give rise to new and known measures multi-partite entanglement. 

\subsection{Prescription on how to construct genuine Rényi multi-entropies $\GM[\mathtt{q}]_n$}

The prescription on how to construct genuine Rényi multi-entropy $\GM[\mathtt{q}]_n$ can be summarized as follows: 
\begin{enumerate}
\item To construct genuine $\mathtt{q}$-partite Rényi multi-entropy, we first need to define $S_n^{(\mathtt{a})}[\cdots]$ for $\mathtt{a} = \mathtt{q}, \cdots , 1$, which are composed of $\mathtt{a}$-partite Rényi multi-entropies. For each $\mathtt{a}$-partite Rényi multi-entropy, we divide the whole system as $\mathtt{a}$-partition and we take a symmetrized combination of Rényi multi-entropies for all subsystems. We will see more concrete examples soon that will clarify what we mean.
  
\item Then we consider the most general linear combination of $S_n^{(\mathtt{a})}[\cdots]$, $\mathtt{a} = \mathtt{q}, \cdots , 1$ for genuine Rényi multi-entropy $\GM[\mathtt{q}]_n$. There are $p(\mathtt{q})-1$ number of parameters for such linear combinations of $S_n^{(\mathtt{a})}[\cdots]$'s, where $p(\mathtt{q})$ is the partition number of the integer $\mathtt{q}$.  The reason why we subtract one is due to our assumption that the whole system is pure, {\it i.e.,} $S_n^{(1)}[\cdots]  = 0$.  
  We set the coefficient of $S_n^{(\mathtt{q})}[\cdots]$ as one by our convention. Thus, there is $p(\mathtt{q})-2$ number of parameters in the linear combination.
  
\item After that, we fix the coefficients of the linear combination by imposing the constraint that the genuine $\mathtt{q}$-partite Rényi multi-entropy vanishes for all states that factorize as eq.~\eqref{genuinecondition}. In general, there will be more coefficients than the number of constraints, and we end up with free parameters in the expression of $\GM[q]_n$. Each of these free parameters corresponds to one additional independent measure of genuine $\mathtt{q}$-partite entanglement that we obtain from our construction.
\end{enumerate}

Since it is easier to understand this prescription through concrete examples, we will explicitly work out the aforementioned recipe for $\mathtt{q}=3,4,5$ below. The reader who would like to know new results immediately can skip Sec.~\ref{howtoq3} and \ref{howtoq4} and jump to Sec.~\ref{howtoq5}. However Sec.~\ref{howtoq3} and \ref{howtoq4} are pedagogical, so the readers who are not familiar with genuine multi-entropy $\GM[\mathtt{q}]_n$ defined first in \cite{Iizuka:2025ioc} are recommended to go through them.

\subsubsection{$\mathtt{q}=3$ case}
\label{howtoq3}

First, we consider the $\mathtt{q}=3$ case. 
One can see that $p(3)=3$ as follows.
\begin{align}
1+1+1=2+1=3.\label{ipq3}
\end{align}
This partition number $p(\mathtt{q})=p(3)$ represents how to divide $\mathtt{q}=3$ subsystems for each Rényi multi-entropy.
Corresponding to each partition, one can define $S_n^{(\mathtt{q})}[\cdots]$, $S_n^{(\mathtt{q}-1)}[\cdots]$, $\dots$, $S_n^{(1)}[\cdots]$ with symmetric permutation of $\mathtt{q}=3$ subsystems $A,B,C$. 
For the $\mathtt{q}=3$ case, 
\begin{align}
S_n^{(3)}[1:1:1]
:=&\;S^{(3)}_n(A:B:C),\\
S_n^{(2)}[2:1]:=&\;S_n^{(2)}(AB:C)+S_n^{(2)}(BC:A)+S_n^{(2)}(CA:B),\\
S_n^{(1)}[3]:=&\;S_n^{(1)}(ABC).
\end{align}
In this way, each partition corresponds to how $\mathtt{q} =3$ subsystems $A,B,C$ are divided. Note that $S_n^{(\mathtt{q})}[\cdots]$, $S_n^{(\mathtt{q}-1)}[\cdots]$, $\dots$, $S_n^{(1)}[\cdots]$ are all symmetric combinations for the subsystems $A,B,C$. The Rényi multi-entropy corresponding to each integer partition of $3$ is summarized in Table \ref{tab:p3}. 

\begin{table}[t]
    \centering
    \begin{tabular}{|c|c|}
        \hline
        \text{ Integer partition of $3$} & Rényi multi-entropy in $\GM[3]_n$\\
        \hline
        $1+1+1$ & $S_n^{(3)}[1:1:1]$\\
        \hline
        $2+1$ & $S_n^{(2)}[2:1]$\\
        \hline
        $3$ & $S_n^{(1)}[3]=0$\\
        \hline
    \end{tabular}
    \caption{Integer partition of $3$ and the corresponding Rényi multi-entropy in $\GM[3]_n$, where $S_n^{(\mathtt{a})}[\dots]$ represents the symmetrized linear combination of the Rényi multi-entropy of three subsystems $A,B,C$.}
    \label{tab:p3} 
\end{table}

To construct the genuine multi-entropy $\GM[3]_n(A:B:C)$ for $\mathtt{q}=3$, we consider the most generic linear combination of these $S_n^{(\mathtt{q})}[\cdots]$, $S_n^{(\mathtt{q}-1)}[\cdots]$, $\dots$, $S_n^{(1)}[\cdots]$. For $\mathtt{q}=3$ case, 
\begin{align}
\label{linearcombiq3}
\GM[3]_n(A:B:C) = c_0 S_n^{(3)}[1:1:1] + c_1 S_n^{(2)}[2:1] + c_2 S_n^{(1)}[3].
\end{align}

We can immediately eliminate two out of the three parameters by demanding that
\begin{enumerate}
\item[(A)] the coefficient of $S_n^{(\mathtt{q})}[1:1: \cdots : 1]$ to be one by our convention. For $\mathtt{q}=3$ case, we set the coefficient of $S_n^{(3)}[1:1:1]$ to be one, {\it i.e.,} $c_0 = 1$.
\item[(B)] $S_n^{(1)}[3]=0$ if the total system is pure. Thus $c_2$ does not appear. In this paper we restrict to the case where the total system is pure\footnote{One can generalize this to the case where total system is mixed in a straightforward fashion.}.
\end{enumerate} 
Thus, there is $p(\mathtt{q})-2=p(3)-2=1$ coefficient left in the linear combination for $\GM[3]_n(A:B:C)$ as in
\begin{align}
\begin{split}
\GM[3]_n(A:B:C)&=S_n^{(3)}[1:1:1] + c_1 S_n^{(2)}[2:1]  
\label{GMn3} 
\\
&=S^{(3)}_n(A:B:C)+c_1\left(S_n^{(2)}(AB:C)+S_n^{(2)}(AC:B)+S_n^{(2)}(BC:A)\right).
\end{split}
\end{align}
Next, we impose $p(\mathtt{q}-1)-1=p(2)-1=1$ constraint so that genuine tripartite Rényi multi-entropy vanishes for states with bipartite entanglement such as
\begin{align}
|\psi\rangle=|\psi_2\rangle_{AB}\otimes |\psi_1\rangle_{C}.
\label{Cfactorout}
\end{align}
The integer partition of $\mathtt{q}-1 =2$, where $p(\mathtt{q}-1) = p(2) = 2$, is given by
\begin{align}
1+1=2.
\end{align}
Corresponding to these integer partitions, the following relationships are valid. First, 
\begin{align}
S_n^{(3)}(A:B:\cancel{C})&=S_n^{(2)}(A\cancel{C}:B) =S_n^{(2)}(B\cancel{C}:A),
\label{q3rs2}
\end{align}
where we use $\cancel{C}$ instead of $C$ to emphasize that $|\psi_1\rangle_{C}$ on $C$ is factorized out as eq.~\eqref{Cfactorout}. 
This relationship\footnote{For this relationship to hold, it is crucial to include a factor ${1}/{n^{\mathtt{q}-2}}$
in the definition by eq.~\eqref{thedefinition}.} corresponds to the integer partition $1+1$ of $2$ for two subsystems $A,B$. The second relationship is 
\begin{align}
 S_n^{(2)}(AB:\cancel{C})&=0.
\label{q3rs}
\end{align}
This relationship corresponds to the integer partition $2$ for two subsystems $A,B$. Again here we use our assumption that the total system $A, B, C$ combined is pure,  thus, for the state given by eq.~\eqref{Cfactorout}, the combined state $AB$ is pure as well. 

 Substituting eqs.~\eqref{q3rs2} and \eqref{q3rs} into eq.~ (\ref{GMn3}), we obtain
\begin{align}
\GM[3]_n(A:B:\cancel{C})=(1+2c_1)S^{(3)}_n(A:B:\cancel{C}).
\end{align}
Imposing that $\GM[3]_n(A:B:\cancel{C})$ is zero regardless of the value of $S^{(3)}_n(A:B:\cancel{C})$, we obtain the constraint equation of $c_1$ as 
\begin{align}
1 + 2 c_1 = 0 \quad \Leftrightarrow \quad c_1=-1/2.
\end{align}
In other words, we impose that the coefficient of $S^{(3)}_n(A:B:\cancel{C})$ to be zero, which  corresponds to the integer partition $1+1$ of $A, B$. There is no additional constraint corresponding to the integer partition $2$ because $S_n^{(2)}(AB:\cancel{C})=0$, again by our assumption that the total system is pure. Thus, we impose $p(\mathtt{q}-1)-1=p(2)-1=1$ constraint. So the number of free parameters in $\GM[3]_n(A:B:C)$ is 
\begin{align}
\begin{split}
&(p(\mathtt{q}) -2) - (p(\mathtt{q}-1) -1) \\
&= 
\underbrace{(p(\mathtt{3}) -2)}_{\mbox{= parameters in eq.~\eqref{GMn3}}} - \underbrace{(p(\mathtt{3}-1) -1)}_{\mbox{=numbers of constraints}} = 0 \,.
\end{split}
\end{align}
Thus, there is no free parameter in $\GM[3]_n(A:B:C)$ and 
in this way, $\GM[3]_n(A:B:C)$ is determined by eq.~(\ref{GMn3}) with $c_1=-1/2$ as
\begin{align}
\begin{split}
\GM[3]_n(A:B:C)&=S_n^{(3)}[1:1:1] -\frac{1}{2} S_n^{(2)}[2:1]\\
&=S^{(3)}_n(A:B:C)-\frac{1}{2}\left(S_n^{(2)}(AB:C)+S_n^{(2)}(AC:B)+S_n^{(2)}(BC:A)\right).
\end{split}
\end{align}
This matches the result of \cite{Penington:2022dhr,Harper:2024ker}. 

\subsubsection{$\mathtt{q}=4$ case}\label{howtoq4}

As another example, let us consider   $\mathtt{q}=4$ case \cite{Iizuka:2025ioc}. One can see that $p(4)=5$ from the following identity
\begin{align}\label{ip4}
1+1+1+1=2+1+1=2+2=3+1=4.
\end{align}
Again, this partition represents how to divide $\mathtt{q}=4$ subsystems $A, B, C, D$ for each Rényi multi-entropy. Corresponding to each partition, one can define $S_n^{(\mathtt{q})}[\cdots]$, $S_n^{(\mathtt{q}-1)}[\cdots]$, $\dots$, $S_n^{(1)}[\cdots]$ with symmetric permutation of $\mathtt{q}=4$ subsystems $A,B,C, D$, as 
\begin{align}
\label{Bone}
    S_n^{(4)}[1:1:1:1]
:=&\;S_n^{(4)}(A:B:C:D),\\
S_n^{(3)}[2:1:1]:=&\;S_n^{(3)}(AB:C:D)+S_n^{(3)}(AC:B:D)+S_n^{(3)}(AD:B:C)\notag\\
& \quad + S_n^{(3)}(BC:A:D)
+S_n^{(3)}(BD:A:C)+S_n^{(3)}(CD:A:B),\\
S_n^{(2)}[2:2]:=&\;S_n^{(2)}(AB:CD)+S_n^{(2)}(AC:BD)+S_n^{(2)}(AD:BC),\\
S_n^{(2)}[3:1]:=&\;S_n^{(2)}(ABC:D)+S_n^{(2)}(ABD:C)
+S_n^{(2)}(ACD:B)\notag\\
&\quad + S_n^{(2)}(BCD:A),\\
S_n^{(1)}[4]:=&\;S_n^{(1)}(ABCD).
\label{Bfive}
\end{align}
The Rényi multi-entropy corresponding to each integer partition of $4$ is summarized in Table \ref{tab:p4}. 
\begin{table}[t]
    \centering
    \begin{tabular}{|c|c|}
        \hline
        \text{ Integer partition of $4$} & Rényi multi-entropy in $\GM[4]_{n}$\\
        \hline
        $1+1+1+1$ & $S_n^{(4)}[1:1:1:1]$\\
        \hline
        $2+1+1$ & $S_n^{(3)}[2:1:1]$\\
        \hline
        $2+2$ & $S_n^{(2)}[2:2]$\\
        \hline
        $3+1$ & $S_n^{(2)}[3:1]$\\
        \hline
        $4$ & $S_n^{(1)}[4]=0$\\
        \hline
    \end{tabular}
    \caption{Integer partition of $4$ and the corresponding Rényi multi-entropy in $\GM[4]_n$, where $S_n^{(\mathtt{a})}[\dots]$ represents linear combination of the Rényi multi-entropy with symmetric permutation of four subsystems $A,B,C,D$.}
    \label{tab:p4} 
\end{table}

To construct genuine multi-entropy for $\mathtt{q}=4$, we consider the linear combination of these objects,
\begin{align}
\begin{split}
\label{linearcombi}
\GM[4]_n(A:B:C:D) &=  S_n^{(4)}[1:1:1:1] + c_1 S_n^{(3)}[2:1:1]\\
&\qquad \qquad + c_2 S_n^{(2)}[2:2] + c_3 S_n^{(2)}[3:1] .  
\end{split}
\end{align}
Similarly to the $\mathtt{q}=3$ case, we set the coefficient of $S_n^{(4)}[1:1:1:1]$ to be one in our convention and assume that the whole system is pure, therefore $S_n^{(1)}[4]=0$ and thus the coefficient in front of $S_n^{(1)}[4]$ does not appear in $\GM[4]_n(A:B:C:D)$. Thus, we have only $p(4)-2=3$ parameters in eq.~\eqref{linearcombi}.

Next we consider the constraints. The genuine quadripartite Rényi multi-entropy $(\mathtt{q}=4)$ must vanish for states with tripartite entanglement such as
\begin{align}
|\psi\rangle=|\psi_3\rangle_{ABC}\otimes |\psi_1\rangle_{D}.\label{3pes}
\end{align}
Note that states with bipartite entanglement can be also expressed by eq.~(\ref{3pes}) with  $|\psi_3\rangle_{ABC}=|\psi_2\rangle_{AB}\otimes|\psi_{1'}\rangle_{C}$ \footnote{Bipartite entangled states $|\psi\rangle=|\psi_2\rangle_{AB}\otimes |\psi_{2'}\rangle_{CD}$ are given by a tensor product of two bipartite entangled states. Since Rényi multi-entropy is additive under the tensor product, $\GM[4]_n(A:B:C:D)$ of $|\psi\rangle=|\psi_2\rangle_{AB}\otimes |\psi_{2'}\rangle_{CD}$ is also zero if we impose that $\GM[4]_n(A:B:C:D)$ of pure states in eq.~(\ref{3pes}) is zero.}. Due to the symmetric construction of $\GM[4]_n(A:B:C:D)$ under permutation of the subsystems, the constraints for the tripartite entanglement state in eq.~(\ref{3pes}), where $|\psi_1\rangle_{D}$ on $D$ is factorized, are sufficient. 

The integer partition of $\mathtt{q}-1=3$ is given by eq.~(\ref{ipq3}).
Corresponding to these integer partitions, the following relationships are valid. 
The first relationship is
\begin{align}
S_n^{(4)}(A:B:C:\cancel{D})
=S_n^{(3)}(A\cancel{D}:B:C) =S_n^{(3)}(B\cancel{D}:A:C)=S_n^{(3)}(C\cancel{D}:A:B),\label{q4rs1}
\end{align}
where we use $\cancel{D}$ to emphasize that $|\psi_1\rangle_{D}$ on $D$ is factorized as $|\psi\rangle=|\psi_3\rangle_{ABC}\otimes |\psi_1\rangle_{D}$, which means that there is no entanglement from the subsystem $D$. This relationship corresponds to the integer partition $1+1+1$ of $3$ for three subsystems $A,B,C$. The second relationship is 
    \begin{align}
    \begin{split}
    &S_n^{(3)}(AB:C:\cancel{D})+S_n^{(3)}(BC:A:\cancel{D})+S_n^{(3)}(CA:B:\cancel{D})\\
    =\;&S_n^{(2)}(AB:C\cancel{D})+S_n^{(2)}(BC:A\cancel{D})+S_n^{(2)}(CA:B\cancel{D})\\
    =\;&S_n^{(2)}(AB\cancel{D}:C)+S_n^{(2)}(BC\cancel{D}:A)+S_n^{(2)}(CA\cancel{D}:B).
    \end{split}\label{q4rs2}
    \end{align}
    This relationship corresponds to the integer partition $2+1$ of $3$ for three subsystems $A,B,C$.
In the genuine Rényi multi-entropy, the Rényi multi-entropy always appears in these symmetric combinations. Finally, 
\begin{align}
S_n^{(2)}(ABC:\cancel{D})=S_n^{(1)}(ABC \cancel{D})=0,\label{q4rs3}
\end{align} which is the identity corresponding to the integer partition $3$ of $3$.
Again this is due to the assumption that the whole $A, B, C, D$ combined system is pure. 

Substituting eqs.~(\ref{q4rs1}), (\ref{q4rs2}), (\ref{q4rs3}) into the genuine Rényi multi-entropy $\GM[4]_n(A:B:C:\cancel{D})$ given by eq.~\eqref{linearcombi}, we obtain
\begin{align}
\begin{split}
&\GM[4]_n(A:B:C:\cancel{D})\\
&=\left(1+3c_1\right)S_n^{(4)}(A:B:C:\cancel{D})\\
&\quad+(c_1+c_2+c_3)\left(S_n^{(3)}(AB:C:\cancel{D})+S_n^{(3)}(AC:B:\cancel{D})+S_n^{(3)}(BC:A:\cancel{D})\right).
\end{split}
\end{align}
Since $S_n^{(4)}(A:B:C:\cancel{D})$ reduces to the tripartite multi-entropy for $A,B,C$, it is in general different from $S_n^{(3)}(AB:C:\cancel{D})+S_n^{(3)}(AC:B:\cancel{D})+S_n^{(3)}(BC:A:\cancel{D})$, which reduces to the linear combinations of bipartite entanglement entropies.  

Therefore, for $\GM[4]_n(A:B:C:\cancel{D})$ to vanish for any tripartite entangled state in eq.~(\ref{3pes}). Each coefficient of $S_n^{(4)}(A:B:C:\cancel{D})$ and $S_n^{(3)}(AB:C:\cancel{D})+S_n^{(3)}(AC:B:\cancel{D})+S_n^{(3)}(BC:A:\cancel{D})$ must vanish:
\begin{align}
1+3c_1&=0,\label{q4cs1}\\
c_1+c_2+c_3&=0.\label{q4cs2}
\end{align}
The constraint eq.~(\ref{q4cs1}) corresponds to the integer partition $1+1+1$ of $\mathtt{q}-1 = 4-1=3$ and the constraint eq.~ \eqref{q4cs2} corresponds to the integer partition $2+1$ of $\mathtt{q}-1 = 4-1=3$. There is no constraint for the integer partition $3$ because of the assumption that the whole system is pure, eq.~(\ref{q4rs3}). Thus, there are $p(\mathtt{q}-1)-1=p(3)-1=2$ constraints. 
Thus the number of free parameters in $\GM[4]_n(A:B:C:D)$ is 
\begin{align}
\begin{split}
&(p(\mathtt{q}) -2) - (p(\mathtt{q}-1) -1) \\
&= 
\underbrace{(p(4) -2)}_{\mbox{= parameters in eq.~\eqref{linearcombi}}} - \underbrace{(p(4-1) -1)}_{\mbox{=numbers of constraints}} = 1 \,,
\end{split}
\end{align}
and the final result for $\GM[4]_n(A:B:C:D)$ is 
\begin{align}
\begin{split}
&\;\;\;\;\;\GM[4]_n(A:B:C:D)\label{GMq4}\\
&=  S_n^{(4)}[1:1:1:1] - \frac{1}{3} S_n^{(3)}[2:1:1] + c_2 S_n^{(2)}[2:2] + \left( \frac{1}{3} - c_2\right) S_n^{(2)}[3:1]  \\
&= S_n^{(4)}(A:B:C:D)
\\
&\quad -\frac{1}{3}\Big(S_n^{(3)}(AB:C:D)+S_n^{(3)}(AC:B:D)
+S_n^{(3)}(AD:B:C)\\
&\quad \;\;\;\;\;\;\;+S_n^{(3)}(BC:A:D)+S_n^{(3)}(BD:A:C)+S_n^{(3)}(CD:A:B)\Big)\\
&\quad +c_2\left(S_n^{(2)}(AB:CD)+S_n^{(2)}(AC:BD)+S_n^{(2)}(AD:BC)\right)\\
&\quad +\left(\frac{1}{3}-c_2\right)\left(S_n^{(2)}(ABC:D)+S_n^{(2)}(ABD:C)+S_n^{(2)}(ACD:B)+S_n^{(2)}(BCD:A)\right).
\end{split}
\end{align}

\subsubsection{$\mathtt{q}=5$ case}\label{howtoq5}

As a further example, let us construct genuine $\mathtt{q}=5$ Rényi multi-entropy, which is a new result. The partition number $p(\mathtt{q})$ for $\mathtt{q}=5$ is $p(5)=7$ as
\begin{align}
1+1+1+1+1=2+1+1+1=2+2+1=3+1+1=3+2=4+1=5.
\end{align}
This partition represents how to divide five subsystems $A,B,C,D,E$ for each Rényi multi-entropy.
We define the genuine $\mathtt{q}=5$ Rényi multi-entropy $\GM[5]_n(A:B:C:D:E)$ by
\begin{align}
\GM[5]_n(A:B:C:D:E)
&:=S_n^{(5)}[1:1:1:1:1]+c_1S_n^{(4)}[2:1:1:1]+c_2S_n^{(3)}[2:2:1]\notag\\
& \qquad +c_3S_n^{(3)}[3:1:1]+c_4S_n^{(2)}[3:2]+c_5S_n^{(2)}[4:1].\label{GMn5}
\end{align}
 Here, $S_n^{(\mathtt{a})}[\dots]$, summarized in Table \ref{tab:p5}, are linear combinations of $S_n^{(\mathtt{a})}$ with symmetric permutation of five subsystems $A,B,C,D,E$, as 
\begin{align}
S_n^{(5)}[1:1:1:1:1]
:=&\;S_n^{(5)}(A:B:C:D:E),\\
S_n^{(4)}[2:1:1:1]:=&\;S_n^{(4)}(AB:C:D:E)+\cdots,\\
S_n^{(3)}[2:2:1]:=&\;S_n^{(3)}(AB:CD:E)+ \cdots ,\\
S_n^{(3)}[3:1:1]:=&\;S_n^{(3)}(ABC:D:E)+ \cdots ,\\
S_n^{(2)}[3:2]:=&\;S_n^{(2)}(ABC:DE)+ \cdots ,\\
S_n^{(2)}[4:1]:=&\;S_n^{(2)}(ABCD:E)+ \cdots ,\\
S_n^{(1)}[5]:=&\;S_n^{(1)}(ABCDE),
\end{align}
where $\cdots$ are terms for symmetrization.
See Appendix \ref{detailGM5} for explicit expressions. In eq.~(\ref{GMn5}), we set the coefficient of $S_n^{(5)}[1:1:1:1:1]$ to be one and use $S_n^{(1)}[5]=0$ because of the assumption that the total system is pure state. Thus, there are $p(5)-2=5$ coefficients $c_1,c_2,c_3,c_4,c_5$ in eq.~(\ref{GMn5}).

\begin{table}[t]
    \centering
    \begin{tabular}{|c|c|}
        \hline
        \text{ Integer partition of $5$} & Rényi multi-entropy in $GM_{n}^{(5)}$\\
        \hline
        $1+1+1+1+1$ & $S_n^{(5)}[1:1:1:1:1]$\\
        \hline
        $2+1+1+1$ & $S_n^{(4)}[2:1:1:1]$\\
        \hline
        $2+2+1$ & $S_n^{(3)}[2:2:1]$\\
        \hline
        $3+1+1$ & $S_n^{(3)}[3:1:1]$\\
        \hline
        $3+2$ & $S_n^{(2)}[3:2]$\\
        \hline
        $4+1$ & $S_n^{(2)}[4:1]$\\
        \hline
         $5$ & $S_n^{(1)}[5]=0$\\
        \hline
    \end{tabular}
    \caption{Integer partition of $5$ and the corresponding Rényi multi-entropy in $\GM[5]_n$, where $S_n^{(\mathtt{a})}[\dots]$ represents linear combination of the Rényi multi-entropy with symmetric permutation of five subsystems $A,B,C,D,E$.}
    \label{tab:p5}
\end{table}

We impose that for quadripartite entangled states,
\begin{align}
\ket{\psi}=\ket{\psi_4}_{ABCD}\otimes\ket{\psi_1}_{E},\label{festate1}
\end{align}
$\GM[5]_n(A:B:C:D:E)$ must vanish. For these states, there are $p(4)-1=4$ relationships between Rényi multi-entropy corresponding to the integer partition of 4 in eq.~(\ref{ip4}). We subtract one as $p(4)-1$ because we have the identity 
\begin{align}
S_n^{(2)}(ABCD:\cancel{E})=S_n^{(1)}(ABCD\cancel{E})=0,\label{idp5}
\end{align}
for pure states in eq.~(\ref{festate1}). This relationship corresponds to the integer partition 4 in eq.~(\ref{ip4}). Here, we use $\cancel{E}$ to emphasize that $|\psi_1\rangle_{E}$ on $E$ is factorized as $|\psi\rangle=|\psi_4\rangle_{ABCD}\otimes |\psi_1\rangle_{E}$.\\

Let us explain other relationships\footnote{Again, for this relationship to hold, it is crucial to include a factor ${1}/{n^{\mathtt{q}-2}}$
in the definition by eq.~\eqref{thedefinition}.}. 

\begin{itemize}
\item
The first relationship, corresponding to the integer partition $1+1+1+1$ of $4$ for four subsystems $A,B,C,D$, is
\begin{align}
\begin{split}
S_n^{(5)}(A:B:C:D:\cancel{E})=\frac{1}{4}\Big(&S_n^{(4)}(A\cancel{E}:B:C:D)+S_n^{(4)}(B\cancel{E}:A:C:D)\\
+\;&S_n^{(4)}(C\cancel{E}:A:B:D)+S_n^{(4)}(D\cancel{E}:A:B:C)\Big).\label{q5rsm1}
\end{split}
\end{align}
\item
The second relationship, corresponding to the integer partition $2+1+1$ of $4$, is
\begin{align}
\begin{split}
&\;S_n^{(4)}(AB:C:D:\cancel{E})+S_n^{(4)}(AC:B:D:\cancel{E})+S_n^{(4)}(AD:B:C:\cancel{E})\\
&+\;S_n^{(4)}(BC:A:D:\cancel{E})+S_n^{(4)}(BD:A:C:\cancel{E})+S_n^{(4)}(CD:A:B:\cancel{E})\\
&=\frac{1}{2}\Big(
\;S_n^{(3)}(AB:C\cancel{E}:D)+S_n^{(3)}(AC:B\cancel{E}:D)+S_n^{(3)}(A\cancel{E}:BC:D)\\
&\qquad + S_n^{(3)}(AB:D\cancel{E}:C)+S_n^{(3)}(AD:B\cancel{E}:C)+S_n^{(3)}(A\cancel{E}:BD:C)\\
&\qquad+\;S_n^{(3)}(AC:D\cancel{E}:B)+S_n^{(3)}(AD:C\cancel{E}:B)+S_n^{(3)}(A\cancel{E}:CD:B)\\
&\qquad+\;S_n^{(3)}(BC:D\cancel{E}:A)+S_n^{(3)}(BD:C\cancel{E}:A)+S_n^{(3)}(B\cancel{E}:CD:A)\Big)\\
&=\;S_n^{(3)}(AB\cancel{E}:C:D)+S_n^{(3)}(AC\cancel{E}:B:D)+S_n^{(3)}(AD\cancel{E}:B:C)\\
&\qquad+\;S_n^{(3)}(BC\cancel{E}:A:D)+S_n^{(3)}(BD\cancel{E}:A:C)+S_n^{(3)}(CD\cancel{E}:A:B).\label{q5rsm2}
\end{split}
\end{align}
\end{itemize}
There are two more relationships. See eqs.~(\ref{q5rs3}) and (\ref{q5rs4}) in Appendix \ref{detailGM5} for details.  Substituting these four relationships and eq.~(\ref{idp5}) into $\GM[5]_n(A:B:C:D:E)$ given by eq.~(\ref{GMn5}), we obtain
\begin{align}
\begin{split}
&\;\;\;\;\;\GM[5]_n(A:B:C:D:\cancel{E})\\
&=(1+4c_1) \,S_n^{(5)}(A:B:C:D:\cancel{E})\\
&\quad +(c_1+2c_2+c_3)\Big(S_n^{(4)}(AB:C:D:\cancel{E})+S_n^{(4)}(AC:B:D:\cancel{E})+S_n^{(4)}(AD:B:C:\cancel{E})\\
&\quad \;\;\;\;\;\;\;\;\;\;\;\;\;\;\;\;\;\;\;\;\;\;\;\;\,+S_n^{(4)}(BC:A:D:\cancel{E})+S_n^{(4)}(BD:A:C:\cancel{E})+S_n^{(4)}(CD:A:B:\cancel{E})\Big)\\
&\quad+(c_2+2c_4)\Big(S_n^{(3)}(AB:CD:\cancel{E})+S_n^{(3)}(AC:BD:\cancel{E})+S_n^{(3)}(AD:BC:\cancel{E})\Big)\\
&\quad+(c_3+c_4+c_5)\Big(S_n^{(3)}(ABC:D:\cancel{E})+S_n^{(3)}(ABD:C:\cancel{E})\\
&\quad \qquad\;\;\;\;\;\;\;\;\;\;\;\;\;\;\;\;\;\;\;\;\;\;\,+S_n^{(3)}(ACD:B:\cancel{E})+S_n^{(3)}(BCD:A:\cancel{E})\Big).
\end{split}
\end{align}
In general, the Rényi multi-entropies $S_n^{(\tilde{\mathtt{q}})}(\dots)$ in $\GM[5]_n(A:B:C:D:\cancel{E})$ take different values because they detect different types of entanglement in $\ket{\psi_4}_{ABCD}$ for various partitions of $ABCD$. 
Since $\GM[5]_n(A:B:C:D:\cancel{E})$ must vanish for any  quadripartite entangled state in eq.~(\ref{festate1}), each coefficient must vanish:
\begin{align}
1+4c_1&=0,\label{q5cs1}\\
c_1+2c_2+c_3&=0,\label{q5cs2}\\
c_2+2c_4&=0,\label{q5cs3}\\
c_3+c_4+c_5&=0\label{q5cs4}.
\end{align}
Thus, we have $p(4)-1=4$ constraints.
The constraint eq.~(\ref{q5cs1}) corresponds to the integer partition $1+1+1+1$ of $\mathtt{q}-1 = 5-1=4$ in eq.~(\ref{ip4}). Similarly, the constraint eq.~\eqref{q5cs2} corresponds to the integer partition $2+1+1$, the constraint eq.~\eqref{q5cs3} corresponds to the integer partition $2+2$, and the constraint eq.~\eqref{q5cs4} corresponds to the integer partition $3+1$. There is no constraint corresponding to $4$ in eq.~(\ref{ip4}) due to the identity eq.~(\ref{idp5}).
Solving the four constraints with $c_5$ as a free parameter, we obtain
\begin{align}
c_1=-\frac{1}{4}, \;\;\;
c_2 = \frac{1 + 4 c_5}{10},\;\;\; c_3 = \frac{1 - 16 c_5}{20}, \;\;\;  c_4 = -\frac{1 + 4 c_5}{20} ,\label{Consts}
\end{align}
and
\begin{align}
\begin{split}
\label{GM5definition}
\GM[5]_n(A:B:C:D:E)
&=S_n^{(5)}[1:1:1:1:1]-\frac{1}{4}S_n^{(4)}[2:1:1:1]  \\
& \quad +\frac{1 + 4 c_5}{10}S_n^{(3)}[2:2:1] +\frac{1 - 16 c_5}{20}S_n^{(3)}[3:1:1] \\
& \quad -\frac{1 + 4 c_5}{20}S_n^{(2)}[3:2]+c_5S_n^{(2)}[4:1].
\end{split}
\end{align}
Therefore, the number of free parameters in $\GM[5]_n(A:B:C:D:E)$ is 
\begin{align}
\begin{split}
&(p(\mathtt{q}) -2) - (p(\mathtt{q}-1) -1) \\
&= 
\underbrace{(p(5) -2)}_{\mbox{= parameters in eq.~\eqref{GMn5}}} - \underbrace{(p(5-1) -1)}_{\mbox{=numbers of constraints}} = 1\,.
\end{split}
\end{align}
In the same way, the genuine Rényi multi-entropy $\GM[{\mathtt{q}}]_n$ for any positive integer $\mathtt{q}\ge 3$ can be easily and straightforwardly constructed.

\subsection{$\GM[4]_n$ and tripartite information $I_{3, n}$ as diagnostics of genuine quadripartite entanglement}

For $\mathtt{q}=4$, genuine multi-entropy $\GM[4]_n$ given by eq.~(\ref{GMq4}) can be expressed as
\begin{align}
\GM[4]_n(A:B:C:D)=S_n^{(4)}[1:1:1:1]-\frac{1}{3}S_n^{(3)}[2:1:1]+\frac{1}{3}S_n^{(2)}[3:1]-c_2I_{3, n},
\end{align}
where the tripartite information $I_{3,n}$  is defined as 
\begin{align}
\label{I3n}
I_{3,n}=S^{(2)}_n[3:1]-S^{(2)}_n[2:2].
\end{align}
For $n = 1$ case, this reduces to the conventional $I_3$, 
\begin{align}
\label{I3n=1}
I_3:= I_{3, n=1}&=S^{(2)}(ABC:D)+S^{(2)}(BCD:A)+S^{(2)}(CDA:B) +S^{(2)}(DAB:C)\notag\\
&\;\;\;\;\;-S^{(2)}(AB:CD)-S^{(2)}(BC:DA)-S^{(2)}(AC:BD).
\end{align}

Since $\GM[4]_n(A:B:C:D)$ with any values of $c_2$ vanishes for all $\tilde{\mathtt{q}}$-partite entangled states with $\tilde{\mathtt{q}}<4$, we obtain
\begin{align}
\frac{\partial}{\partial c_2} \left( \GM[4]_n(A:B:C:D) \right) =-I_{3, n}=0,\label{I30}
\end{align}
for all $\tilde{\mathtt{q}}$-partite entangled states with $\tilde{\mathtt{q}}<4$. For example, let us consider the tripartite entangled state in eq.~(\ref{3pes}). For this state, by using eqs.~(\ref{q4rs2}) and (\ref{q4rs3}),
one can explicitly confirm eq.~(\ref{I30}). Therefore, for $\mathtt{q}=4$, we have $(p(4) -2) - (p(3) -1)+1=2$ independent diagnostics to detect genuine quadripartite entanglement: The tripartite information $I_{3, n}$ and 
\begin{align}
\GM[4]_n(A:B:C:D)|_{c_2=0}=S_n^{(4)}[1:1:1:1]-\frac{1}{3}S_n^{(3)}[2:1:1]+\frac{1}{3}S_n^{(2)}[3:1]   \,.
\end{align}
It was already pointed out by \cite{Balasubramanian:2014hda} that the tripartite information $I_3 :=  I_{3,n = 1}$ can be used to detect genuine quadripartite entanglement. Here we obtained the same conclusion from a different viewpoint for the $n \to 1$ case: using the genuine multi-entropy for $\mathtt{q}=4$, eq.~(\ref{I30}).
Our construction of $\GM[4]_n(A:B:C:D)$ naturally leads to $I_{3, n}$ as a diagnostic for genuine quadripartite entanglement for the generic R\'enyi case. In addition, $\GM[4]_n(A:B:C:D)$ gives another diagnostic $\GM[4]_n(A:B:C:D)|_{c_2=0}$ for genuine quadripartite entanglement.
To verify that these two quantities indeed serve as independent measures of quadripartite entanglement, and to provide some physical intuition for them, we evaluated $I_{3}$ and $\GM[4]_{n=2}(A:B:C:D)\big|_{c_2=0}$ for both GHZ and various other four-qubit states in Appendix~\ref{app:fourqubit}.

\subsection{$\GM[5]_n$ and new diagnostics of genuine pentapartite entanglement}
\label{newdiagnosticsforq5}

The expression of $\GM[5]_n(A:B:C:D:E)$ can be reorganized as follows.
\begin{align}
\GM[5]_n(A:B:C:D:E)&=\GM[5]_n(A:B:C:D:E)|_{c_5=0}\nonumber \\
& \quad +c_5\frac{\partial}{\partial c_5}
\left( \GM[5]_n(A:B:C:D:E) \right),\label{q5GMc5}\\
\GM[5]_n(A:B:C:D:E)|_{c_5=0}&=S_n^{(5)}[1:1:1:1:1]-\frac{1}{4}S_n^{(4)}[2:1:1:1]+\frac{1}{10}S_n^{(3)}[2:2:1]\notag\\
&\quad +\frac{1 }{20}S_n^{(3)}[3:1:1]-\frac{1 }{20}S_n^{(2)}[3:2],\label{q5GM1}\\
\frac{\partial}{\partial c_5}
\left( \GM[5]_n(A:B:C:D:E) \right)&=\frac{ 2}{5}S_n^{(3)}[2:2:1]
-\frac{4}{5}S_n^{(3)}[3:1:1]-\frac{1}{5}S_n^{(2)}[3:2]+S_n^{(2)}[4:1].\label{q5GM2}
\end{align}
As well as eq.~(\ref{I30}), we obtain
\begin{align}
&\frac{\partial}{\partial c_5}
\left( \GM[5]_n(A:B:C:D:E) \right)=0\,,\label{I40}
\end{align}
for all $\tilde{\mathtt{q}}$-partite entangled states with $\tilde{\mathtt{q}}<5$. For instance, consider the quadripartite entangled state in eq.~(\ref{festate1}). For this state, by using eqs.~\eqref{q5rs2},\eqref{q5rs3},\eqref{q5rs4},\eqref{p5id},
one can explicitly confirm eq.~(\ref{I40}). Therefore, for $\mathtt{q}=5$, we have $( p(5)-2 ) -  (p(4)-1)+1=2$ independent diagnostics to detect genuine pentapartite entanglement: 
\begin{align}
\label{q5twodiagnostics}
\GM[5]_n(A:B:C:D:E)|_{c_5=0} \quad \mbox{and} \quad \frac{\partial}{\partial c_5}
\left( \GM[5]_n(A:B:C:D:E) \right),
\end{align}
given by eqs.~\eqref{q5GM1} and \eqref{q5GM2} respectively. 
Note that one cannot construct such diagnostics for $\mathtt{q}=5$ by using only bipartite measures $S_n^{(2)}[3:2]$ and $S_n^{(2)}[4:1]$. Therefore, by using the Rényi multi-entropy with $\mathtt{q}\ge2$, one can construct new diagnostics of genuine pentapartite entanglement. 

Later in Sec.~\ref{q5holography}, we will examine each of these independent measures of pentapartite entanglement in a holographic setting.

\subsection{Black hole genuine $\mathtt{q}=5$ multi-entropy curve}

As a generalization of $\mathtt{q}=3,4$ results in \cite{Iizuka:2025ioc}, we study the black hole genuine $\mathtt{q}=5$ multi-entropy curve for several values of $c_5$. Here, we approximate an evaporating black hole and its Hawking radiation with a Haar-random state, where the total system is divided into four Hawking radiation subsystems $A=\text{R1}$, $B=\text{R2}$, $C=\text{R3}$, $D=\text{R4}$, and an evaporating black hole subsystem $E=\text{BH}$. To evaluate $\GM[5](\text{R1:R2:R3:R4:BH})$, which is $n \to 1$ limit of $\GM[5]_n(\text{R1:R2:R3:R4:BH})$, we use the asymptotic behaviors of multi-entropy at $d_\text{R}\ll d_\text{BH}$ and $d_\text{R}\gg d_\text{BH}$ studied by \cite{Iizuka:2024pzm}, where we set $\dim\mathcal{H}_\text{R1}=\dim\mathcal{H}_\text{R2}=\dim\mathcal{H}_\text{R3}=\dim\mathcal{H}_\text{R4}=d_\text{R}$, $\dim\mathcal{H}_\text{BH}=d_\text{BH}$.

\begin{figure}[t]
    \centering
    \includegraphics[width=15cm]{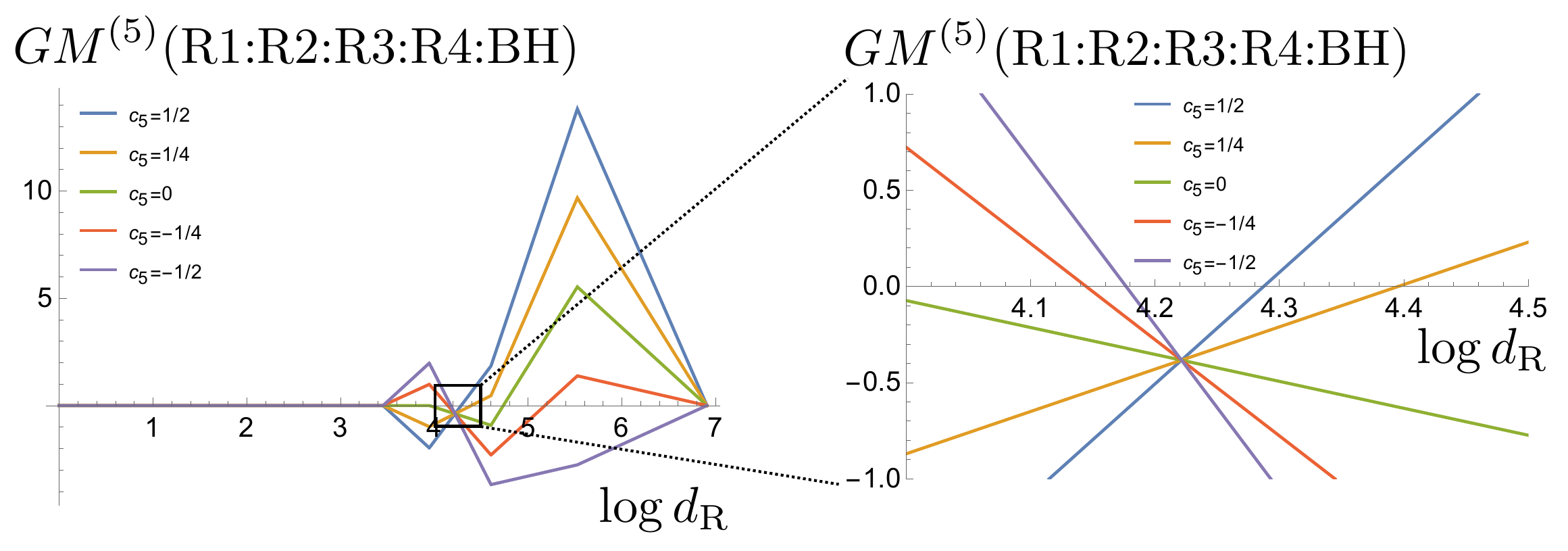}
    \caption{Asymptotic behaviors of $\GM[5](\text{R1:R2:R3:R4:BH})$ for a Haar-random state with several values of $c_5$, where $\dim\mathcal{H}_\text{R1}=\dim\mathcal{H}_\text{R2}=\dim\mathcal{H}_\text{R3}=\dim\mathcal{H}_\text{R4}=d_\text{R}$, $\dim\mathcal{H}_\text{BH}=d_\text{BH}$, and we fix $d_{\rm Total} = d_{\rm BH}d_{\rm R}^{4}=10^{12}$. The right figure is an enlarged plot around $\log d_\text{R}\sim4.2$.}
    \label{fig:GMq5Random}
\end{figure}

Figure \ref{fig:GMq5Random} shows the black hole genuine $\mathtt{q}=5$ multi-entropy curve for several values of $c_5$, where the total dimension is fixed as $d_{\rm Total} = d_{\rm BH}d_{\rm R}^{4}=10^{12}$. 
Similar to the $\mathtt{q}=3, 4$ cases \cite{Iizuka:2025ioc}, $\mathtt{q}=5$ genuine multi-entropy is almost zero until the Page time. 
One can see that both positive and negative regions of $\GM[5](\text{R1:R2:R3:R4:BH})$ exist. In particular, at $\log d_\text{R} \approx 4.22$, $\GM[5](\text{R1:R2:R3:R4:BH})$ is negative regardless of the value of $c_5$. Due to the $c_5$-dependence of the expression in eq.~(\ref{q5GMc5}), this $c_5$-independent point is given by a solution of  
\begin{align}
\frac{\partial}{\partial c_5} \left( \GM[5](A:B:C:D:E) \right) =0 \quad \left(\mbox{at $d_\text{R} \approx 4.22$}\right). 
\end{align}
This result means that there is no parameter region of $c_5$ where $\GM[5](\text{R1:R2:R3:R4:BH})$ is always nonnegative, which is different from the $\mathtt{q}=3,4$ results in \cite{Iizuka:2025ioc}. 
This suggests that the sign of $\GM[\mathtt{q}]$ may not carry a definite physical meaning. We will comment further on this point in the next subsection. 

\subsection{Generic natures of genuine multi-entropy $\GM[\mathtt{q}]_n$}
\label{N1diagnosticsforgqe}
\subsubsection{The number of parameters $N(\mathtt{q})$  in genuine multi-entropy $\GM[\mathtt{q}]_n$}
\label{howmanyNq}

\begin{figure}[t]
    \centering
    \includegraphics[width=10cm]{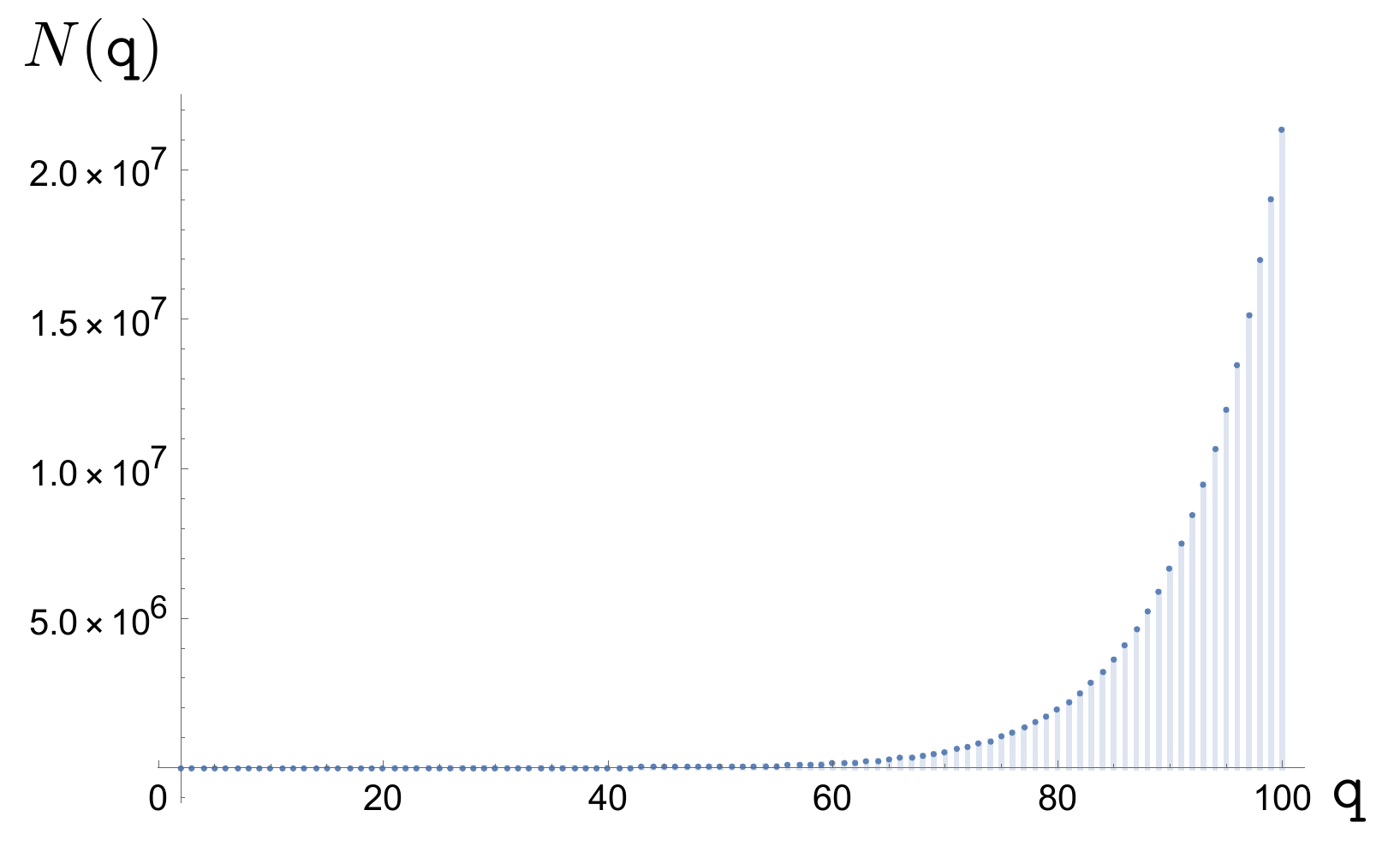}
    \caption{$\mathtt{q}$-dependence of $N(\mathtt{q})$ given by eq.~(\ref{Nq}).}
    \label{fig:Nq}
\end{figure}

Generically, we impose constraints so that genuine $\mathtt{q}$-partite Rényi multi-entropy vanishes for states with $(\mathtt{q}-1)$-partite entanglement such as
\begin{align}
|\psi\rangle=|\psi_{\mathtt{q}-1}\rangle_{A_1A_2\dots A_{\mathtt{q}-1}}\otimes |\psi_1\rangle_{A_\mathtt{q}}.
\end{align}
Such constraints are determined by relationships between $S_n^{(\mathtt{a})}$ and $S_n^{(\mathtt{a}-1)}$ for $\mathtt{a}=3,4,\dots,\mathtt{q}$, and the number of these constraints is given by $p(\mathtt{q}-1)-1$, where $p(\mathtt{q}-1)$ represents how to divide $\mathtt{q}-1$ subsystems $A_1,A_2,\dots,A_{\mathtt{q}-1}$.
Thus, from $p(\mathtt{q})-2$ coefficients in $\GM[\mathtt{q}]_n$, $p(\mathtt{q}-1)-1$ parameters are determined by the constraints and we are left with $N(\mathtt{q})$ parameters where 
\begin{align}
N(\mathtt{q}) = ( p(\mathtt{q})-2 ) -  (p(\mathtt{q}-1)-1) =p(\mathtt{q}) -  p(\mathtt{q}-1 ) -1  \,.
\label{Nq} 
\end{align}
The explicit values of $N(\mathtt{q})$ for small $\mathtt{q}$ are
\begin{align}
    N(3)=0, \;\;\; N(4)=1, \;\;\; N(5)=1, \;\;\; N(6)=3, \;\;\; N(7)=3, 
\end{align}
which are consistent with the expressions of genuine $\mathtt{q}$-partite Rényi multi-entropy for $\mathtt{q}=3,4,5$.  We plot the $\mathtt{q}$-dependence of $N(\mathtt{q})$ in Figure \ref{fig:Nq}, which increases exponentially as the number of subsystems $\mathtt{q}$ increases due to the Hardy–Ramanujan asymptotic formula \cite{Hardy:1918kpu}
\begin{align}
p(\mathtt{q})\sim\frac{1}{4\mathtt{q}\sqrt{3}}e^{\pi\sqrt{\frac{2\mathtt{q}}{3}}} \quad (\mathtt{q}\to\infty)\,.
\end{align}

The large number of free parameters in genuine $\mathtt{q}$-partite Rényi multi-entropy can be interpreted as follows. It is well known that, in three-qubit systems, there are two inequivalent classes of pure states with genuine tripartite entanglement \cite{Dur:2000zz}, where the GHZ state and the W state are two remarkable representatives of them. As the number of subsystems $\mathtt{q}$ increases, so does the number of such inequivalent classes of pure states with genuine multi-partite entanglement. The increase of the number of free parameters in genuine $\mathtt{q}$-partite Rényi multi-entropy would be related to the existence of many such inequivalent classes.

\subsubsection{$N(\mathtt{q})+1$ diagnostics for genuine $\mathtt{q}$-partite entanglement from $\GM[{\mathtt{q}}]_n$ }
As given by eq.~(\ref{Nq}), genuine $\mathtt{q}$-partite Rényi multi-entropy has $N(\mathtt{q})$ free parameters. Suppose that we determine these parameters $a_l$ for $l=1,\dots,N(\mathtt{q})$ by solving the constraints on each coefficient\footnote{These free parameters $a_l$ are given by linear combinations of $c_i$ in the previous subsections.} in $\GM[{\mathtt{q}}]_n$. Since $\GM[{\mathtt{q}}]_n$ is a linear expression of $a_l$, we can express $\GM[{\mathtt{q}}]_n$ as
\begin{align}
\label{linearcombGMa}
\GM[{\mathtt{q}}]_n=\GM[{\mathtt{q}}]_n|_{a_l=0}+\sum_{l=1}^{N(\mathtt{q})} a_l \frac{\partial}{\partial a_l} \left( \GM[{\mathtt{q}}]_n\right) . 
\end{align}
For any values of $a_l$, this expression vanishes for all $\tilde{\mathtt{q}}$-partite entangled states with $\tilde{\mathtt{q}}<\mathtt{q}$. Therefore, we obtain $N(\mathtt{q})+1$ diagnostics  $\{D_l\}$ for genuine $\mathtt{q}$-partite entanglement: 
\be
\frac{\partial}{\partial a_l}\left( \GM[{\mathtt{q}}]_n\right) \quad \mbox{(for $l=1,\dots,N(\mathtt{q})$)}
\quad 
\mbox{and} \quad \GM[{\mathtt{q}}]_n|_{a_l=0} .
\ee
These diagnostics are all independent because we impose the independent $p(\mathtt{q}-1)-1$ constraints corresponding to the integer partition of $\mathtt{q}-1$. We emphasize that our construction of these diagnostics is {\it quite generic and systematic for any positive integer $\mathtt{q}\ge3$}. Therefore, no matter how large $\mathtt{q}$ is, one can construct several kinds of diagnostics to detect genuine $\mathtt{q}$-partite entanglement by using genuine $\mathtt{q}$-partite Rényi multi-entropy $\GM[{\mathtt{q}}]_n$.

In this way, it is natural to regard $\GM[{\mathtt{q}}]_n$ not as a single privileged measure but as a general linear combination of independent diagnostics $\{D_l\}$ that vanish on all $\tilde{\mathtt{q}}< \mathtt{q}$ states as \eqref{linearcombGMa}. Accordingly, the coefficients $\{a_l\}$ have no intrinsic physical content beyond a basis choice; different bases are related by invertible linear transformations, and an overall rescaling may even flip the sign. From this perspective, we do not impose global non-negativity: a negative value simply reflects the chosen linear combination, rather than the absence of genuine $\mathtt{q}$-partite entanglement.

\section{Genuine multi-entropy in holography}
\label{sec:holographicGM}
\begin{figure}[t]
  \centering
  \includegraphics[scale=.33]{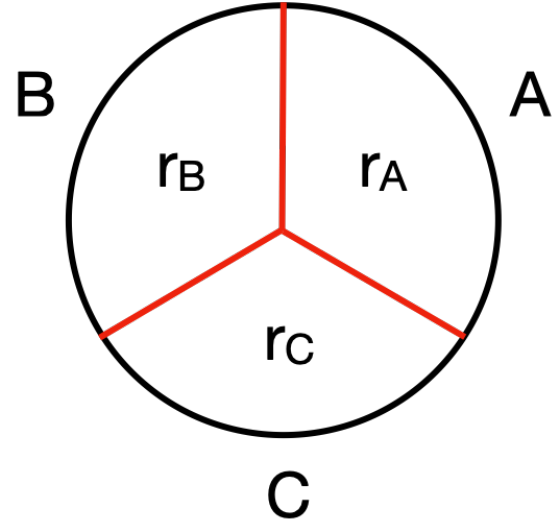}
  \hspace{2cm}
  \includegraphics[scale=.33]{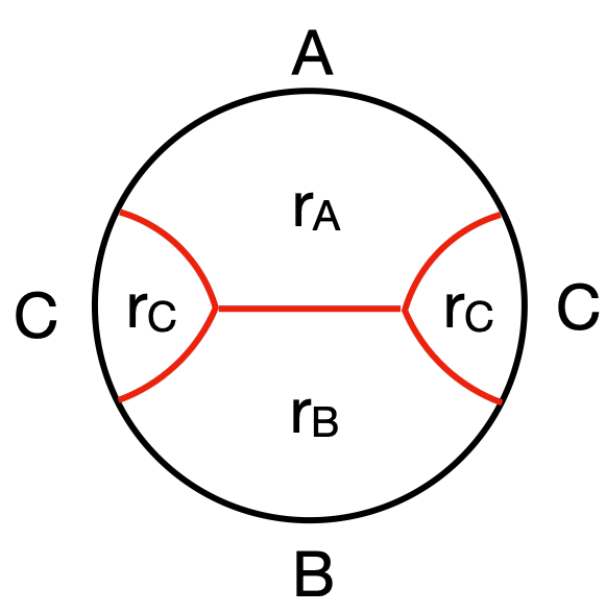}
  \caption{(Left) A minimal triway cut. (Right) When the boundary regions are disconnected (in this example it is region C), the minimal triway cut may be given by a higher-partite cut instead of a 3-cut. Note that in both of the examples given, the cut interfaces (red lines) always meet at equiangular trivalent vertices. This is a generic feature for all minimal multiway cuts.}
  \label{fig:triway}
\end{figure}

In this section, we shift our focus to holography.
Consider a constant time slice $\Sigma$ of a static d-dimensional AdS space. We divide its asymptotic boundary $\partial\Sigma$ into $\mathtt{q}\in \mathbb{N}$ boundary subregions $R_1,R_2,\cdots,R_{\mathtt{q}}$. We want to evaluate the genuine multi-entropy $\GM[q](R_1:R_2:\cdots:R_\mathtt{q})$ on the CFT state dual to $\Sigma$. We have seen in Sec.~\ref{sec:GMq} that $\GM[q](R_1:R_2:\cdots:R_\mathtt{q})$ can be decomposed into a linear combination of multi-entropies $\S[q]$ with $\mathtt{p}\le \mathtt{q}$. We will make use of the fact that $\S[p]$ has a proposed holographic dual\footnote{Since the definition of multi-entropy involves an analytic continuation in $n$, the argument consists of a proposed dual of Renyi multi-entropies and an analytically continuation $n\to1$. Actually checking the proposal is tricky due to the continuation. There are known counter examples for the proposed Renyi dual for $n\ge3$ in AdS/CFT \cite{Penington:2022dhr}, but it might still be true that the $n\to1$ proposal is valid. On the other hand, no such violations are known for holographic tensor networks \cite{Hayden:2016cfa,Pastawski:2015qua}. The absense of such violation allows the required $n\to1$ analytic continuation to be carried out properly, thus ensuring the validity of the holographic proposal of the multi-entropy. It has been argued  that these tensor network models are dual to ``fixed area states'' \cite{Akers:2018fow,Dong:2018seb} in AdS/CFT, so the holographic proposal is most likely still true provided we restrict to these fixed area states even if it fails for generic holographic states.}  as the area of (minimal) \emph{multiway cuts} \cite{Gadde:2022cqi,Gadde:2023zzj}, which we now define:

A \emph{$\mathtt{p}$-cut} is the division of the bulk $\Sigma$ into $\mathtt{p}$ connected subregions $r_1\sqcup r_2 \sqcup \cdots \sqcup r_\mathtt{p} = \Sigma$. Similarly, a \emph{$\mathtt{p}$-multiway cut} (or a \emph{$\mathtt{p}$-way cut}) is a division $r_1\sqcup r_2 \sqcup \cdots \sqcup r_\mathtt{p} = \Sigma$ where each $r_i$ is homologous to the corresponding boundary subregion $R_i$. Note that the subregions $r_i$ in the multiway cut need not be connected. It can contain disconnected components when some boundary subregions $R_i$ are disconnected. See Fig.~\ref{fig:triway} for examples.

Given a $\mathtt{p}$-cut $(r_1,r_2,\cdots,r_\mathtt{p})$ on $\Sigma$, the area of the cut is the $(d-2)$-dimensional volume of the interface separating the cut regions $r_i$, which we will denote as $\A[p](r_1:r_2:\cdots:r_\mathtt{p})$. For $\mathtt{p}$-way cuts based on the boundary region $R_1,R_2,\cdots,R_\mathtt{p}$, the \emph{minimal $\mathtt{p}$-way cut} minimizes the area over all the possible bulk divisions:
\begin{equation}
  \A[p](R_1:R_2:\cdots:R_\mathtt{p}) \equiv \min_{\{r_i\}} \A[p](r_1:r_2:\cdots:r_\mathtt{p}), \quad R_i \subset r_i.
\end{equation}

We now define the holographic multi-entropy $\S[p]$ as the area of this minimal $\mathtt{p}$-way cut:
\begin{equation}
  \label{eq:Sp_holographic}
  \S[p](R_1:R_2:\cdots:R_\mathtt{p}) = \frac{1}{4G_N} \A[p](R_1:R_2:\cdots:R_\mathtt{p}).
\end{equation}
This equivalence between the multi-entropy and minimal mutliway cuts was first suggested in \cite{Gadde:2022cqi} and further expanded upon in \cite{Gadde:2023zzj,Gadde:2024taa}. See also \cite{Harper:2024ker} for partial results on verifying the proposal in 2d CFTs.

Our main goal in this section is to show that the holographic genuine multi-entropies are non-zero for generic partitions of the boundary\footnote{By \eqref{eq:GM_holographic} we mean that $\GM[q]= O(1/G_N)$ unless one fine-tunes the sizes of $R_i$ to some special values.}, {\it i.e.}, 
\begin{equation}
  \label{eq:GM_holographic}
  \GM[q](R_1:R_2:\cdots:R_\mathtt{q}) = O\left(\frac{1}{G_N}\right) \quad (\text{for holographic states})
\end{equation}
in vacuum AdS. For $\mathtt{q}>3$, $\GM[q]$ can be decomposed into different independent measures of multi-partite entanglement. We expect that \eqref{eq:GM_holographic} holds for at least one of them. Note that it already implies the existence of large multi-partite entanglement in holographic CFT states.

For the ease of computation we will restrict ourselves to $d=3$, and where the bulk constant time slice $\Sigma$ is isometric to the hyperbolic disk. 
For this particular setting, it has been proven \cite{7f59d6a7-717a-33f0-b433-de2c5f4f8dd4,Cotton2002} that the solutions of the minimal multiway cuts are given by joined segments of geodesics that separate the boundary regions apart. Furthermore, the geodesics can only intersect each other at equiangular trivalent vertices. See examples in Fig.~\ref{fig:triway}.
Configurations where four or more geodesics intersecting at the same vertex are never minimal.

Despite only working with $d=3$ in this section, we nonetheless expect a similar result to hold in higher dimensions. This is closely related to the existence of the nontrivial trivalent vertices (or their higher-dimensional equivalence) of the minimal surfaces. See Sec.~\ref{sec:dis} and Ref.\cite{Iizuka:2025bcc} for discussions and implications on the holographic error corrections. Note that for spacetimes with an black hole or more than one asymptotic boundaries there may be large parametric regions where $\GM[q]=0$, which we briefly touch on in Sec.~\ref{sec:dis}. 

In what follows we will give explicit examples of \eqref{eq:GM_holographic} for $\mathtt{q} = 3,4,5$, both analytically and numerically in the $d=3$ setting. The derivations of some results in this section are rather technical, so we relegate them to Appendix~\ref{app:analytic} for interested readers.

One peculiar property we observe from the examples considered in this section is that the holographic genuine multi-entropies $\GM[q]$ are all \emph{UV-finite} and does not depend on the cutoff scale $\epsilon$.
  We present a proof of this property in Appendix \ref{app:UV_cancellation}. Note that our proof requires complete knowledge of the linear coefficients of $\S[p]$ in genuine multi-entropies, which we have only worked out for $\mathtt{q}\le 5$ in this paper. It is however reasonable to believe that UV-finiteness is a universal and remains true for all $\mathtt{q}>5$. This property has interesting implications on the structure of entanglement in holographic CFTs which we elaborate more in Sec.~\ref{sec:dis}.

\subsection{$\mathtt{q}=3$ case}
\label{holographicq=3}
As a quick warmup, let's consider the case where the boundary $\partial\Sigma$ is divided into three subregions $(A,B,C)$. 
The holographic $\mathtt{q}=3$ genuine multi-entropy is given by
\begin{equation}
  \label{eq:GM3_holo}
  4G_N\GM[3](A:B:C) = \A[3](A:B:C) - \frac{1}{2}\Bigl(\mathcal{A}(A) + \mathcal{A}(B) + \mathcal{A}(C) \Bigr),
\end{equation}
where we have abbreviated $\mathcal{A}(A) \equiv \A[2](A:A^c) = \A[2](A:BC)$ for 2-cuts.

\begin{figure}[t]
  \centering
  \includegraphics[width=.9\linewidth]{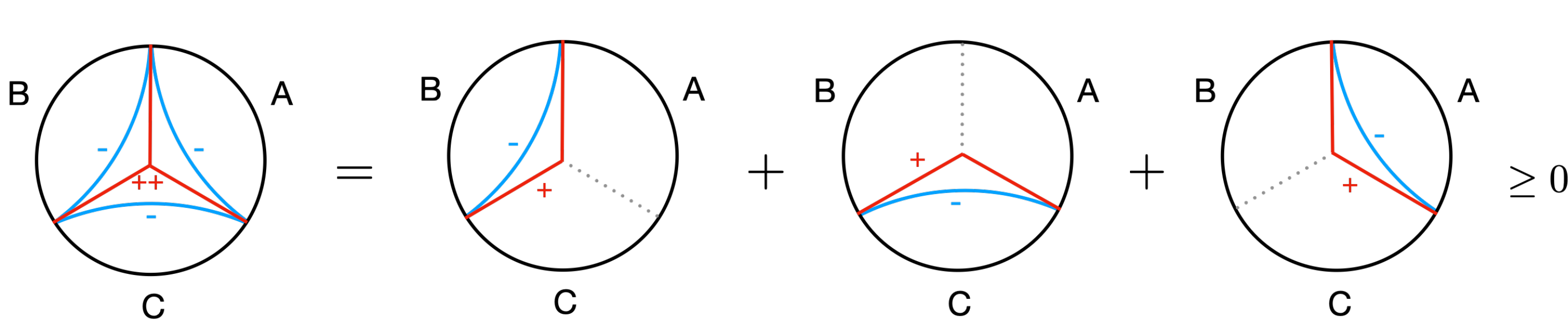}
  \caption{A geometrical proof of $\GM[3]\ge0$ \cite{Iizuka:2025ioc}. Geodesics colored in red come with plus signs and geodesics colored in blue come with minus signs. We split the red triway cut into three 2-cuts and combine with the corresponding three blue minimal cuts. Applying triangle inequality on each term of the RHS we can show that $\GM[3]\ge 0$.}
  \label{fig:tri_decomp}
\end{figure}

It is straightforward to argue that the linear combination of \eqref{eq:GM3_holo} is positive semidefinite. As shown in Fig.~\ref{fig:tri_decomp}, the 3-cut $\A[3](A:B:C)$ can be split into the sum of three 2-cuts $r_A,r_B,r_C$, each homologous to one boundary region. We have $\mathcal{A}(r_A)\ge \mathcal{A}(A)$ since the latter is given by a minimal surface. For the other two regions, we proceed in the same fashion. Hence we have shown that $\GM[3] > 0$ for holographic states as long as $A,B,C$ are connected subregions \cite{Iizuka:2025ioc}.

In fact, it is easy to evaluate \eqref{eq:GM3_holo} analytically from simple hyperbolic geometric identities. One finds\footnote{See Appendix \ref{app:analytic} for details. The same result has been derived previously in \cite{Harper:2024ker}.}
\begin{equation}
  \GM[3](A:B:C) = \frac{3}{4G_N}\log\frac{2}{\sqrt{3}},
\end{equation}
independent of how we arranged the boundary subregions, as long as $A,B,C$ are connected subregions.
The constant behavior of $\GM[3]$ is a simple consequence of conformal symmetry. Since we know that $\GM[q]$ are UV-finite quantities which does not depend on the cutoff scale, $\GM[q]$ must be conserved under a global conformal transformation. In the case of three boundary points, we can use such transformation to lock the three boundary points to any configurations. Therefore the value of $\GM[3]$ cannot depend on the boundary points and has to be a constant.

\subsection{$\mathtt{q}=4$ case}
\label{holographicq=4}
We now move on the the quadripartite case where the boundary is divided into four subregions $(A,B,C,D)$.
The form of $\GM[4]$ is given by eq.~\eqref{GMq4}, which we reproduce here:
\begin{align}
  \GM[4](A:B:C:D)|_a &= \S[4][1:1:1:1] - \tfrac{1}{3}\S[3][2:1:1] + a\S[2][2:2] + (\tfrac{1}{3}-a)\S[2][3:1] \nonumber \\
    \label{eq:GM4_I3}
  &=\GM[4](A:B:C:D)|_{a=\frac{1}{3}} - (\tfrac{1}{3}-a) I_3(A:B:C).
\end{align}
There is one free parameter $a:= c_2$. 
Here
\begin{equation}
\label{anotherdiago}
  \GM[4](A:B:C:D)|_{a=\frac{1}{3}} = \S[4][1:1:1:1] - \tfrac{1}{3}\S[3][2:1:1] + \tfrac{1}{3}\S[2][2:2]
\end{equation}
is independent of $a$ and
\begin{equation}
  I_3(A:B:C) = \S[2][2:2] - \S[2][3:1]
\end{equation}
is the tripartite information.  We will see both quantities are UV-finite by construction. Note that if any of the two is non-zero, \eqref{eq:GM4_I3} will be nonzero for some $a$ and thus we can deduce that there are non-trivial quadripartite entanglement in our state of interest. 

It is known that the tripartite information is negative, $I_3(A:B:C)\le0$ for holographic states \cite{Hayden:2011ag}, thus the second term on the RHS of \eqref{eq:GM4_I3} is semipositive when $a\ge1/3$.
We will show that the other contribution in \eqref{eq:GM4_I3} is also positive holographically, namely
\begin{equation}
  \label{eq:GM4_holo_ineq}
  \GM[4](A:B:C:D)|_{a= \frac{1}{3}} \ge 0, \quad \text{(for holographic states)}.
\end{equation}
This implies that the quadripartite genuine multi-entropy $\GM[4](A:B:C:D)|_a$ is semipositive for $a\ge 1/3$.
Note that this semipositive property is for holographic states, and $\GM[4](A:B:C:D)|_{a\ge \frac{1}{3}}$ for generic states can be negative such as four-qubit examples in Appendix \ref{app:fourqubit}.

We now present a geometric proof for \eqref{eq:GM4_holo_ineq} in the setting where there is only one asymptotic boundary and all the subregions $(A,B,C,D)$ are connected. However, we have checked \eqref{eq:GM4_holo_ineq} on discrete geometries defined by simple graphs. This led us to believe that the inequality holds quite generally. For our record we write out the full expression of the holographic $\GM[4](A:B:C:D)|_{a= \frac{1}{3}}$ explicitly:
\begin{align}
  \label{eq:GM4a}
  \begin{split}
   &4G_N\GM[4](A:B:C:D)|_{a=\frac{1}{3}} \equiv \A[4](A:B:C:D) \\
&\quad-\frac{1}{3}\Big(\A[3](AB:C:D)+\A[3](AC:B:D) +\A[3](AD:B:C) \\
&\quad\quad+\A[3](BC:A:D)+\A[3](BD:A:C)+\A[3](CD:A:B)\Big)\\
&\quad+\frac{1}{3}\Big(\mathcal{A}(AB)+\mathcal{A}(AC)+\mathcal{A}(AD)\Big) .
  \end{split}
\end{align}

We begin by splitting \(\mathcal{A}^{(4)}(A:B:C:D)\) into the following linear combination
\begin{equation}
  \begin{matrix}
     \includegraphics[scale=0.33]{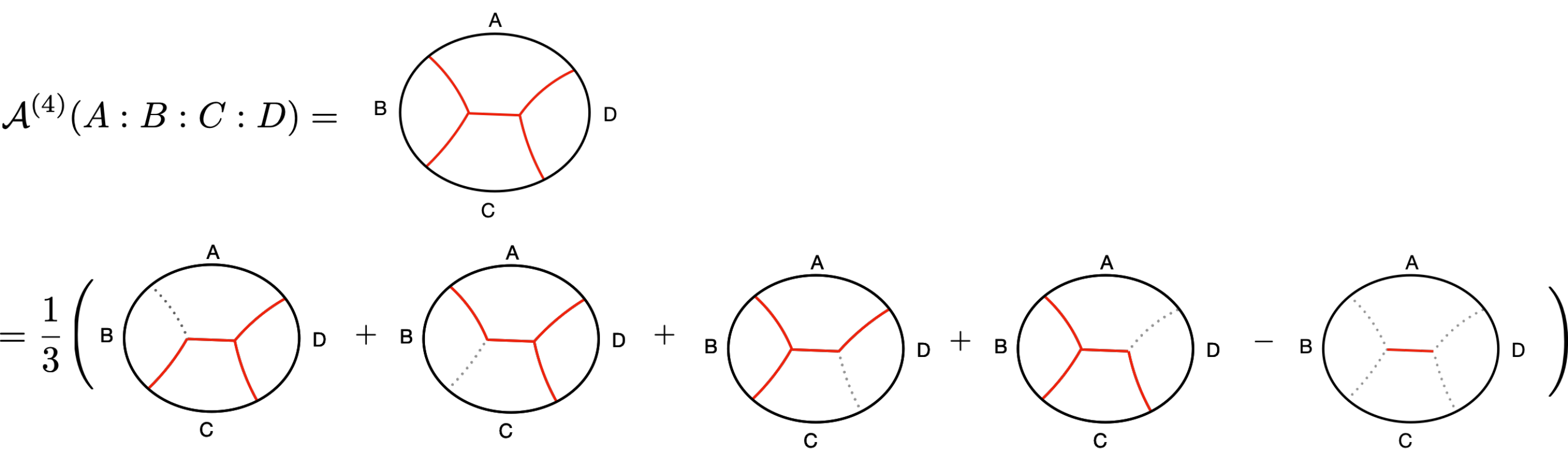}.
  \end{matrix}
\end{equation}
On the other hand, we have the following inequality regarding the tripartite multi-entropy:
\begin{equation}
  \begin{matrix}
   \includegraphics[scale=0.33]{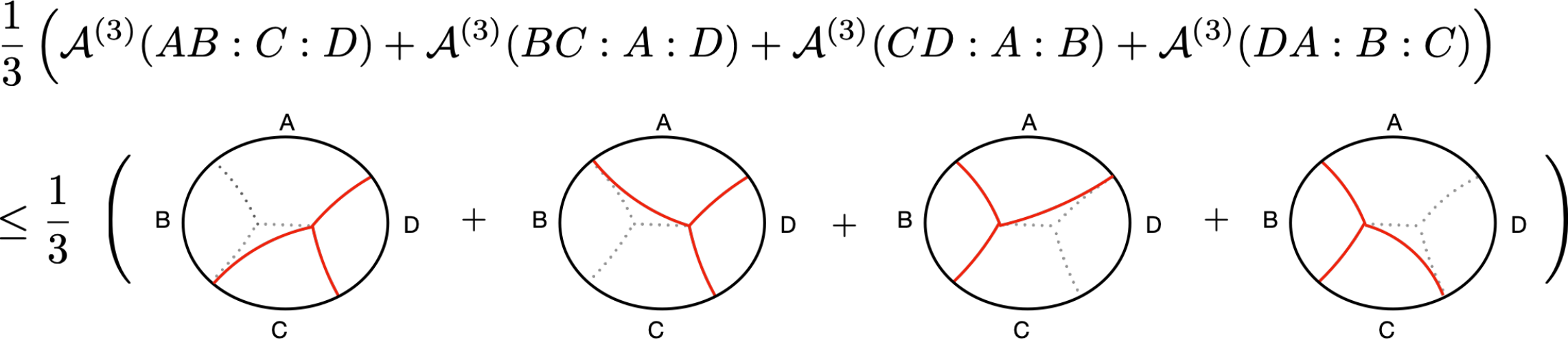},
  \end{matrix}
\end{equation}
which involves four out of the six \(\mathcal{A}^{(3)}\) triway cuts.
Taking the difference, we obtain the following bound
\begin{equation}
  \begin{matrix}
   \includegraphics[width=\linewidth]{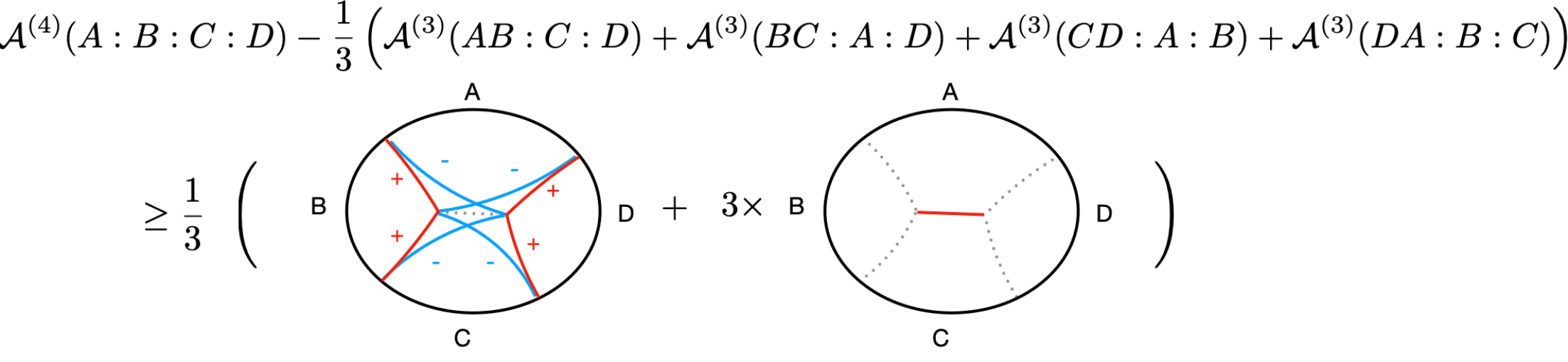},
  \end{matrix}
\end{equation}
where we used the blue color to indicate that the area of the surface contributes negatively.
Note that since 1) the geodesic legs are always perpendicular to the boundary 2) the contribution from blue and red legs originating from any boundary point comes with equal but opposite signs, the divergent contribution to the multi-entropy cancels out. This is one evidence that the $\mathtt{q}=4$ genuine multi-entropy is a UV-finite quantity, which we prove in Appendix \ref{app:UV_cancellation}. 

We have yet to deal with the remaining two \(\mathcal{A}^{(3)}\) involving disconnected regions, {\it i.e.}, \(\mathcal{A}^{(3)}(AC:B:D)+\mathcal{A}^{(3)}(BD:A:C)\).
The claim is that in the current configuration, at least one of them will be equal to \(\mathcal{A}(AC)\) or $\mathcal{A}(BD)$.
To see this, note that since the two regions \(AC\) and \(BD\) are disconnected, at least one of the entanglement wedge \({\rm EW}(AC)\) and \({\rm EW}(BD)\) will be connected.
Without loss of generality let's assume that it is \({\rm EW}(AC)\). The RT surface for \(S(AC)\) is then also a minimal surface \(\mathcal{A}^{(3)}(AC:B:D)\), so we have
\begin{equation}
  \mathcal{A}^{(3)}(AC:B:D) = \mathcal{A}(AC) = \mathcal{A}(BD).
\end{equation}
Therefore,
\begin{equation}
  \begin{matrix}
   \includegraphics[scale=0.37]{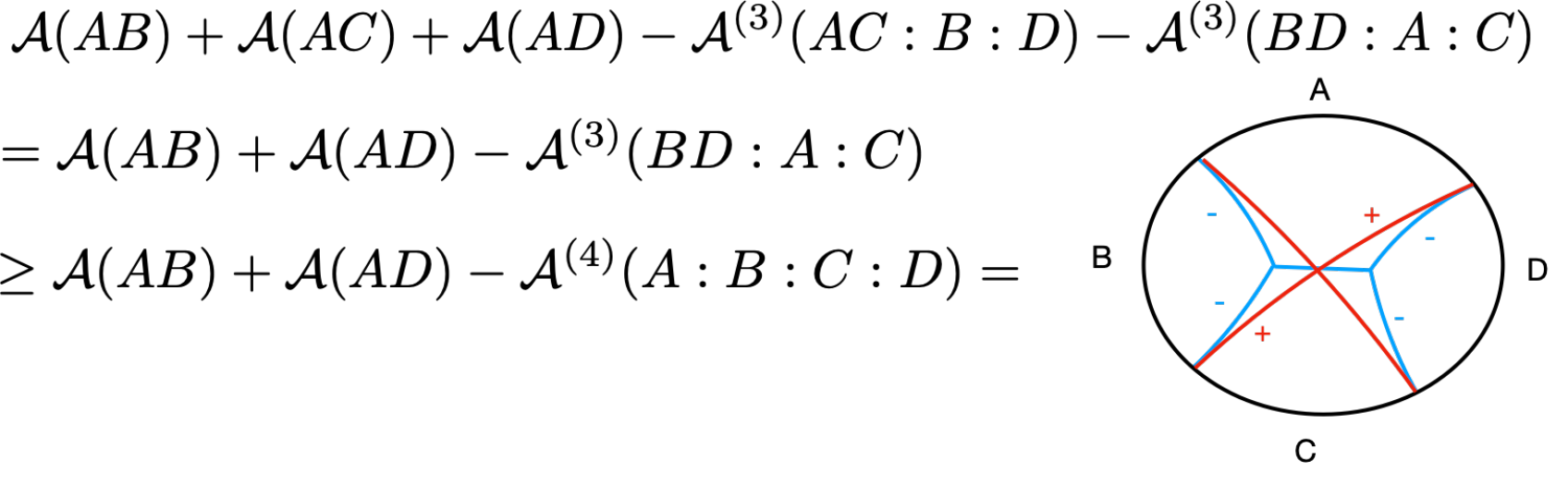},
  \end{matrix}
\end{equation}
where we used the relation \(\mathcal{A}^{(4)}(A:B:C:D)\ge \mathcal{A}^{(3)}(BD:A:C)\) in the last line.

Putting what we have so far together, we now have a bound for \(\rm GM^{(4)}\):
\begin{equation}
  \begin{matrix}
   \includegraphics[scale=0.35]{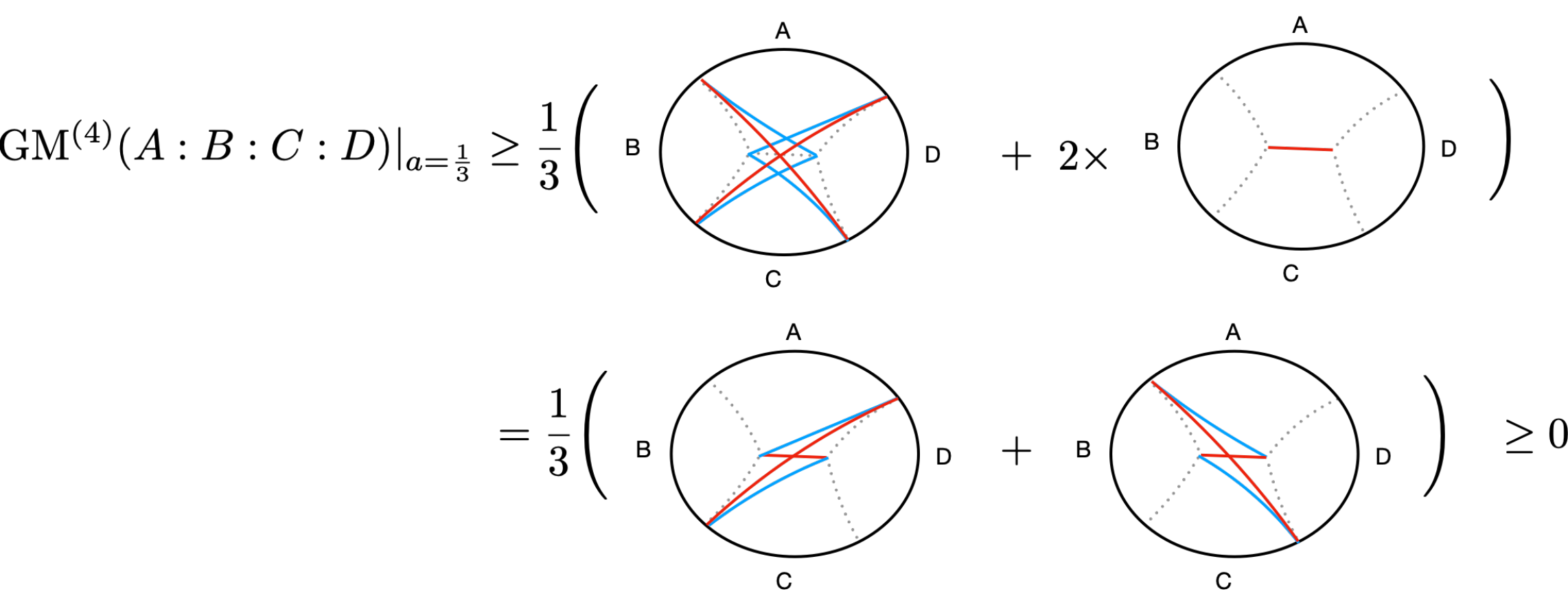} 
  \end{matrix}
\end{equation}
What we have done is to write $\GM[4]|_{a=\frac{1}{3}}$ as the sum of two different geometries as in shown in the second line.
The contribution of each of the geometry is positive simply by triangle inequality. 
Therefore, we have proven \eqref{eq:GM4_holo_ineq}.

To verify our claim, we analytically calculated the holographic genuine multi-entropy for vacuum AdS$_3$ and show that it is indeed positive in general. It is given by
\begin{align}
  \label{eq:GM4analytic_m}
  \GM[4](\eta)|_{a=\frac{1}{3}}  =\begin{cases}
     \frac{1}{4G_N}\left(\frac{3}{4}\log\left(\frac{1+\sqrt{1-\eta}}{2}\right)-\frac{2}{3}\log\sqrt{1-\eta}\right),\quad &0 \le \eta < \eta^*, \\
     \frac{1}{4G_N}\left(2\log\frac{2}{\sqrt{3}}+\frac{1}{3}\log \eta+2\log\left(\frac{1+\sqrt{1-\eta}}{2}\right)-\frac{4}{3}\log(1-\eta)\right), &\eta^* \le \eta \le \frac{1}{2},
 \end{cases}
\end{align}
where $\eta=\frac{(t_1-t_2)(t_3-t_4)}{(t_1-t_3)(t_2-t_4)}$ is the cross-ratio for the four boundary points\footnote{We choose a cyclic ordering of $(t_1,t_2,t_3,t_4)$ such that $0\le\eta\le1$, where $t_i = e^{i \theta_i}$.} $(t_1,t_2,t_3,t_4)$ separating the subregions $(A,B,C,D)$ and $\eta^*\approx0.265$. For $\frac{1}{2}<\eta\le1$, simply replace $\eta\to1-\eta$ in \eqref{eq:GM4analytic_m}. See Fig.~\ref{fig:GM4} for a plot. Detailed derivation of \eqref{eq:GM4analytic_m} is given in Appendix \ref{app:analytic}.
\begin{figure}[t]
  \centering
  \includegraphics[width=.5\textwidth]{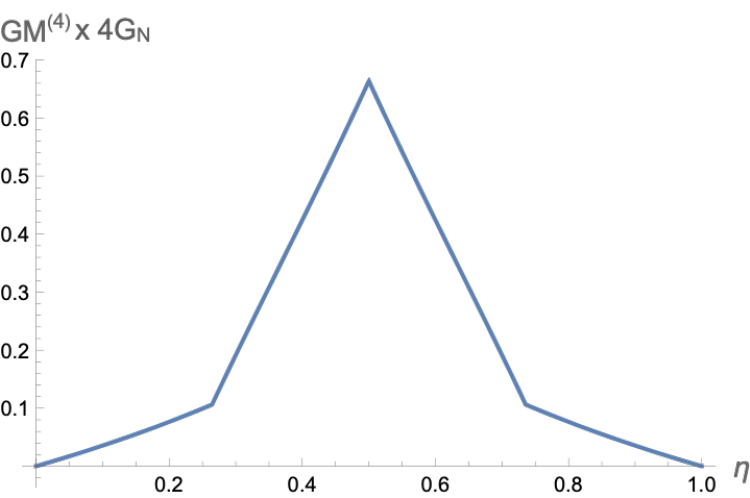}
  \caption{The holographic values of $\GM[4]|_{a=1/3}$ in vacuum AdS$_3$, plotted against the conformal cross-ratio $\eta$ of the four points separating the boundary regions. We see explicitly that $\GM[4]|_{a=1/3}>0$ except for when $\eta=0,1$, where we are in the coincident limit and one of the regions vanishes.}
  \label{fig:GM4}
\end{figure}

\subsection{$\mathtt{q}=5$ case}
\label{q5holography}
\begin{figure}[t]
  \centering
  \includegraphics[width=.49\linewidth]{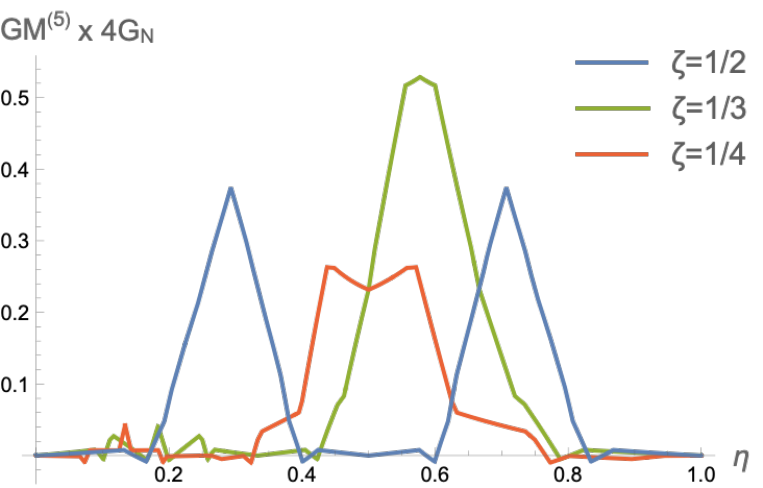}
  \includegraphics[width=.49\linewidth]{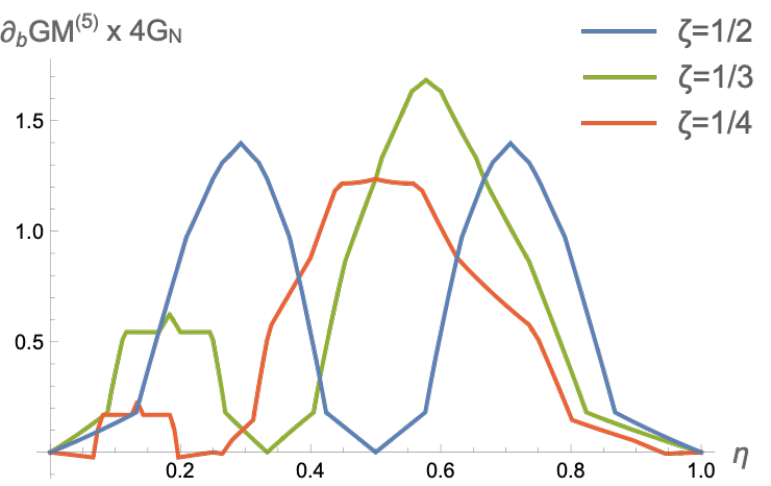}
  \caption{The numerical evaluation of: $\GM[5]|_{b=0}$ (left); $\partial_b\GM[5]$ (right) , in vacuum AdS$_3$. There are five boundary separation points, which leaves us with two remaining degrees of freedom, which we take to be the the conformal cross ratios $(\eta,\zeta)$ of the first two points with respect to the remaining three. More explicitly, $\eta=\frac{(t_1-t_3)(t_4-t_5)}{(t_1-t_4)(t_3-t_5)}$ and $\zeta=\frac{(t_2-t_3)(t_4-t_5)}{(t_2-t_4)(t_3-t_5)}$ where $(t_1,t_2,t_3,t_4,t_5)$ are the five boundary points separating $(A,B,C,D,E)$.}
  \label{fig:GM5}
\end{figure}

For pentapartite divisions, the genuine multi-entropy is given by eq.~\eqref{GM5definition}:
\begin{align}
  \label{eq:GM5_holo}
  \begin{split}
   \GM[5](A:B:C:D:E) &= \S[5][1:1:1:1:1] - \frac{1}{4}\S[4][2:1:1:1] + \frac{1+4b}{10}\S[3][2:2:1] \\
  &\qquad + \frac{1-16b}{20}\S[3][3:1:1] - \frac{1+4b}{20}\S[2][3:2] + b\S[2][4:1].
  \end{split}
\end{align}
There is one free parameter $b := c_5$. While we don't have a geometric proof or an analytical expression for $\GM[5]$ for generic subregions, we do except the quantity defined in \eqref{eq:GM5_holo} to be non-zero in general. We give evidence for our claim by numerically evaluating two diagnostics given by \eqref{q5twodiagnostics}, {\it i.e.,} $\GM[5]|_{b=0}$ and the part of \eqref{eq:GM5_holo} that depends on $b$, 
\begin{equation}
  \label{eq:GM5b_holo}
\partial_b \GM[5]  = \frac{1}{5} \left( 2\S[3][2:2:1]  - 4\S[3][3:1:1] - \S[2][3:2] + 5\S[2][4:1]  \right)
\end{equation}
and showing that they are non-zero in general. See Fig.~\ref{fig:GM5}. As we mentioned in Sec. \ref{newdiagnosticsforq5}, despite only consisting of lower partite multi-entropies, \eqref{eq:GM5b_holo} is also a valid diagnostic of pentapartite entanglement (Same as $I_3$, a quantity comprised of only $\S[2]$, can be used to diagnose quadripartite entanglement).

In contrast to the $\mathtt{q}=4$ case, we observe that the value of $\GM[5]$ can take negative values on the hyperbolic disk. See Fig.~\ref{fig:GM5}. This is still fine since a non-zero value of genuine multi-entropy already signifies the presence of multi-partite entropy. In $\mathtt{q}=3,4$ examples above, the positivity of $\GM[3]$ and $\GM[4]|_a$ ($a \ge 1/3$) can be thought of as additional constraints holographic states have to satisfy, but for $\mathtt{q}\ge 5$ we do not have such constraints. Note that even for $\GM[4]|_a$, if we take $a \to - \infty$, $\GM[4]|_a$ becomes negative since $I_3 < 0$ for holographic states\footnote{This is similar to the story of the holographic entropy cone\cite{Bao:2015bfa}. It is interesting to investigate the analogy more for $\GM[q]$.}. 
This is consistent with our viewpoint that $\GM[\mathtt{q}]_n$ is not a single privileged measure but a general linear combination of independent diagnostics that vanish on all $\tilde{\mathtt{q}}<\mathtt{q}$ states, as we discussed in Sec. \ref{N1diagnosticsforgqe}. 

\begin{figure}[t]
  \centering
  \includegraphics[width=.6\linewidth]{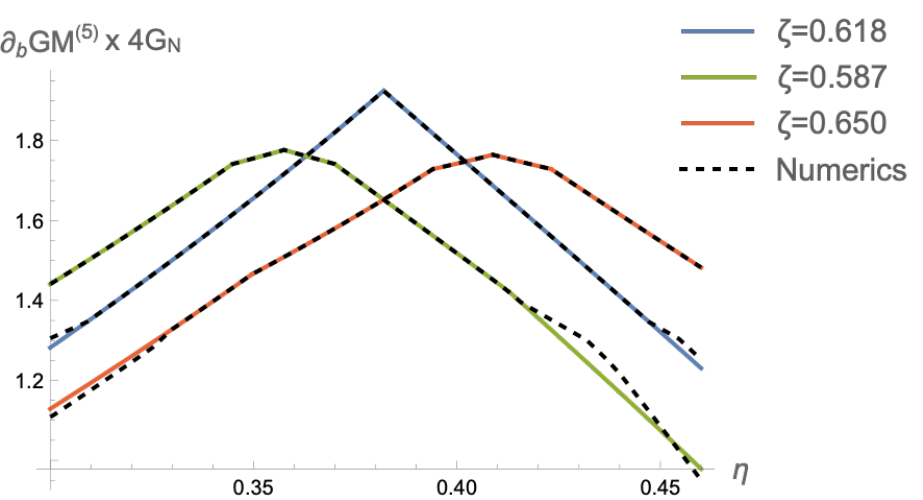}
  \caption{We plot the analytic expression for $\partial_b\GM[5]$ (given in \eqref{eq:GM5b_analytic_full} and \eqref{eq:GM5b_analytic_restricted}) versus the direct numerical evaluation near the parametric region where the sizes of the boundary regions are similar in size. The analytic formulas are shown in solid colored lines and the numerics are shown in dashed lines. The analytic formulas are exact as long as we are not too far from the equipartite case. In terms of the cross-ratios the equipartite point is $(\eta,\zeta) = (\frac{3-\sqrt{5}}{2},\frac{-1+\sqrt{5}}{2})\approx(0.382,0.618)$.}
  \label{fig:GM5b_analytic_cr}
\end{figure}

To close this section, we analytically computed the value of \eqref{eq:GM5b_holo} for the case where the five boundary regions are of equal sizes. Its value is given by
\begin{equation}
 \frac{\partial}{\partial b}\GM[5](|A|=|B|=|C|=|D|=|E|) = \frac{1}{4G_N}\left(4\log\sin(2\pi/5)-4\log\sin(\pi/5)\right).
\end{equation}
We also computed the analytic expression for \eqref{eq:GM5b_holo} where the sizes of the five boundary regions are close to the equipartite case and checked that it matches with the numerics. See Fig.~\ref{fig:GM5b_analytic_cr}. The full analytic expression is lengthy and we list them in \eqref{eq:GM5b_analytic_full} and \eqref{eq:GM5b_analytic_restricted} of Appendix \ref{app:analytic}, along with their derivation.

Thus we conclude that for $\mathtt{q}=3, 4, 5$, the genuine multi-entropy is nonzero in general. These results lead us to the conjecture that the genuine multi-entropy is nonzero for all the values of $\mathtt{q} \ge 3$.

\section{Discussion}
\label{sec:dis}

As we mentioned in the introduction, the grand motivation for this project is to understand how genuine multi-partite entanglement plays a role in the emergence of a holographic bulk, especially the Anti-de Sitter vacuum. To this end, we have two goals in this paper. First, to define a good diagnostics for detecting genuine $\mathtt{q}$-partite entanglement and second, to study that measure under the holographic setting. For the first goal, by generalizing the construction of genuine Rényi multi-entropy $\GM[\mathtt{q}]_n$ proposed in the previous paper \cite{Iizuka:2025ioc}, we succeed in giving a general prescription on how to construct $\GM[\mathtt{q}]_n$ systematically for any $\mathtt{q}$.
The upshot is that the construction naturally involves the partition number $p(\mathtt{a})$ of integer $\mathtt{a}$.
For general $\mathtt{q}$, we found that $\GM[\mathtt{q}]_n$ contains $N(\mathtt{q}) = p(\mathtt{q}) - p(\mathtt{q} -1) -1$ number of parameters as shown in eq.~\eqref{Nq}. Furthermore, these give $N(\mathtt{q}) + 1$ number of new diagnostics for genuine $\mathtt{q}$-partite entanglement.
Especially for the $\mathtt{q}=4$ case, we reproduce the known diagnostics $I_3$ for quadripartite entanglement pointed out by  \cite{Balasubramanian:2014hda}. 
For the second goal, we studied $\GM[\mathtt{q}]$ for $\mathtt{q}=3, 4, 5$ in AdS$_3$ vacuum in holography and showed that these are nonzero and not small, {\it i.e.,} order of ${\mathcal{O}}{\left(1/G_N\right)}$ as eq.~\eqref{eq:GM_holographic} both analytically and numerically.

These studies led us to the conjecture that {\it for holography CFT states, genuine $\mathtt{q}$-partite entanglement is generically not only nonzero but also of the order of the central charges for all range of the values of $ \mathtt{q} \ge 3$.}\footnote{More precisely, we expect this to the large $\mathtt{q}$ up to the cutoff scale.} 
We expect this genuine multi-partite entanglement to be crucial for the holographic bulk reconstruction.
\vspace{1mm}

Some discussions leading to our conjecture are summarized below.

\begin{itemize}
\item As is pointed out by \cite{Almheiri:2014lwa}, quantum error correction is expected to be deeply involved in the emergence of bulk, and as we pointed out in our related essay \cite{Iizuka:2025bcc}, these quantum error correction encoding maps involve multi-partite entanglement in general. 
For example, the encoding map where we encode one qubit into four \cite{Grassl:1996eh} is given by
 \begin{align}
 \label{zerobaronebar}
   \begin{split}
     \ket{0} &\to \frac{1}{\sqrt{2}} \left(\ket{0000}+\ket{1111}\right) \equiv \ket{\bar{0}}, \;\;\;
     \ket{1} \to \frac{1}{\sqrt{2}} \left(\ket{1010}+\ket{0101}\right) \equiv \ket{\bar{1}}.
   \end{split}
 \end{align}
 and the encoding map where we encode one qubit into five \cite{Laflamme:1996iw} is given by
 \begin{align}
     &\ket{\bar{0}} = \frac{1}{4}\Big(\ket{00000}+\ket{10010}+\ket{01001}+\ket{10100}+\ket{01010}-\ket{11011}-\ket{00110}-\ket{11000}\nonumber\\
     & \quad -\ket{11101}-\ket{00011}-\ket{11110}-\ket{01111}-\ket{10001}-\ket{01100}-\ket{10111}+\ket{00101}\Big), \nonumber\\
     &\ket{\bar{1}} = \frac{1}{4}\Big(\ket{11111}+\ket{01101}+\ket{10110}+\ket{01011}+\ket{10101}-\ket{00100}-\ket{11001}-\ket{00111}\nonumber\\
     & \quad -\ket{00010}-\ket{11100}-\ket{00001}-\ket{10000}-\ket{01110}-\ket{10011}-\ket{01000}+\ket{11010}\Big).
 \end{align}
These encoding maps make use of a highly multi-partite entanglement structure, which can be verified by examining the genuine (Rényi) multi-entropy for these code states.
 
\item As is pointed out in \cite{Almheiri:2014lwa, Iizuka:2025bcc}, a defining feature of the holographic bulk reconstruction is that there can be bulk regions such that it is reconstructable from the entire boundary but not when one possesses only a portion of the boundary system. We refer to such bulk region the \emph{reconstruction shadow}, as illustrated in Fig.~\ref{fig:shadow}. As we divide the boundary into finer and finer subregions, the reconstruction shadow grows in size and engulf more and more UV region in the bulk. What this tells us is that the reconstruction of deeper bulk operators is harder than bulk operators closer to the boundary and requires access to more copies of the boundary subregions.
 Together with our results in Sec.~\ref{sec:holographicGM} that the holographic CFT state must possess genuine $\mathtt{q}$-partite entanglement, we are led to believe that the reconstruction of deeper bulk operators must rely more on higher-partite entanglement between finer boundary subregions. In other words, higher $\mathtt{q}$-partite entanglement is crucial to see the deeper central region in the bulk. In terms of a purely boundary point of view, this translates into the statement that higher $\mathtt{q}$-partite entanglements are increasingly more important in the IR, since it is known that the radial direction in the bulk AdS can be interpreted as an RG flow of the boundary CFT \cite{Susskind:1998dq,deBoer:2000cz,Heemskerk:2010hk}.
  \begin{figure}[ht]
    \centering
    \includegraphics[width=.9\linewidth]{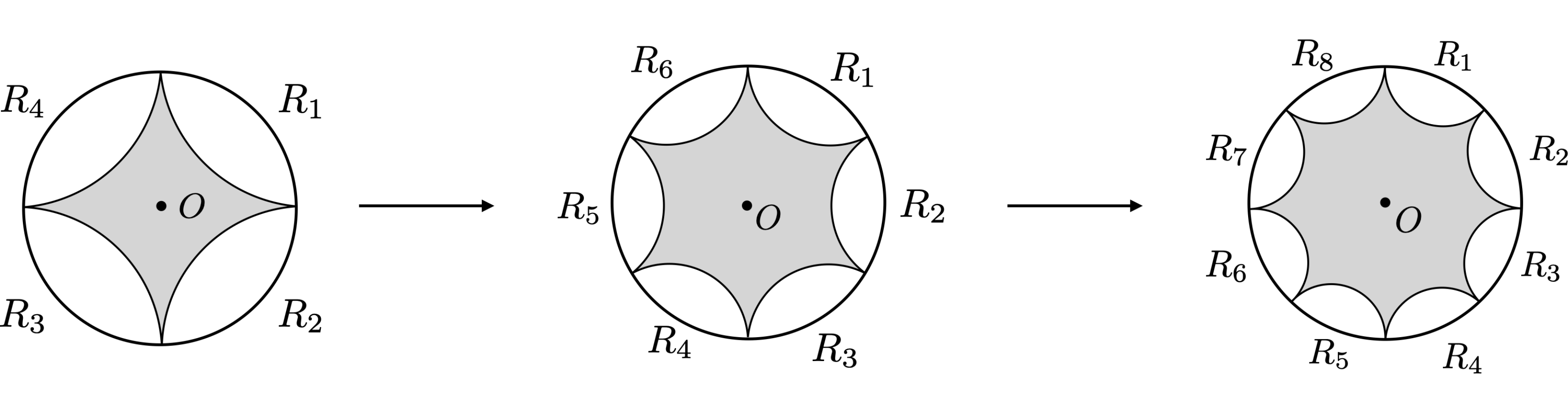}
    \caption{Boundary subregions $R_i$ and their entanglement wedges (shown as white). The gray region represents the ``reconstruction shadow'', where it is not possible to reconstruct the interior operator $O$ with access to only one of the boundary subregion. As we divide the boundary into finer divisions, the reconstruction shadow becomes larger and occupies more bulk UV region. This figure is taken from \cite{Iizuka:2025bcc}.}
    \label{fig:shadow}
  \end{figure}

\item The importance of higher-partite entanglement in holography can also be seen from the inequality (\ref{eq:hpt_bound_new}) for AdS$_3$, which gives a lower bound on the difference $\S[q](R_1:R_2:\cdots:R_\mathtt{q})-\frac{1}{2}\sum_\mathtt{k} \S[2](R_\mathtt{k})$ based on the number of vertices in the minimal multiway cut that is dual to $\S[q]$.
  The idea is that the difference (\ref{excludebicoombi}) of the multi-entropy and bipartite entanglement entropy in the inequality (\ref{eq:hpt_bound_new}) is sensitive to higher-partite entanglement.
  Since the number of vertices in the $\mathtt{q}$-way cut grows with the number of subregions $\mathtt{q}$, the difference (\ref{excludebicoombi}) becomes large when $\mathtt{q}$ is large, which implies a large amount of higher-partite entanglement in holography.
  
\item For holographic states with more than one asymptotic boundaries or where there is a black hole in the bulk dual, there may be large parametric regions where the genuine multi-entropy is zero. For example, we observed that the genuine multi-entropy becomes almost zero before the Page time in the black hole genuine multi-entropy curves \cite{Iizuka:2025ioc}, see also Fig.~\ref{fig:GMq5Random}. 
This is the situation where one degree of freedom (in this case, black hole) is much bigger than the rest combined. 
Closely related is the bulk configuration of a multi-boundary wormhole in the limit where the size of one of its asymptotic boundaries is larger compared to the others combined, as shown in Fig.~\ref{fig:wormhole}. 
However, the fact that the genuine multi-entropy becomes almost zero if one subsystem is much bigger than the rest combined, does not mean that multi-partite entanglements are not important in these states. In contrast, one simply has not divided the boundary fine enough to ``see'' multi-partite entanglement. Indeed, as one further divides the boundaries, the minimal $\mathtt{q}$-way cut becomes non-trivial again, and likewise we expect $\GM[q]>0$.
  \begin{figure}[ht]
    \centering
    \includegraphics[width=.33\linewidth]{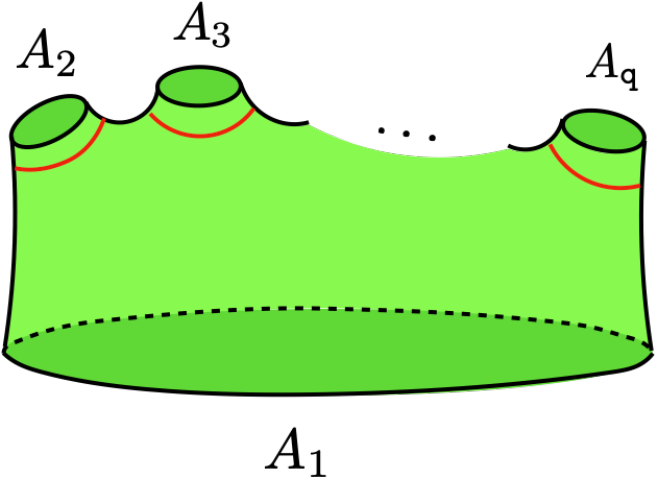}
    \hspace{2cm}
    \includegraphics[width=.33\linewidth]{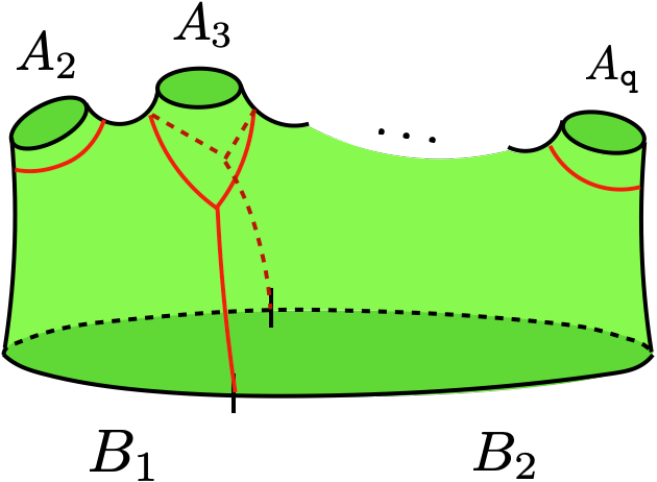}
    \caption{(Left) A multi-boundary wormhole. $A_1,A_2,\cdots,A_\mathtt{q}$ represents the size of the corresponding asymptotic boundaries. When $A_1 \gg A_2+A_3+\cdots+A_\mathtt{q}$, the genuine multi-entropy $\GM[q](A_1:A_2:\cdots:A_\mathtt{q})$ vanishes. The red surfaces indicates the minimal $\mathtt{q}$-way cut. (Right) If we divide the boundary region $A_1=B_1\sqcup B_2$, the new minimal cut becomes nontrivial and thus $\GM[q+1](B_1:B_2:A_2:\cdots:A_\mathtt{q})>0$.}
    \label{fig:wormhole}
  \end{figure}

\item Based on our computations of $\mathtt{q}=3,4,5$ examples in Appendix \ref{app:UV_cancellation}, we expect that the holographic genuine multi-entropy $\text{GM}^{(\mathtt{q})}$ for $\mathtt{q}\ge3$ is UV-divergent free. Since $\text{GM}^{(\mathtt{q})}$ is given by a symmetric linear combination of $S^{(\mathtt{q})}, S^{(\mathtt{q}-1)},\dots,S^{(2)}$, one can obtain an ``irreducible representation decomposition'' of the multi-entropy $S^{(\mathtt{q})}(R_1:R_2:\cdots :R_\mathtt{q})$ by the genuine multi-entropy as
\begin{align}
S^{(\mathtt{q})}(A_1:A_2:\cdots :A_\mathtt{q})=f(\text{GM}^{(\mathtt{q})},\text{GM}^{(\mathtt{q}-1)},\dots,\text{GM}^{(3)})+g(S^{(2)}),
\end{align}
where $f(\text{GM}^{(\mathtt{q})},\text{GM}^{(\mathtt{q}-1)},\dots,\text{GM}^{(3)})$ is a linear combination of the genuine multi-entropy with $\mathtt{q}\ge3$, and $g(S^{(2)})$ is a linear combination of the bipartite entanglement entropy $S^{(2)}$. For example, the $\mathtt{q}=4$ example is given by eq.~(5.3) of \cite{Iizuka:2025ioc}
\begin{align}
\label{quadripartite2}
S^{(4)}(A:B:C:D)&=\text{GM}^{(4)}(A:B:C:D)+\frac{1}{3}\text{GM}^{(3)}[2:1:1]\notag\\
&\;\;\;\;-\left(a-\frac{1}{3}\right)S^{(2)}(AB:CD)[2:2]+\left(a+\frac{1}{6}\right)S^{(2)}[3:1],
\end{align}
where $\text{GM}^{(3)}[2:1:1]$ is a symmetric linear combination of $\text{GM}^{(3)}$ for the integer partition $2+1+1$.
Since $\GM[q]$ is UV-finite, the divergence of $\S[q]$ can only come from contributions from bi-partite entanglement $S^{(2)}$.

It is known that the leading divergent piece of entanglement entropy in QFT ground states satisfies an \emph{area law}, which comes from the contribution of short-ranged entanglement across the entangling surface. By arguing that $\GM[q]$ is divergence-free, we have indirectly shown that the this divergent entanglement in QFT ground state (or in holographic CFT states) is always \emph{bipartite} in nature. 

\item From the viewpoint of holography, the UV divergence comes from the boundary near the entangling surfaces. Indeed, such UV-divergent contributions of all the multiway cuts in holographic genuine multi-entropy cancel out. See Appendix 
  \ref{app:UV_cancellation}. Moreover, the multiway-cut surfaces always lie within the reconstructable shadow. This also suggests a connection between bulk reconstructions of operators near the center of bulk and higher-partite entanglement detected by genuine multi-entropy. See Fig.~\ref{fig:tricut}.

\end{itemize}

\begin{figure}[t]
    \centering
    \includegraphics[width=0.3\linewidth]{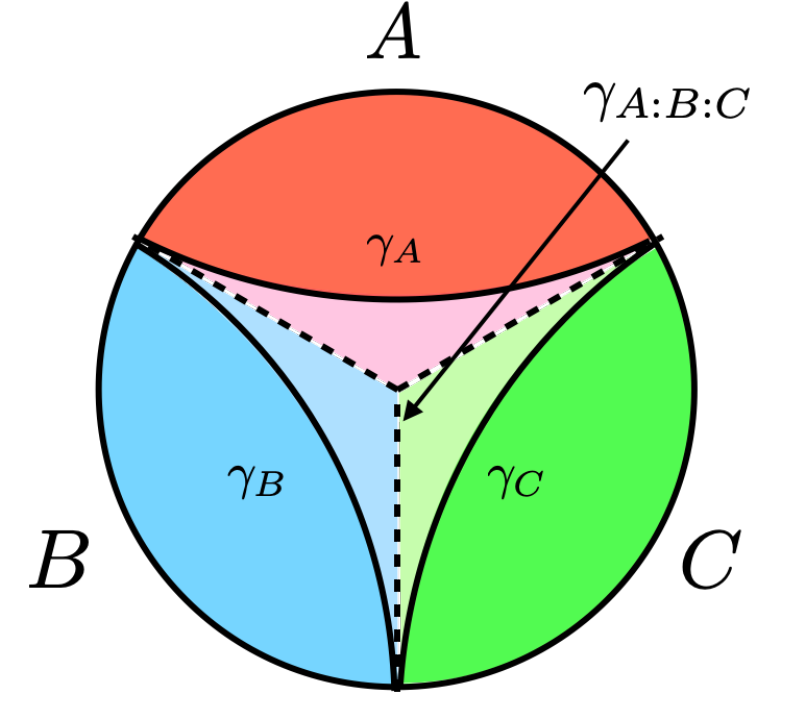}
    \caption{The multi-way cut $\gamma_{A:B:C}$ for the computation of $\GM[3]$. A bulk local operator in the white region, bounded by three minimal surfaces $\gamma_i$, cannot be reconstructed from any single boundary subsystems. This figure is taken from \cite{Iizuka:2025bcc}. }    \label{fig:tricut}
\end{figure}

Finally, we conclude this discussion with some possible future directions. Together with our accompanying essay \cite{Iizuka:2025bcc}, we have studied genuine multi-entropy in holographic settings as well as quantum error correction codes for the qubit systems. However, we are missing the understanding of genuine multi-entropy in the boundary quantum field theories. It is important to understand the multi-partite entanglement structure in various field theories including the holographic QFT's. For previous literature for the the study of boundary multi-entropies, see for example, \cite{Harper:2024ker, Liu:2024ulq}.  Another very important nature to be investigated is if the genuine multi-entropy shows the monotonicity nature under the LOCC so that the genuine multi-entropy can be a good quantum `measure'. See \cite{Gadde:2023zni, Gadde:2024jfi} as well for this direction. In relation to holography, it is also interesting to further study the nature of genuine multi-entropies for various wormhole settings, including the thermo-field double/eternal black hole settings \cite{Maldacena:2001kr} or multi-boundary wormhole settings such as  \cite{Skenderis:2009ju, Balasubramanian:2014hda, Marolf:2015vma, Anegawa:2020lzw, Emparan:2020ldj, Balasubramanian:2024ysu}. 
By investigating the role of genuine multi-partite entanglement in these various situations, we hope to gain a better understanding of quantum entanglement in many-body systems and holography.

\section*{Acknowledgements}
N.I. would like to thank NYU Abu Dhabi for its hospitality, where part of this work was carried out. 
S.L. would like to thank Ahmed Almheiri for insightful discussions and comments. The work of N.I. was supported in part by JSPS KAKENHI Grant Number 18K03619, MEXT KAKENHI Grant-in-Aid for Transformative Research Areas A “Extreme Universe” No. 21H05184, and NSTC of Taiwan Grant Number 114-2112-M-007-025-MY3. 

\appendix


\section{A new lower bound for $\S[q]$ in $\rm AdS_3$}
\label{app:lowerbound}

In this appendix, we give a proof for the inequality \eqref{eq:hpt_bound_new}, a positivity bound involving $\mathtt{q}$-partite multi-entropy and symmetrized (bipartite) entanglement entropy. For the ease of referencing we reproduce the inequality here:
\begin{equation}
\label{eq:hpt_bound}
\S[q](R_1:R_2:\cdots:R_\mathtt{q}) - \frac{1}{2}\left(\sum_\mathtt{k} \S[2] (R_\mathtt{k}) \right) \ge (\# \text{ of vertices in \texttt{q}-way cut}) \times \frac{3}{4G_N}\log\frac{2}{\sqrt{3}}
\end{equation}
We work in AdS$_3$/CFT$_2$, assuming that the holographic dual of the multi-entropy is given by minimal multiway cuts\footnote{We refer the reader to Sec.~\ref{sec:holographicGM} for details on multiway cuts as an holographic dual, as well as some notations we use in this appendix.}. We work with the unit where $\ell_{\rm AdS}=1$ in this appendix.

\begin{figure}[t]
    \centering
    \includegraphics[width=\linewidth]{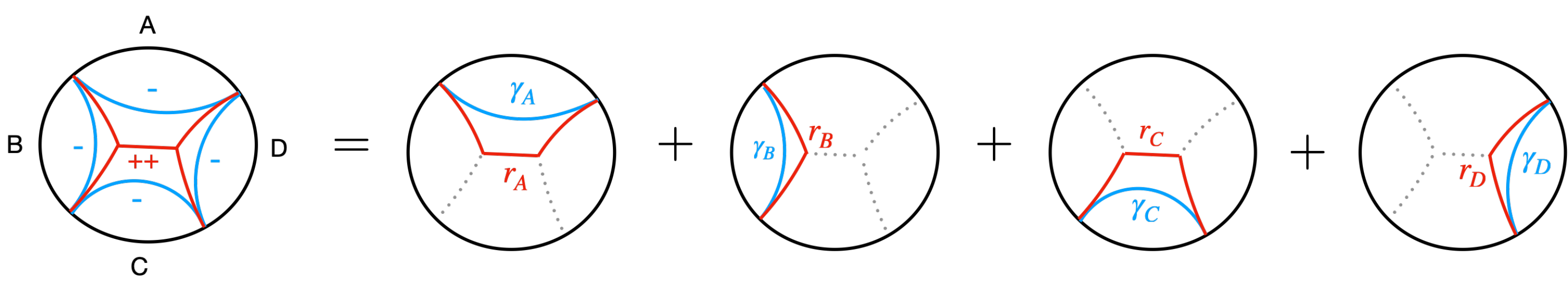}
    \caption{A pictorial proof for \eqref{eq:hpt_bound_weak} for $\mathtt{q}=4$. We split the 4-way cut corresponding to $\S[4](A:B:C:D)$ into four contributions $r_{A,B,C,D}$ and pair each of them up with the corresponding RT surface $\gamma_{A,B,C,D}$. The difference $\mathcal{A}(r_\mathtt{k})-\mathcal{A}(\gamma_\mathtt{k})$ is always positive by triangle inequality.}
    \label{fig:quad_decomp}
\end{figure}

Our proof is a straightforward generalization of the one given in \cite{Hayden:2021gno}.
Let us first recall how to prove the weaker bound
\begin{equation}
  \label{eq:hpt_bound_weak}
  \S[q](R_1:\cdots:R_\mathtt{q}) - \frac{1}{2}\left(\sum_\mathtt{k} S^{(2)}(R_\mathtt{k}) \right)\ge0.
\end{equation}
The geometric dual of \eqref{eq:hpt_bound_weak} is given by the difference of a \(\mathtt{q}\)-way cut and \(\mathtt{q}\) 2-way cuts, as shown in Fig.~\ref{fig:quad_decomp}.
We split the \(\mathtt{q}\)-way cut \(\mathcal{A}(R_1:\cdots:R_\mathtt{q})\) into \(\mathtt{q}\) 2-cuts \(r_\mathtt{k}\), each homologous with \(R_\mathtt{k}\). We then pair each of them with the corresponding minimal RT surface \(\gamma_\mathtt{k}\) of \(R_\mathtt{k}\). For each pair we have
\begin{equation}
\label{eq:triangle_ineq}
 \mathcal{A}(r_\mathtt{k}) - \mathcal{A}(\gamma_\mathtt{k}) \ge 0
\end{equation}
by the triangle inequality. This is true for all \(\mathtt{k}\) and the positivity of \eqref{eq:hpt_bound_weak} simply follows.

\begin{figure}[t]
  \centering
  \includegraphics[width=.4\linewidth]{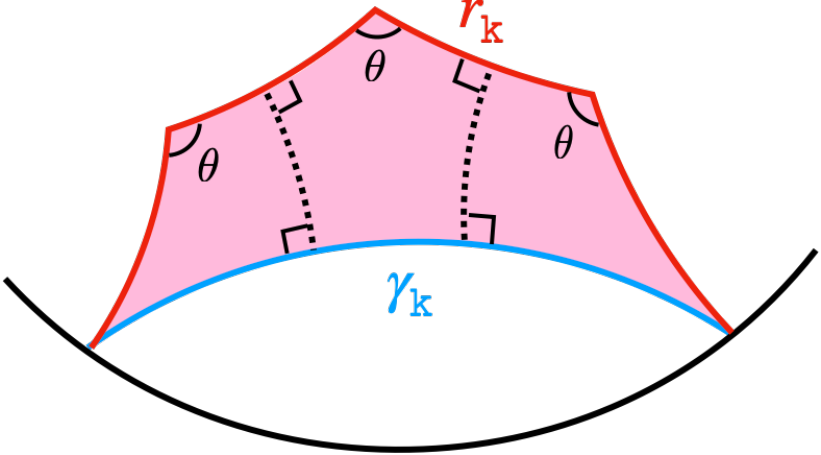}
  \caption{We divide the region bound between $r_\mathtt{k}$ and $\gamma_\mathtt{k}$ into $n$ pentagons. Each pentagon has four right angles and a fifth angle given by $\theta=2\pi/3$.}
  \label{fig:segments}
\end{figure}

We now show that \eqref{eq:triangle_ineq} can be strengthened to
\begin{equation}
\label{eq:hpt_bound_seg}
\mathcal{A}(r_\mathtt{k}) - \mathcal{A}(\gamma_\mathtt{k}) \ge (\# \text{ of kinks in } r_\mathtt{k}) \times \frac{1}{4G_N}\log\frac{4}{3}.
\end{equation}
First we note that the kinks in \(r_\mathtt{k}\) each corresponds to an intersection vertex in the \(\mathtt{q}\)-way cut, which can only be equiangular and trivalent. This fixes the angle of all kinks to be \(\theta=2\pi/3\).
Suppose there are \(n\) kinks on \(r_\mathtt{k}\). We divide the region bound by \(r_\mathtt{k}\) and \(\gamma_\mathtt{k}\) into \(n\) pentagons as shown in Fig.~\ref{fig:segments}, where each of the pentagon\footnote{It may seem like the first and last ``pentagon'' only have four edges. However they should really be viewed as "degenerate" pentagons where two of the boundary vertices coincide. Our argument still applies for this shape.} have four right angles. The fifth angle of the pentagon is \(\theta=2\pi/3\).

\begin{figure}[t]
  \centering
  \includegraphics[width=.3\linewidth]{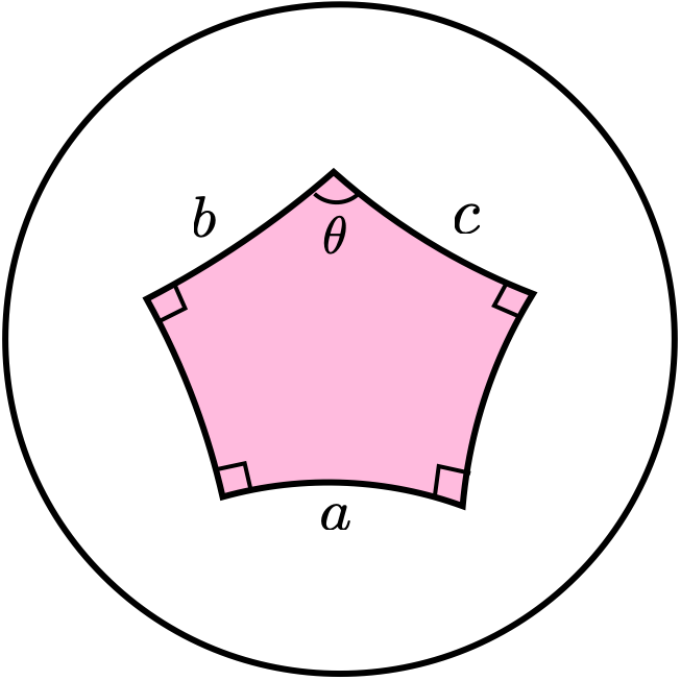}
  \caption{Each hyperbolic pentagon in Fig.~\ref{fig:segments} satisfies a law of cosine relating the angle $\theta$ and lengths of the sides $a,b,c$ as labeled.}
  \label{fig:pentagon}
\end{figure}

For any such hyperbolic pentagon with four right angles, there is a ``law of cosine'' relating its side lengths and the angle \(\theta\), as in\footnote{A proof of this formula can be found in reference on hyperbolic geometry, \emph{e.g.} Ch.~VI.3.2 of \cite{Fenchel:Hyperbolic_Geometry}.}
\begin{equation}
  \cosh a = \sinh b \sinh c - \cosh b \cosh c \cos \theta,
\end{equation}
where \(b,c\) are the lengths of sides adjacent to \(\theta\) and \(a\) is the length of the unique side that is opposite to \(\theta\). See Fig~\ref{fig:pentagon}.
In our case \(\cos \theta=\cos(3\pi/2) = -1/2\), hence
\begin{equation}
\label{eq:penta}
\cosh a = \sinh b \sinh c + \frac{1}{2}\cosh b \cosh c.
\end{equation}
The idea is that we can use \eqref{eq:penta} to bound the difference \(b+c-a\):
\begin{equation}
\label{eq:b+c-a}
b+c-a \ge \log \frac{4}{3}.
\end{equation}
We will prove \eqref{eq:b+c-a} later. For now let's assume this inequality is true. Going back to Fig.~\ref{fig:segments}, we have one  pentagon for each kink on \(r_\mathtt{k}\), and for each pentagon we have an inequality \eqref{eq:b+c-a}. Adding up the contribution from all \(n\) pentagons we obtain \eqref{eq:hpt_bound_seg}.

To prove \eqref{eq:hpt_bound}, we simply sum \eqref{eq:hpt_bound_seg} over all \(\mathtt{k} = 1,\cdots,\mathtt{q}\). Since all the vertices in the minimal \(\mathtt{q}\)-way cut are trivalent, the total number of vertices is simply the total number of kinks divided by three. Therefore, we obtain
\begin{align}
  \begin{split}
    &\S[q](R_1:\cdots:R_\mathtt{q}) - \frac{1}{2}\left(\sum_\mathtt{k=1}^{\mathtt{q}} S^{(2)}(R_\mathtt{k}) \right)\\
    =& ~(\# \text{ of kinks in \texttt{q}-way cut}) \times \frac{1}{4 G_N}\times\frac{1}{2}\log\frac{4}{3}, \\
    =& ~(\# \text{ of vertices in \texttt{q}-way cut}) \times \frac{1}{4 G_N}\times 3\log\frac{2}{\sqrt{3}},
  \end{split}
\end{align}
and we finish the proof.

We now show how to obtain \eqref{eq:b+c-a} from the laws of cosine. Substituting \eqref{eq:penta} into \(b+c-a\) we get
\begin{equation}
  b+c-a = b+c-\cosh^{-1}\big(\sinh b \sinh c + \tfrac{1}{2}\cosh b \cosh c\big).
\end{equation}
Introducing the variables \(u=\sinh b\) and \(v=\sinh c\), we can write this equation as
\begin{equation}
  \label{eq:bca_uv}
  b+c-a = \sinh^{-1} u +\sinh^{-1}v -\cosh^{-1}\left(uv + \tfrac{1}{2}\sqrt{1+u^2}\sqrt{1+v^2}\right).
\end{equation}
We now minimize RHS of \eqref{eq:bca_uv} across all valid parameter regions of \(u,v\).
First, we must have \(u,v\ge0\), which follows from the positivity condition \(b,c\ge0\). Second, the argument of \(\cosh^{-1}\) must be greater than \(q\), which leads to the constraint \(uv+\frac{1}{2}\sqrt{1+u^2}\sqrt{1+v^2} \ge 1\).

It turns out that the gradient of \eqref{eq:bca_uv} only vanishes at \(u,v = \pm i\), which is outside the region of interest. Thus the minimal value of \(b+c-a\) must be realized on the boundary of the parametric region, i.e. the region bound by \(u=v=0\) and \(uv+\frac{1}{2}\sqrt{1+u^2}\sqrt{1+v^2} = 1\).
With a little bit of algebra one can show that the minimum is attained when \(u,v\to \infty\), where
\begin{equation}
  \lim_{u,v\to\infty}\left(\sinh^{-1} u +\sinh^{-1}v -\cosh^{-1}[uv + \tfrac{1}{2}\sqrt{1+u^2}\sqrt{1+v^2}]\right) = \log\frac{4}{3},
\end{equation}
and thus we prove \eqref{eq:b+c-a}.

It is illustrating to see how this bound works in actual examples. Analytic expressions of minimal $\mathtt{q}$-cuts for connected boundary subregions in vacuum AdS$_3$ are computed in Appendix \ref{app:analytic} for $\mathtt{q}=3, 4$.
For \(\mathtt{q}=3\), the value of \eqref{eq:hpt_bound} is completely fixed by conformal symmetry. We simply have
\begin{equation}
  \S[3](A:B:C)-\frac{1}{2}(S(A)+S(B)+S(C)) = \frac{3}{4G_N}\log\frac{2}{\sqrt{3}}
\end{equation}
which saturates our bound.

\begin{figure}[ht]
  \centering
  \includegraphics[width=.48\linewidth]{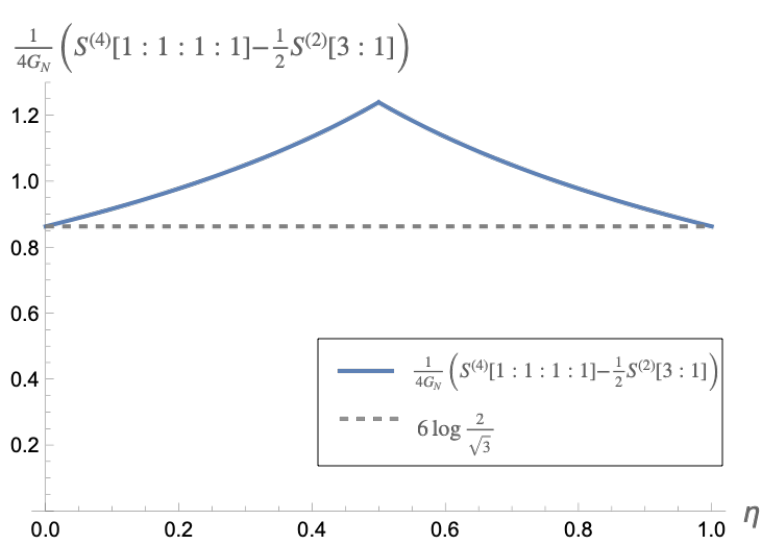}
  \includegraphics[width=.5\linewidth]{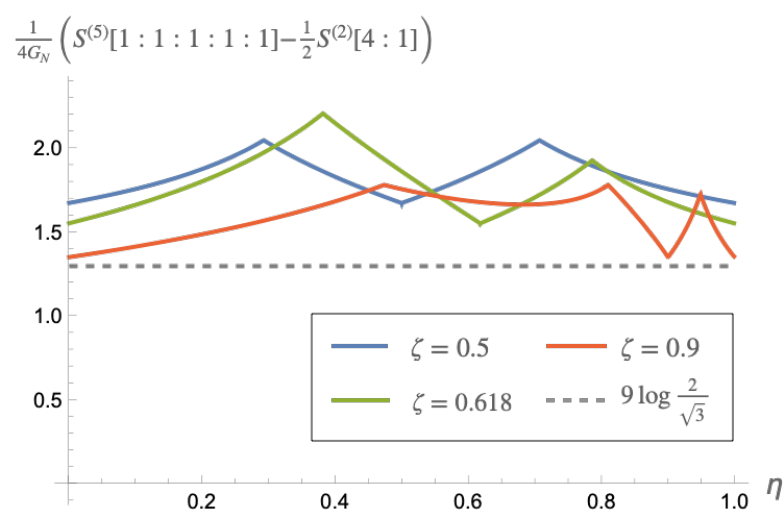}
  \caption{The difference between the holographic $\mathtt{q}$-partite multi-entropy $\S[q]$ and one half of the the symmetrized bipartite entanglement entropy $\frac{1}{2}\sum_{\mathtt{k}=1}^\mathtt{q}S(A_\mathtt{k})$ for $\mathtt{q}=4$ (left) and $\mathtt{q}=5$ (right), along with the predicted minimum value given by \eqref{eq:hpt_bound}. For the right plot, $(\eta,\zeta)$ are the cross-ratios of the two boundary points with respect to the remaining three. More explicitly, $\eta=\frac{(t_1-t_3)(t_4-t_5)}{(t_1-t_4)(t_3-t_5)}$ and $\zeta=\frac{(t_2-t_3)(t_4-t_5)}{(t_2-t_4)(t_3-t_5)}$ where $(t_1,t_2,t_3,t_4,t_5)$ are the five boundary points separating the region $(A,B,C,D,E)$.}
  \label{fig:Sq_bound}
\end{figure}

For \(\mathtt{q}=4\), the generic minimal multiway cut has two trivalent vertices.
Using the expression \eqref{eq:A4}, we find that
\begin{align}
  \label{eq:A4-A2}
  \begin{split}
  &S^{(4)}(A:B:C:D)-\frac{1}{2}(S(A)+S(B)+S(C)+S(D))  \\
  =&\begin{cases}
    \frac{1}{4G_N}\left( 6\log\frac{2}{\sqrt{3}} + 2\log(\tfrac{1}{2}(1+\sqrt{1-\eta}))-\log(1-\eta) \right), \quad &0\le\eta\le1/2, \\
    \frac{1}{4G_N}\left(6\log\frac{2}{\sqrt{3}} + 2\log(\tfrac{1}{2}(1+\sqrt{\eta}))-\log(\eta)\right), \quad &1/2<\eta\le1, \\
  \end{cases}    
  \end{split}
\end{align}
where $\eta=\frac{(t_1-t_2)(t_3-t_4)}{(t_1-t_3)(t_2-t_4)}$ is the conformal cross-ratio  of the four boundary points $(t_1,t_2,t_3,t_4)$ separating $(A,B,C,D)$. We plot \eqref{eq:A4-A2} in Fig.~\ref{fig:Sq_bound} (left).
For \(\mathtt{q}=5\) we have evaluated \eqref{eq:hpt_bound} numerically for minimal 5-way cuts and we verify that the bound \eqref{eq:hpt_bound} is again satisfied. The result is plotted in Fig.~\ref{fig:Sq_bound} (right).

\section{Detailed expressions for $\GM[5]_n(A:B:C:D:E)$}
\label{detailGM5}

The linear combinations of $S_n^{(\mathtt{a})}$ in Table (\ref{tab:p5}) are defined by
\begin{align}
S_n^{(5)}[1:1:1:1:1]
:=&\;S_n^{(5)}(A:B:C:D:E),\\
S_n^{(4)}[2:1:1:1]:=&\;S_n^{(4)}(AB:C:D:E)+S_n^{(4)}(AC:B:D:E)+S_n^{(4)}(AD:B:C:E)\notag\\
+&\;S_n^{(4)}(AE:B:C:D)
+S_n^{(4)}(BC:A:D:E)+S_n^{(4)}(BD:A:C:E)\notag\\
+&\;S_n^{(4)}(BE:A:C:D)+S_n^{(4)}(CD:A:B:E)+S_n^{(4)}(CE:A:B:D)\notag\\
+&\;S_n^{(4)}(DE:A:B:C),\\
S_n^{(3)}[2:2:1]:=&\;S_n^{(3)}(AB:CD:E)+S_n^{(3)}(AC:BD:E)+S_n^{(3)}(AD:BC:E)\notag\\
+&\;S_n^{(3)}(AB:CE:D)+S_n^{(3)}(AC:BE:D)+S_n^{(3)}(AE:BC:D)\notag\\
+&\;S_n^{(3)}(AB:DE:C)+S_n^{(3)}(AD:BE:C)+S_n^{(3)}(AE:BD:C)\notag\\
+&\;S_n^{(3)}(AC:DE:B)+S_n^{(3)}(AD:CE:B)+S_n^{(3)}(AE:CD:B)\notag\\
+&\;S_n^{(3)}(BC:DE:A)+S_n^{(3)}(BD:CE:A)+S_n^{(3)}(BE:CD:A)\\
S_n^{(3)}[3:1:1]:=&\;S_n^{(3)}(ABC:D:E)+S_n^{(3)}(ABD:C:E)+S_n^{(3)}(ABE:C:D)\notag\\
+&\;S_n^{(3)}(ACD:B:E)+S_n^{(3)}(ACE:B:D)+S_n^{(3)}(ADE:B:C)\notag\\
+&\;S_n^{(3)}(BCD:A:E)+S_n^{(3)}(BCE:A:D)+S_n^{(3)}(BDE:A:C)\notag\\
+&\;S_n^{(3)}(CDE:A:B),\\
S_n^{(2)}[3:2]:=&\;S_n^{(2)}(ABC:DE)+S_n^{(2)}(ABD:CE)+S_n^{(2)}(ABE:CD)\notag\\
+&\;S_n^{(2)}(ACD:BE)+S_n^{(2)}(ACE:BD)+S_n^{(2)}(ADE:BC)\notag\\
+&\;S_n^{(2)}(BCD:AE)+S_n^{(2)}(BCE:AD)+S_n^{(2)}(BDE:AC)\notag\\
+&\;S_n^{(2)}(CDE:AB),\\
S_n^{(2)}[4:1]:=&\;S_n^{(2)}(ABCD:E)+S_n^{(2)}(ABCE:D)+S_n^{(2)}(ABDE:C)\notag\\
+&\;S_n^{(2)}(ACDE:B)+S_n^{(2)}(BCDE:A),\\
S_n^{(1)}[5]:=&\;S_n^{(1)}(ABCDE).
\end{align}

We impose that $\GM[5]_n(A:B:C:D:E)$ of quadripartite entangled states, such as
\begin{align}
\ket{\psi}=\ket{\psi_1}_{ABCD}\otimes\ket{\psi_2}_{E},\label{festate}
\end{align}
vanishes. 
For these states, there are $p(4)-1=4$ relationships between Rényi multi-entropy corresponding to the integer partition of 4 in eq.~(\ref{ip4}) as follows.
\begin{itemize}
\item
The first relationship, corresponding to the integer partition $1+1+1+1$ of $4$ for four subsystems $A,B,C,D$, is
\begin{align}
\begin{split}
S_n^{(5)}(A:B:C:D:\cancel{E})=\frac{1}{4}\Big(&S_n^{(4)}(A\cancel{E}:B:C:D)+S_n^{(4)}(B\cancel{E}:A:C:D)\\
+\;&S_n^{(4)}(C\cancel{E}:A:B:D)+S_n^{(4)}(D\cancel{E}:A:B:C)\Big).\label{q5rs1}
\end{split}
\end{align}
\item
The second relationship, corresponding to the integer partition $2+1+1$ of $4$, is
\begin{align}
\begin{split}
&S_n^{(4)}(AB:C:D:\cancel{E})+S_n^{(4)}(AC:B:D:\cancel{E})+S_n^{(4)}(AD:B:C:\cancel{E})\\
&+ S_n^{(4)}(BC:A:D:\cancel{E})+S_n^{(4)}(BD:A:C:\cancel{E})+S_n^{(4)}(CD:A:B:\cancel{E})\\
=& \frac{1}{2}\Big(
 S_n^{(3)}(AB:C\cancel{E}:D)+S_n^{(3)}(AC:B\cancel{E}:D)+S_n^{(3)}(A\cancel{E}:BC:D)\\
&\quad +S_n^{(3)}(AB:D\cancel{E}:C)+S_n^{(3)}(AD:B\cancel{E}:C)+S_n^{(3)}(A\cancel{E}:BD:C)\\
& \quad +S_n^{(3)}(AC:D\cancel{E}:B)+S_n^{(3)}(AD:C\cancel{E}:B)+S_n^{(3)}(A\cancel{E}:CD:B)\\
& \quad +S_n^{(3)}(BC:D\cancel{E}:A)+S_n^{(3)}(BD:C\cancel{E}:A)+S_n^{(3)}(B\cancel{E}:CD:A)\Big)\\
=&\; S_n^{(3)}(AB\cancel{E}:C:D)+S_n^{(3)}(AC\cancel{E}:B:D)+S_n^{(3)}(AD\cancel{E}:B:C)\\
& \quad +S_n^{(3)}(BC\cancel{E}:A:D)+S_n^{(3)}(BD\cancel{E}:A:C)+S_n^{(3)}(CD\cancel{E}:A:B).\label{q5rs2}
\end{split}
\end{align}
\item
The third relationship, corresponding to the integer partition $2+2$ of $4$, is
\begin{align}
\begin{split}
&S_n^{(3)}(AB:CD:\cancel{E})+S_n^{(3)}(AC:BD:\cancel{E})+S_n^{(3)}(AD:BC:\cancel{E})\\
=\frac{1}{2}\Big(&S_n^{(2)}(AB\cancel{E}:CD)+S_n^{(2)}(AC\cancel{E}:BD)+S_n^{(2)}(AD\cancel{E}:BC)\\
+\;&S_n^{(2)}(BC\cancel{E}:AD)+S_n^{(2)}(BD\cancel{E}:AC)+S_n^{(2)}(CD\cancel{E}:AB)\Big).\label{q5rs3}
\end{split}
\end{align}
\item
The fourth relationship, corresponding to the integer partition $3+1$ of $4$, is
\begin{align}
\begin{split}
&\;\;\;\;\;S_n^{(3)}(ABC:D:\cancel{E})+S_n^{(3)}(ABD:C:\cancel{E})+S_n^{(3)}(ACD:B:\cancel{E})
+S_n^{(3)}(BCD:A:\cancel{E})\\
&=S_n^{(2)}(ABC:D\cancel{E})+S_n^{(2)}(ABD:C\cancel{E})+S_n^{(2)}(ACD:B\cancel{E})
+S_n^{(2)}(BCD:A\cancel{E})\\
&=S_n^{(2)}(ABC\cancel{E}:D)+S_n^{(2)}(ABD\cancel{E}:C)
+S_n^{(2)}(ACD\cancel{E}:B)+S_n^{(2)}(BCD\cancel{E}:A).\label{q5rs4}
\end{split}
\end{align}
\end{itemize}
In addition, there is the identity 
\begin{align}
S_n^{(2)}(ABCD:\cancel{E})=S_n^{(1)}(ABCD\cancel{E})=0,\label{p5id}
\end{align}
for pure states in eq.~(\ref{festate}) corresponding to the integer partition 4 in eq.~(\ref{ip4}).

\section{Analytic formulas for geometric genuine multi-entropies}
\label{app:analytic}
In this appendix we study analytic expressions for generic multl-way cuts on the hyperbolic disk.
The hyperbolic disk is defined to be the interior of unit disk on the complex plane, with metric given by
\begin{equation}
  ds^2 = \frac{4dzd\bar{z}}{(1-|z|^2)^2}.
\end{equation}
Given any two points $z_1,z_2$ inside the unit circle, the distance between them is given by
\begin{equation}
  d(z_1,z_2) = 2\tanh^{-1}\left|\frac{z_1-z_2}{1-z_1\bar{z}_2}\right|.
\end{equation}
Now suppose that we fix $z_1=1-\epsilon$ ($\epsilon$ being the UV cutoff) and an interior point $z_2 = r e^{i\theta}$, the (regularized) distance between $z_1=1$ and $z_2$ is
\begin{equation}
  d(1,z) = \log\left(\frac{2}{\epsilon}\right) -\log\left(\frac{1-z\bar{z}}{(1-z)(1-\bar{z})}\right) + O(\epsilon).
\end{equation}
If $z_2=e^{i\theta}$ is also on the boundary, the regularized distance is
\begin{equation}
  d(1,e^{i\theta}) = 2\log\left(\frac{2}{\epsilon}\right) +2\log(\sin (\theta/2)) + O(\epsilon).
\end{equation}

We denote area of the minimal $k$-way cut to be $\mathcal{A}^{(k)}(\theta_1:\cdots:\theta_k)$, where $\theta_i$ is the angular size of the boundary region and $\sum_i\theta_i=2\pi$.
In the remaining of this appendix we will give an analytic expression for $\mathcal{A}^{(3)}$ and $\mathcal{A}^{(4)}$, as well as several genuine multi-entropies.
To simplify expressions, we work in the units where we set $4G_N=1$ in this appendix. 

\subsection*{Triway cut}
We now give an analytic expression for the area of the minimal triway cut $\mathcal{A}^{(3)}(\theta_1:\theta_2:\theta_3)$. 
Suppose the three boundary points are located at $(e^{i\phi_1},e^{i\phi_2}, e^{i\phi_3})$ respectively and the trivalent junction is located at $z$.
By using a $SL(2;\mathbb{C})$ transformation
\begin{equation}
  z \to w(z)=\frac{a+bz}{c+dz}
\end{equation}
we can map these three points into $(1,e^{2\pi i/3},e^{4\pi i/3})$, where all of the regions have equal size.
Since we know that the trivalent for this symmetric configuration is simply $w=0$, the point $z$ can be simply read out from the inverse transformation $z(w=0)$. One finds that
\begin{equation}
  z = e^{i\phi_1}\frac{-e^{i\phi_2}-e^{2\pi i/3}e^{i\phi_3}+e^{\pi i/3}e^{i(\phi_2+\phi_3)}}{-e^{\pi i/3}+e^{2\pi i/3}e^{i\phi_2}+e^{i\phi_3}}.
\end{equation}
Therefore, we find that the area of the minimal triway cut is given by
\begin{align}
  \label{eq:A3}
  \begin{split}
    \mathcal{A}^{(3)}(\theta_1:\theta_2:\theta_3) &= d(e^{i\phi_1},z) + d(e^{i\phi_2},z) + d(e^{i\phi_3},z) \\
  &= 3\log(\tfrac{2}{\epsilon}) + 3\log(\tfrac{2}{\sqrt{3}})+\log(\sin \tfrac{\theta_1}{2})+\log(\sin \tfrac{\theta_2}{2})+\log(\sin \tfrac{\theta_3}{2}).
  \end{split}
\end{align}

\subsubsection*{Tripartite genuine multi-entropy $\GM[3]$}
As a simple exercise, we can use it to find the formula of $\GM[3]$ using \eqref{eq:A3}
It is given by the difference between $\mathcal{A}^{(3)}$ and three simple 2-cuts $\mathcal{A}^{(2)}$. It is easy to verify that
\begin{align}
  \begin{split}
      \GM[3](\theta_1,\theta_2,\theta_3) &\equiv \mathcal{A}(\theta_1:\theta_2:\theta_3) - \frac{1}{2}\left(\mathcal{A}(\theta_1)+\mathcal{A}(\theta_2)+\mathcal{A}(\theta_3)\right)\\
  &= 3\log\frac{2}{\sqrt{3}}.
  \end{split}
\end{align}
This expression is constant regardless of the boundary configurations.
This is expected behavior, since a global conformal transformation can be used to uniquely fix three points on the complex plane. As a result there is no residual degree of freedom left for tripartite genuine multi-entropy in a CFT.

\subsection*{Four-way cut}
\begin{figure}[t]
  \centering
  \includegraphics[width=.7\linewidth]{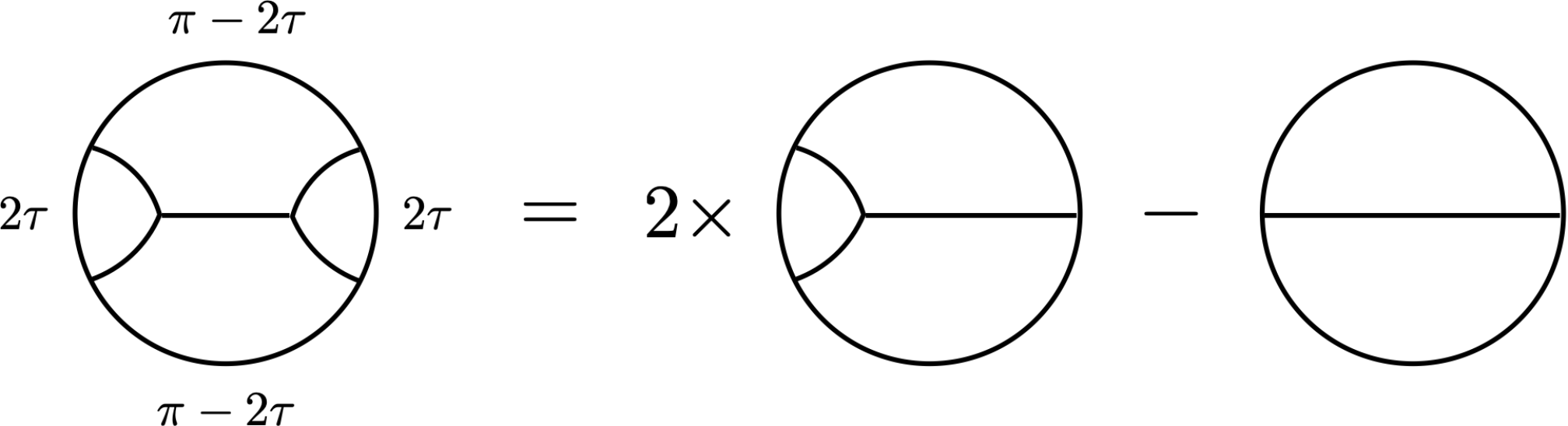}
  \caption{In the symmetric settings considered here, the minimal 4-way cut can be decomposed into the difference of triway cuts and the radius.}
  \label{fig:S4sub}
\end{figure}

We now evaluate the area of the minimal 4-way cut $\mathcal{A}^{(4)}(\theta_1,\theta_2,\theta_3,\theta_4)$.
We use a similar strategy here. Suppose that the four boundary points are located at $(e^{i\phi_1},e^{i\phi_2}, e^{i\phi_3}, e^{i\phi_4})$.
By utilizing an $SL(2;\mathbb{C})$ transformation we can bring the four boundary points to
\begin{equation}
  (t_1,t_2,t_3,t_4) = (e^{-i\tau},e^{i\tau},e^{i(\pi-\tau)},e^{i(\pi+\tau)}), \quad\tau \le \pi/4,
\end{equation}
so the entire configuration is symmetric upon reflection across the real and imaginary axis.
The minimal surface for this configuration has two trivalent intersections.
One can easily notice that this solution can be decomposed as two tri-way cuts minus the radius, see Fig.~\ref{fig:S4sub}.
In terms of equations this reads
\begin{align}
  \begin{split}
   \mathcal{A}^{(4)}(2\tau:\pi-2\tau:2\tau:\pi-2\tau) &= 2\mathcal{A}^{(3)}(\pi-\tau:2\tau:\pi-\tau)-\mathcal{A}^{(2)}(\pi) \\
   &= 4\log(\tfrac{2}{\epsilon})+6\log(\tfrac{2}{\sqrt{3}})+4\log(\sin(\tfrac{\pi-\tau}{2}))+2\log(\sin \tau).
  \end{split}
\end{align}

The angle $\tau$ is related to the conformal cross-ratio $\eta$ by
\begin{equation}
  \eta = \frac{(t_1-t_2)(t_3-t_4)}{(t_1-t_3)(t_2-t_4)} = \sin^2\tau.
\end{equation}
So we can also express the 4-way cut in terms of the cross-ratio as\footnote{Strictly speaking, \eqref{eq:A4} is not correct since it involves the regulator $\epsilon$ which is not invariant under $SL(2;\mathbb{C})$ transformations. However, when used to construct genuine multi-entropies $\GM[q]$, the $\epsilon$ dependence always cancels if a particular regularization scheme is fixed.}
\begin{equation}
  \label{eq:A4}
  \mathcal{A}^{(4)}(\eta) = 4\log(\tfrac{2}{\epsilon})+6\log(\tfrac{2}{\sqrt{3}})+\log \eta+2\log(\tfrac{1}{2}(1+\sqrt{1-\eta})).
\end{equation}
This formula is valid when $\eta\le1/2$. When $1/2<\eta\le1$ we simply replace $\eta\to1-\eta$ in \eqref{eq:A4}.

\subsubsection*{Quadripartite genuine multi-entropy $\GM[4]$}
Using \eqref{eq:A4} we can write down an explicit formula for $\GM[4]$ as a function of the cross-ratio. We work with $a=1/3$ here, since the coefficient proportional to $(a-\frac{1}{3})$ is simply the tripartite information $I_3$, which is given by linear combinations of 2-cuts.
We continue to work with the bulk configuration given by Fig.~\ref{fig:S4sub}. We have
\begin{equation}
  \GM[4]|_{a=1/3}(A:B:C:D) = \S[4][1:1:1:1] - \frac{1}{3}\S[3][2:1:1] + \frac{1}{3}\S[3][2:2],
\end{equation}
where we label $|A|=|C|=2\tau$ and $|B|=|D|=\pi-2\tau$. Note that $|A|,|C|\le|B|,|D|$ since $\tau\le\pi/4$.
Individual term breakdown:
\begin{itemize}
\item $\S[4][1:1:1:1]$: \\
  This is simply given by \eqref{eq:A4}.
  
\item $\S[3][2:1:1]$: \\
  There are in total 6 terms. It can be further split into two kinds of contributions depending on whether the boundary is connected or not.
  First we have four terms that look like $\S[3](AB:C:D)$, whose value is simply given by a 3-cut:
  \begin{align}
    &\S[3](AB:C:D) = \S[3](BC:D:A) =\S[3](CD:A:B) = \S[3](DA:B:C) \nonumber \\
    = ~&\A[3](2\tau,\pi-2\tau,\pi).
  \end{align}

  We also have the remaining two disconnected contributions, $\S[3](AC:B:D)$ and $\S[3](BD:A:C)$. Each of them has two possible candidate minimal surfaces: One given by two RT surfaces $S(B)+S(D)$ and the other given by a 4-cut, see Fig.~\ref{fig:GM4saddle}. In our current setting where $\eta\le 1/2$, it turns out that the solution given by two RT surfaces always wins in $\S[3](BD:A:C)$ (but not $\S[3](AC:B:D)$). Hence we have
  \begin{align}
    \S[3](BD:A:C)&=2\A[2](2\tau), \\
    \S[3](AC:B:D)&=\min(2\A[2](\pi-2\tau),\A[4](\eta)).
  \end{align}
  The cross-over point for $\S[3](AC:B:D)$ happens when $2\A[2](\pi-2\tau)=\A[4](\eta)$, which is roughly $\eta^*\approx0.265$.
  \begin{figure}[t]
    \centering
    \includegraphics[width=.6\linewidth]{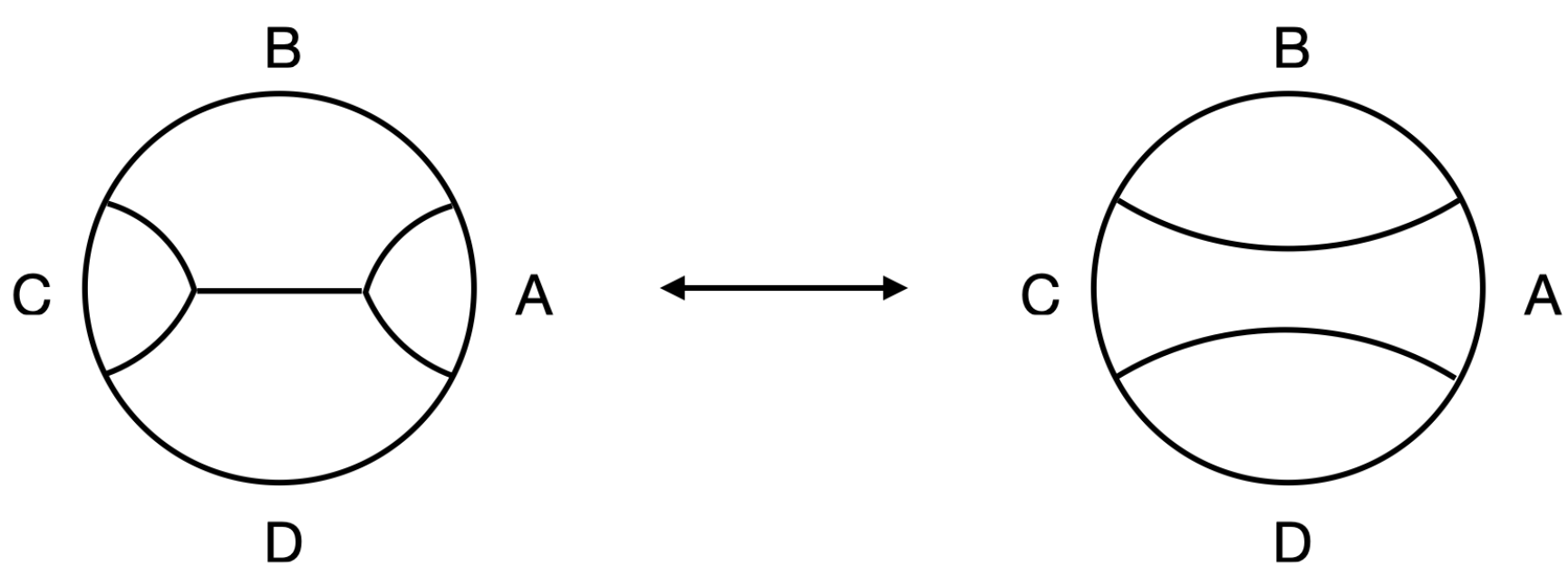}
    \caption{The two different possible bulk candidates for $\S[3](AC:B:D)$. The right configuration dominates when $\eta>\eta^*\approx0.265$.}
    \label{fig:GM4saddle}
  \end{figure}
  
\item $\S[2][2:2]$: \\
  There are three terms. The two terms featuring connected subregions are simply given by
  \begin{equation}
    \S[2](AB:CD)=\S[2](BC:AD) = \A[2](\pi).
  \end{equation}
  The remaining one is given by the smaller of the two possible RT surfaces:
  \begin{equation}
    \label{eq:S2ACBD}
    \S[2](AC:BD)=2\A[2](2\tau).
  \end{equation}
\end{itemize}

\begin{figure}[t]
  \centering
  \includegraphics[width=.5\linewidth]{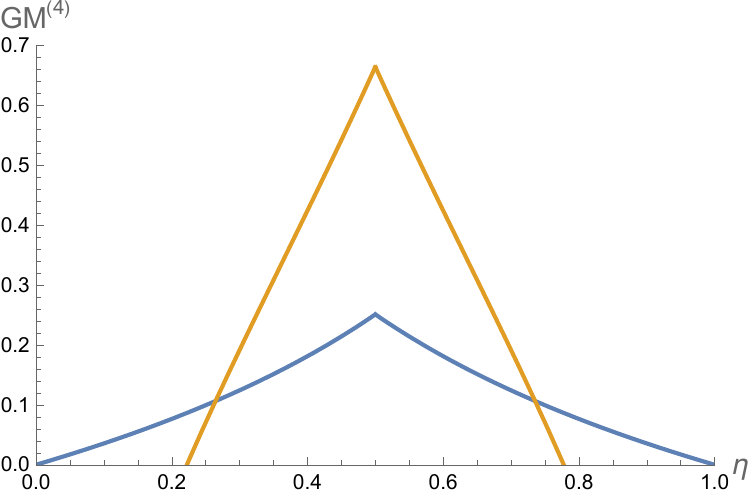}
  \caption{The analytic solution of the 4-partite genuine multi-entropy (Eq. \ref{eq:GM4analytic}), plotted against the cross-ratio $\eta$ of the four boundary points. The blue and orange lines represent different phases. Note that $\GM[4]|_{a=1/3}$ is positive semidefinite, as proven using geometric method in Sec.~\ref{holographicq=4}.}
  \label{fig:GM4analytic}
\end{figure}

Having listed all the individual contributions, we are now ready to calculate the value of $\GM[4]$. Using results from \eqref{eq:A4} to \eqref{eq:S2ACBD} we obtain
\begin{align}
  \label{eq:GM4analytic}
  \GM[4](\eta)|_{a=1/3} 
   &= \A[4](\eta)+\frac{1}{3}\left(2\A[2](\pi)-4\A[3](2\tau,\pi-2\tau,\pi)- \min(2\A[2](\pi-2\tau),\A[4](\eta))\right), \nonumber \\  
   &=\begin{cases}
     \frac{2}{3}\left(2\log\left(\frac{1+\sqrt{1-\eta}}{2}\right)-\log\sqrt{1-\eta}\right),\quad &0 \le \eta < \eta^*, \\
     2\log\frac{2}{\sqrt{3}}+\frac{1}{3}\log \eta+2\log\left(\frac{1+\sqrt{1-\eta}}{2}\right)-\frac{4}{3}\log(1-\eta),\quad &\eta^* \le \eta \le 1/2.
 \end{cases}
\end{align}
For $1\ge\eta>1/2$ simply replace $\eta\to1-\eta$.
We see that the cutoff dependence completely cancels out, as expected. We plot the two branches of \eqref{eq:GM4analytic} in Fig.~\ref{fig:GM4analytic} for reference.

\subsection*{Pentapartite genuine multi-entropy $\GM[5]$ (symmetric case)}
We close this appendix by showing a computation for the $\mathtt{q}=5$ genuine multi-entropy for the case where all the five boundary subregions are of the same size. For the sake of simplicity we will only focus on the coefficient in front of the parameter $b$, which only involves linear combinations of $\S[2]$ and $\S[3]$, namely:
\begin{equation}
  \label{eq:GM5b}
  5 {\partial_b}\GM[5](A:B:C:D:E) = 2S^{(3)}[2:2:1]-4S^{(3)}[3:1:1]-S^{(2)}[3:2]+5S^{(2)}[4:1].
\end{equation}
 Nevertheless by showing that \eqref{eq:GM5b} is non-zero, we confirm that there is non-trivial 5-partite entanglement presence in holography.
Individual term breakdown:
\begin{itemize}
\item $S^{(2)}[4:1]$: \\
  They all have the form of $S(A:BCDE)$ given by a single 2-cut. The total contribution is simply
  \begin{equation}
    S^{(2)}[4:1] = 5\A[2](2\pi/5).
  \end{equation}
  
\item $S^{(2)}[3:2]$: \\
  There are two different kinds of contributions. The first one is of the connected type $\S[2](AB:CDE)$, where the value is again simply given by a single 2-cut :
  \begin{equation}
    S(AB:CDE) = 5\A[2](4\pi/5).
  \end{equation}
  The other is the disconnected contribution of type $\S[2](AC:BDE)$. The minimal configuration is given by two 2-cuts:
  \begin{equation}
    S(AC:BDE) = 10\A[2](2\pi/5).
  \end{equation}
  Hence we have
  \begin{equation}
    \S[2][3:2] = 5\A[2](4\pi/5) + 10\A[2](2\pi/5).
  \end{equation}

\item $S^{(3)}[3:1:1]$: \\
  There are also two different kinds of contributions. There is the connected type, i.e. $S^{(3)}(ABC:D:E)$, where the minimal contribution is a simple triway cut:
  \begin{equation}
    \S[3](ABC:D:E) = \A[3](2\pi/5,2\pi/5,6\pi/5).
  \end{equation}
  The other one is the disconnected type such as $S^{(3)}(ABD:C:E)$.
  There are two possible candidates for each disconnected type, same as in Fig.~\ref{fig:GM4saddle}. The first is given by $S(C)+S(E)=2\A[2](2\pi/5)$.
  The second is given by the 4-cut $\A[4](AB:C:D:E)$. However one can argue that the 4-cut cannot dominate.
  To see this, note that the conformal cross-ratio of the configuration is given by $\eta=\frac{1}{2} (3-\sqrt{5}) \approx 0.38>\eta^*$.
  From the discussion in the previous section (in $\S[3][2:1:1]$) we know that in this phase the 2-cut solution wins. 
  Hence we have
  \begin{equation}
    \S[3](ABD:C:E) = 2\A[2](2\pi/5),
  \end{equation}
  and by cyclic permutation,
  \begin{equation}
    \S[3][3:1:1] = 5\A[3](2\pi/5,2\pi/5,6\pi/5) + 10\A[2](2\pi/5).
  \end{equation}

\item $S^{(3)}[2:2:1]$: \\
  This is the most complicated class with three different subtypes. The first one is of the form $S^{(3)}(AB:CD:E)$, with solution given by a  triway cut:
  \begin{equation}
    \S[3](AB:CD:E) = \A[3](4\pi/5,4\pi/5,2\pi/5).
  \end{equation}
  The second type is where one subregion is disconnected, i.e. $S^{(3)}(AB:CE:D)$.
  Similar to the case of $S^{(3)}(ABD:C:E)$ above, there are two candidate saddles, featuring either two 2-cuts, or a single 4-cut. We can check again that the 4-cut solution does not dominate from the cross-ratio $\eta=\frac{1}{2}(\sqrt{5}-1)\approx0.62$. Hence
  \begin{equation}
    S^{(3)}(AB:CE:D) = S(AB)+S(D) = \A[2](4\pi/5)+\A[2](2\pi/5).
  \end{equation}

  The last type is where both subregions are disconnected, {\it i.e.,} $S^{(3)}(AC:BD:E)$. There are three candidate saddles in this case, as illustrated in Fig.~\ref{fig:GM5saddle}. The first two are of the from a 2-cut and a 3-cut:
  \begin{equation}
    S(B)+S^{(3)}(ABC:D:E)\quad  \text{or}\quad  S(C)+S^{(3)}(EAB:C:D)
  \end{equation}
  They are equal in the symmetric configuration considered here.
  The third candidate is given by a 5-cut $\A[5](A:B:C:D:E)$.
  In contrary to 4-multiway cuts, we do not have an analytical expression for the area of a generic 5-cut.
  However, as explained in Fig.~\ref{fig:GM5saddle}, it is easy to see that in the particular symmetric setting we consider here, the 5-cut solution does not dominate. Hence
\begin{align}
    \S[3][2:2:1] &= 5\A[3](4\pi/5,4\pi/5,2\pi/5) + 5\A[3](2\pi/5,2\pi/5,6\pi/5) \nonumber \\
    &\qquad + 5\A[2](4\pi/5)+10\A[2](2\pi/5).
\end{align}

  \begin{figure}[t]
    \centering
    \includegraphics[width=.9\linewidth]{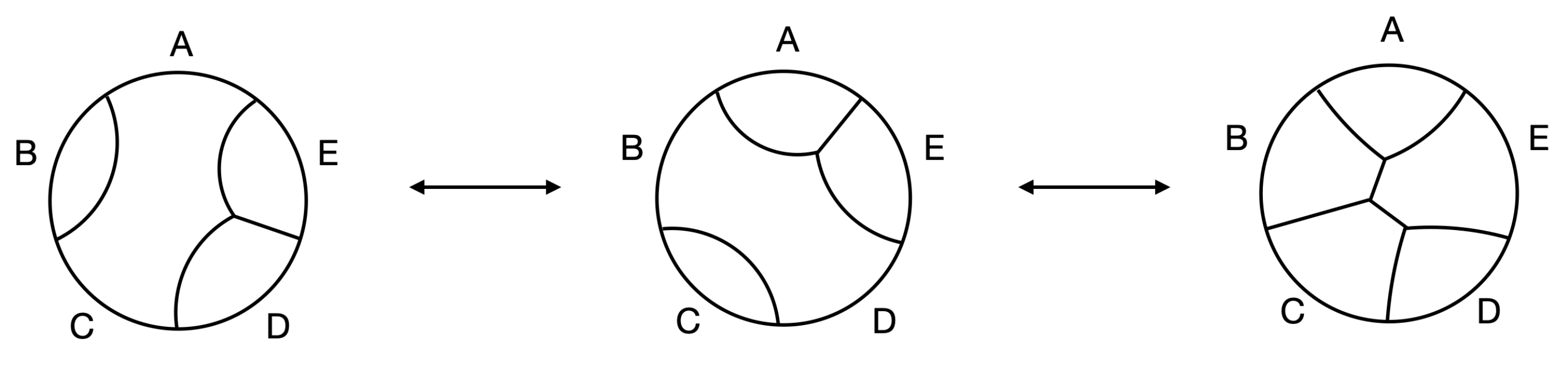}
    \caption{The three possible bulk candidates for $\S[3](AC:BD:E)$. In the our setting where all the five regions have equal size, we can show that the 5-cut does not dominate. This can be seen from starting from the 5-cut above and the by removing the geodesic line separating region $B$ and region $E$ (or equivalently between $C$ and $E$). After this the solution is manifestly a 2-cut separating $A:BCDE$ and a 3-cut separating $ABE:C:D$. Hence we have $\A[5](A:B:C:D:E)> \A[2](A)+\A[3](ABE:C:D)$. The RHS has the same minimal value as the left two configurations since all regions are symmetric. Thus the 5-cut solution cannot be dominant here.}
    \label{fig:GM5saddle}
  \end{figure}

\end{itemize}

Therefore, we see that in this particular setting where all five boundary subregions are of the same size, the minimal surfaces in \eqref{eq:GM5b} are given by a combination of 2-cuts and/or 3-cuts. We would like to remind the reader that it is not true more general settings.
Now we can put everything together. One finds that \eqref{eq:GM5b} is equal to
\begin{align}
  \label{eq:GM5b_equipartite}
    &5\partial_b\GM[5](A:B:C:D:E)=2S^{(3)}[2:2:1]-4S^{(3)}[3:1:1]-S^{(2)}[3:2]+5S^{(2)}[4:1] \nonumber\\
  &=10\A[3](4\pi/5,4\pi/5,2\pi/5)-10\A[3](2\pi/5,2\pi/5,6\pi/5)+5\A[2](4\pi/5)-5\A[2](2\pi/5) \nonumber \\
  &=20\log(\sin(2\pi/5))-20\log(\sin(\pi/5)),
\end{align}
which again does not depend on $\epsilon$ and is manifestly positive.

In fact we can extend the result of \eqref{eq:GM5b_equipartite} to nearby parametric regions, given that we stay close enough to the equipartite configuration so that the dominant saddles we considered above does not change. Since we can use a $SL(2)$ transformation to fix any three points of the five boundary points, we will use it to fix three of them to $(1,e^{6\pi i/5},e^{8\pi i/5})$. We place the remaining two points at $e^{i\alpha}$ and $e^{i\beta}$ with $6\pi/5>\beta>\alpha>0$. The area of the minimal cuts can be listed accordingly:
\begin{align}
  \begin{split}
  \S[2][4:1] &= 2\A[2](\tfrac{2\pi}{5}) + \A[2](\alpha) + \A[2](\tfrac{6\pi}{5}-\beta) + \A[2](\beta-\alpha), \\
  \S[2][3:2] &= \A[2](\tfrac{4\pi}{5}) + \A[2](\tfrac{2\pi}{5}+\alpha) + \A[2](\beta) +\A[2](\tfrac{6\pi}{5}-\alpha) +\A[2](\tfrac{8\pi}{5}-\beta) \\
    &\quad+ 2\S[2][4:1], \\
  \S[3][3:1:1] &= \A[3](\tfrac{2\pi}{5},\tfrac{2\pi}{5},\tfrac{6\pi}{5}) + \A[3](\tfrac{2\pi}{5},\alpha,\tfrac{8\pi}{5}-\alpha) + \A[3](\alpha,\beta-\alpha,2\pi-\beta) \\
    &\quad+ \A[3](\beta-\alpha,\tfrac{6\pi}{5}-\beta,\tfrac{4\pi}{5}+\alpha) + \A[3](\tfrac{6\pi}{5}-\beta,\tfrac{2\pi}{5},\tfrac{2\pi}{5}+\beta)  \\
             &\quad+2\S[2][4:1], \\
  \S[3][2:2:1] &= \A[3](\tfrac{2\pi}{5},\tfrac{2\pi}{5}+\alpha,\tfrac{6\pi}{5}-\alpha) + \A[3](\tfrac{2\pi}{5},\beta,\tfrac{8\pi}{5}-\beta) + \A[3](\alpha,\tfrac{6\pi}{5}-\alpha,\tfrac{4\pi}{5}) \\
             &\quad+ \A[3](\beta-\alpha,\tfrac{8\pi}{5}-\beta,\tfrac{2\pi}{5}+\alpha) + \A[3](\tfrac{6\pi}{5}-\beta,\tfrac{4\pi}{5},\beta)  \\
    &\quad+\S[2][3:2]-\S[2][4:1]+\mathcal{M}^{(3)}[3:1:1],
  \end{split}
\end{align}
where $\mathcal{M}^{(3)}$ picks the smaller configuration of the two left bulk candidates in Fig.~\ref{fig:GM5saddle}. Written out more explicitly:
\begin{align}
  \begin{split}
    \mathcal{M}^{(3)}[3:1:1] &\equiv \min\big(\A[2](B)+\A[3](ABC:D:E),\A[2](C)+\A[3](BCD:E:A)\big) + \text{(cyclic perm)}, \\
  &=15\log\tfrac{2}{\sqrt{3}}+\min\big(\mathcal{A}(B)+\mathcal{A}(D)+\mathcal{A}(DE),\mathcal{A}(A)+\mathcal{A}(C)+\mathcal{A}(AE)\big) + \text{(cyclic perm)}.
  \end{split}  
\end{align}
Plugging these expressions into \eqref{eq:GM5b} we get
\begin{align}
  \label{eq:GM5b_analytic_full}
  \begin{split}
    &5\partial_b\GM[5](\alpha,\beta)=2S^{(3)}[2:2:1]-4S^{(3)}[3:1:1]-S^{(2)}[3:2]+5S^{(2)}[4:1]\\
    =&\log(4+\tfrac{8}{\sqrt{5}})+12\log(\sin\tfrac{\beta}{2})-12\log(\sin\tfrac{\alpha}{2})-12\log(\sin\tfrac{\beta-\alpha}{2})\\
    +&2\log(\sin(\tfrac{\alpha}{2}+\tfrac{\pi}{5}))+2\log(\sin(\tfrac{\beta}{2}+\tfrac{\pi}{5}))+2\log(\cos(\tfrac{\alpha}{2}-\tfrac{\pi}{10}))-12\log(\cos(\tfrac{\beta}{2}-\tfrac{\pi}{10})) \\
    +&2\big( \min(m_1,m_2)+\min(m_2,m_3)+\min(m_3,m_4)+\min(m_4,m_5) + \min(m_5,m_1) \big).
  \end{split}  
\end{align}
where
\begin{align}
  \begin{split}
    m_1 &= \log(\sin\tfrac{\alpha}{2})+\log(\sin\tfrac{\beta}{2})+\log(\sin\tfrac{\beta-\alpha}{2}), \\
    m_2 &= \tfrac{1}{2}\log(2+\tfrac{2}{\sqrt{5}})+2\log(\sin\tfrac{\alpha}{2})+\log(\sin(\tfrac{\alpha}{2}+\tfrac{\pi}{5}))+\log(\cos(\tfrac{\beta}{2}-\tfrac{\pi}{10})), \\
    m_3 &= \tfrac{1}{2}\log(\tfrac{5+\sqrt{5}}{8})+4\log(\sin\tfrac{\beta-\alpha}{2}), \\
    m_4 &= \tfrac{1}{2}\log(2+\tfrac{2}{\sqrt{5}})+2\log(\sin\tfrac{\alpha}{2})+\log(\sin(\tfrac{\beta}{2}+\tfrac{\pi}{5}))+\log(\cos(\tfrac{\beta}{2}-\tfrac{\pi}{10})), \\
    m_5 &= \log(\cos(\tfrac{\alpha}{2}-\tfrac{\pi}{10}))+\log(\cos(\tfrac{\beta}{2}-\tfrac{\pi}{10}))+\log(\sin\tfrac{\beta-\alpha}{2}).
  \end{split}
\end{align}

\begin{figure}[t]
  \centering
  \includegraphics[width=.6\linewidth]{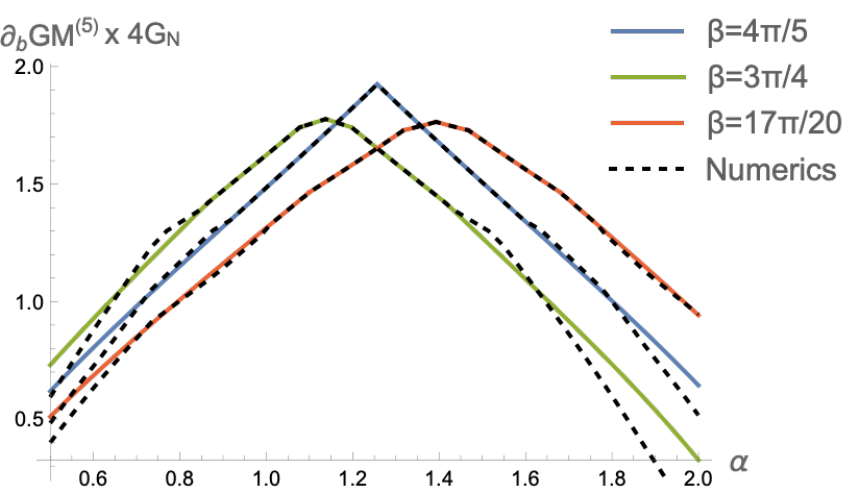}
  \caption{The analytical expression of $\partial_b\GM[5]$ (Eqn.~\ref{eq:GM5b_analytic_full}), plotted against the numerical evaluation. The colored lines are analytic predictions and the dashed curves are the numerics. We see that \eqref{eq:GM5b_analytic_full} is exact in a parametric regime close to the equipartite case $(\alpha,\beta)=(2\pi/5,4\pi/5)$, as expected. Fig.~\ref{fig:GM5b_analytic_cr} is the same plot but parametrized in terms of the conformal cross-ratios of the two points.}
  \label{fig:GM5b_analytic}
\end{figure}

We can obtain a cleaner expression if we fix $\beta=4\pi/5$. In this case \eqref{eq:GM5b_analytic_full} simplifies to
\begin{align}
  \label{eq:GM5b_analytic_restricted}
  \begin{split}
  \partial_b\GM[5](\alpha,\beta=\tfrac{4\pi}{5}) 
  =\begin{cases}
    \log(161+72\sqrt{5})+2\log(\sin\tfrac{\alpha}{2})+6\log(\sin(\tfrac{\alpha}{2}+\tfrac{\pi}{5})) &\\
    \quad+2\log(\cos(\tfrac{\alpha}{2}-\tfrac{\pi}{10}))-10\log(\cos(\tfrac{a}{2}+\tfrac{\pi}{10})), \quad\quad &\alpha\le2\pi/5, \\    \log(161+72\sqrt{5})-10\log(\sin\tfrac{\alpha}{2})+2\log(\sin(\tfrac{\alpha}{2}+\tfrac{\pi}{5}))  &\\
    \quad+6\log(\cos(\tfrac{\alpha}{2}-\tfrac{\pi}{10}))+2\log(\cos(\tfrac{a}{2}+\tfrac{\pi}{10})), \quad\quad &\alpha>2\pi/5.
  \end{cases}    
  \end{split}
\end{align}
We plot \eqref{eq:GM5b_analytic_full} and \eqref{eq:GM5b_analytic_restricted} against numerical evaluations of $\partial_b\GM[5]$ in Fig.~\ref{fig:GM5b_analytic}.

\section{UV cancellations in $\GM[q]$}
\label{app:UV_cancellation}
In this appendix, we show that \(\mathtt{q}\ge 3\) holographic genuine multi-entropies $\GM[q]$ receive no contribution from the cutoff $\epsilon$ at the entangling surface. Or to put it differently, we show that $\GM[q]$ is a UV-finite quantity.

\begin{figure}[t]
  \centering
  \includegraphics[width=.25\linewidth]{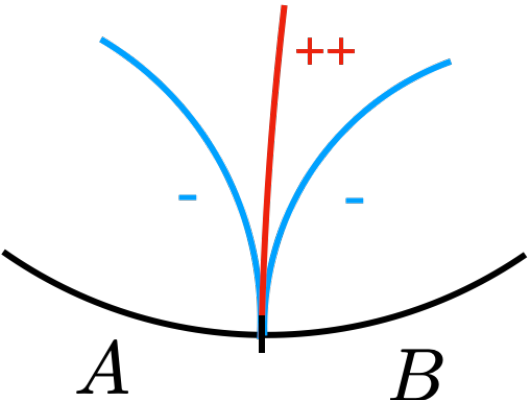}
  \caption{An illustration of the cancellation of the UV divergence in holographic genuine multi-entropy $\GM[q]$. Here we focus on the contribution from the minimal multiway cut of $\S[p]$ with $\mathtt{p}\leq \mathtt{q}$ close to the entangling surface of boundary subregions $A$ and $B$. The contribution consists of different minimal surfaces, all perpendicular to the asymptotic boundary of AdS. They can come with positive contribution (shown in red), or negative contribution (shown in blue). The claim is that the total divergent contribution of blue and red legs near the boundary cancels out.}
  \label{fig:UV_cancellation}
\end{figure}

Since \(\GM[q]\) is symmetric under permutation, it suffices to look at a specific entangling surface, say the contribution at interface of two boundary regions \(A:B\).
The basic idea is that since \(\GM[q]\) is given by minimal surfaces in the bulk, the divergence of the \(\GM[q]\) near the interface \(A:B\) is given by\footnote{For $d=3$ there is a log divergence $\log (1/\epsilon)$ instead of a power law divergence. Likewise for $d>4$ there may be subleading divergences. We only focus on the cancellation of the leading term in this appendix.}
\begin{equation}
  \label{eq:arealaw}
  \A[q](A:B:\cdots) \sim \frac{1}{\epsilon^{d-3}}\text{Area} (\partial A \cap \partial B) + \cdots
\end{equation}
where $d$ is the spacetime dimension of the bulk.
This should be thought of as an simple extension of the area law of holographic entanglement entropy into higher-partite scenarios.
Geometrically, \eqref{eq:arealaw} comes from the fact that bulk minimal area surfaces must be perpendicular to the asymptotic boundary. The UV piece of the $\A[q]$ comes from close-to-boundary regions since the AdS metric becomes divergent at the boundary, which is the same for all perpendicular surfaces.
This then allows us to compare the UV contribution of \(S^{(\mathtt{q})}\) at different \(\mathtt{q}\) and show that they cancel with each other. See Fig.~\ref{fig:UV_cancellation} for an illustration.

In what follows we will show that the UV contribution of genuine multi-entropy \(\GM[q]\) from the entangling surface \(A:B\) vanishes from the constraint equations imposed by $\GM[q]=0$ for factorizable states. We assume $d=3$ so the leading divergence to the multi-entropy scales as $\log (2/\epsilon)$. We work out explicit calculations leading to the cancellation for \(\mathtt{q} = 3,4,5\). While we currently do not have a proof for generic \(\mathtt{q}\), we expect it to be true based on the examples we checked.

\subsection*{Case of $\mathtt{q}=3$}
We have
\begin{equation}
  \GM[3](A:B:C) = S^{(3)}(A:B:C) - \frac{1}{2}(S(A)+S(B)+S(C)).
\end{equation}
Focusing on the contribution from \(A:B\), we get
\begin{equation}
  S^{(3)}(A:B:C)|_{A:B}-\frac{1}{2}(S(A)+S(B))
\end{equation}
as \(S(C)\) receives no contribution from the \(A:B\) interface.
Since in the UV limit
\begin{equation}
  S^{(3)}(A:B:C)|_{A:B} = S^{(2)}(A) = S^{(2)}(B) = \log (2/\epsilon) + O(\epsilon^0),
\end{equation}
from which we see that the total contribution from \(A:B\) is \(O(\epsilon^0)\).

\subsection*{Case of $\mathtt{q}=4$}
We have
\begin{equation}
  \GM[4](A:B:C:D) = S^{(4)}[1:1:1:1] + c_1 S^{(3)}[2:1:1] + c_2 S^{(3)}[2:2] + c_3 S^{(3)}[3:1],
\end{equation}
where the coefficients $(c_1,c_2,c_3)$ satisfy the following relations
\begin{equation}
  \label{eq:GM4constraints}
  1+3c_1 = c_1+c_2+c_3 = 0.
\end{equation}

We now focus on the contributions from the interface \(A:B\) in $\GM[4]$.
They are given by multi-entropies in which \(A\) and \(B\) occupy different partitions (for example from \(S^{(4)}(A:B:CD)\) but not \(S^{(4)}(AB:C:D)\)).
We list the contributions below:
\begin{itemize}
\item For \(S^{(4)}[1:1:1:1]\), there is a single contribution \(S^{(4)}(A:B:C:D)\).
\item For \(S^{(3)}[2:1:1]\), there are 5 terms contributing
  \begin{align}
    \begin{split}
      &S^{(3)}(A:B:CD) + S^{(3)}(AC:B:D) + S^{(3)}(AD:B:C)\\
      + &S^{(3)}(BC:A:D) + S^{(3)}(BD:A:C).
    \end{split}
\end{align}
\item For \(S^{(2)}[2:2]\), there are 2 terms contributing
  \begin{equation}
    S^{(2)}(AC:BD) + S^{(2)}(AD:BC).
  \end{equation}
\item for \(S^{(2)}[3:1]\), there are also 2 terms contributing
  \begin{equation}
    S^{(2)}(A:BCD)+S^{(2)}(B:ACD).
  \end{equation}
\end{itemize}

Hence, all the terms that contribute to \(A:B\) are
\begin{align}
  \begin{split}
&S^{(4)}(A:B:C:D) + c_1\Big(S^{(3)}(A:B:CD) + S^{(3)}(AC:B:D)\\
 & \quad +S^{(3)}(AD:B:C) + S^{(3)}(BC:A:D) + S^{(3)}(BD:A:C)\Big) \\
 &\quad+ c_2\Big(S^{(2)}(AC:BD) + S^{(2)}(AD:BC)) + c_3(S(A)+S(B)\Big) \\
&= \log(2/\epsilon) \times (1+5c_1+2c_2+2c_3) + O(\epsilon^0).    
  \end{split}
\end{align}
Note that the coefficient multiplying the logarithm is zero by the constraint equation \eqref{eq:GM4constraints}:
\begin{equation}
  1+5c_1+2c_2+2c_3 = (1+3c_1) + 2(c_1+c_2+c_3) = 0.
\end{equation}
Hence \(\GM[4]\) is UV finite for all \((c_1,c_2,c_3)\) satisfying the constraint.

Also, since we know that $\GM[4]$ can be linearly decomposed as
\begin{equation}
  \GM[4](A:B:C:D)|_a = \GM[4](A:B:C:D)|_{c_3=0} + c_3I_3
\end{equation}
for any given \(c_3\), each individual terms in the decomposition must also be UV finite.
Indeed, it is known that the tripartite information \(I_3\) is UV finite.
One can check that $\GM[4](A:B:C:D)|_{c_3=0}$ is also UV finite from its explicit expression \eqref{eq:GM4analytic} in AdS$_3$.

\subsection*{Case of $\mathtt{q}=5$}
We have
\begin{align}
\GM[5](A:B:C:D:E) &= S^{(5)}[1:1:1:1:1] + c_1S^{(4)}[2:1:1:1] + c_2S^{(3)}[2:2:1] \nonumber\\
 &+ c_3S^{(3)}[3:1:1] + c_4S^{(3)}[3:2] + c_5S^{(2)}[4:1].
\end{align}
The constraint equations are
\begin{equation}
  \label{eq:GM5_constraints}
  1+4c_1 = c_1 + 2c_2 + c_3  = c_2 +2c_4 =  c_3+c_4+c_5 = 0.
\end{equation}

Again, focusing on the contribution on \(A:B\) we find that the contributions are:
\begin{itemize}
\item For \(\S[5][1:1:1:1:1]\) there is 1 term  \(S^{(5)}(A:B:C:D:E)\).

\item For \(\S[4][2:1:1:1]\) there are 9 terms
  \begin{align}
    \begin{split}
&S_n^{(4)}(AC:B:D:E)+S_n^{(4)}(AD:B:C:E)+S_n^{(4)}(AE:B:C:D) \\
+&S_n^{(4)}(BC:A:D:E)+S_n^{(4)}(BD:A:C:E)+S_n^{(4)}(BE:A:C:D)\\
+&S_n^{(4)}(CD:A:B:E)+S_n^{(4)}(CE:A:B:D)+S_n^{(4)}(DE:A:B:C).      
    \end{split}
  \end{align}
  
\item For \(\S[3][2:2:1]\) there are 12 terms
  \begin{align}
    \begin{split}
& S_n^{(3)}(AC:BD:E)+S_n^{(3)}(AD:BC:E)+S_n^{(3)}(AC:BE:D)\\
+&S_n^{(3)}(AE:BC:D)+S_n^{(3)}(AD:BE:C)+S_n^{(3)}(AE:BD:C)\\
+&S_n^{(3)}(AC:DE:B)+S_n^{(3)}(AD:CE:B)+S_n^{(3)}(AE:CD:B)\\
+&S_n^{(3)}(BC:DE:A)+S_n^{(3)}(BD:CE:A)+S_n^{(3)}(BE:CD:A).\\      
    \end{split}
  \end{align}
  
\item For \(\S[3][3:1:1]\) there are 7 terms
  \begin{align}
    \begin{split}
&\;S_n^{(3)}(ACD:B:E)+S_n^{(3)}(ACE:B:D)+S_n^{(3)}(ADE:B:C)\\
+&\;S_n^{(3)}(BCD:A:E)+S_n^{(3)}(BCE:A:D)+S_n^{(3)}(BDE:A:C)\\
+&\;S_n^{(3)}(CDE:A:B).\\      
    \end{split}
  \end{align}
  
\item For \(\S[2][3:2]\) there are 6 terms
  \begin{align}
    \begin{split}
    &\;S_n^{(2)}(ACD:BE)+S_n^{(2)}(ACE:BD)+S_n^{(2)}(ADE:BC)\\
+&\;S_n^{(2)}(BCD:AE)+S_n^{(2)}(BCE:AD)+S_n^{(2)}(BDE:AC).
    \end{split}
  \end{align}
  
\item For \(\S[2][4:1]\) there are 2 terms
  \begin{align}
    &\;S_n^{(2)}(ACDE:B)+S_n^{(2)}(BCDE:A).
  \end{align}
\end{itemize}

Hence the total divergent contribution to \(A:B\) is 
\begin{align}
\log(2/\epsilon) \times (1+9c_1+12c_2+7c_3+6c_4+2c_5) + O(\epsilon^0)
\end{align}
Similar to previous examples, we see this is identically zero by the constraint equations \eqref{eq:GM5_constraints}, since we have the following contribution
\begin{equation}
\begin{alignedat}{4}
&1&&+9c_1&&+12c_2&&+7c_3+6c_4+2c_5 \\
=\;&1&&+4c_1 && &&\\
& &&+5c_1&&+10c_2&&+5c_3  \\
& && &&+~2c_2&&+0c_3+4c_4 \\
& && && &&+2c_3+2c_4+2c_5 = 0    
\end{alignedat}
\end{equation}
Hence we see that \(\GM[5]\) is UV finite.

\section{Genuine quadripartite multi-entropy for four-qubit states}\label{app:fourqubit}

\subsection{Genuine quadripartite multi-entropy of GHZ state}
\label{sec:GM4_GHZ}
In this section, we compute genuine quadripartite multi-entropy of GHZ state in a four-qubit system, whose behavior is totally different from genuine quadripartite multi-entropy in  holographic settings. We also explicitly compute genuine quadripartite Rényi multi-entropy $(n=2)$ of various four-qubit states.

GHZ state in $\mathtt{q}$-partite qubit systems is given by
\begin{align}
|\text{GHZ}_\mathtt{q}\rangle&=\frac{1}{\sqrt{2}}(|\underbrace{0 \dots 0}_{\mathtt{q}}\rangle+|\underbrace{1 \dots 1}_{\mathtt{q}}\rangle).
\end{align}
Replica partition function for $(n,\mathtt{q})$-multi-entropy of GHZ state is
\begin{align}
Z_n^{(\mathtt{q})}=\frac{1}{2^{n^{\mathtt{q}-1}-1}}.
\end{align}
Thus $(n,\mathtt{q})$-multi-entropy of GHZ state is given by
\begin{align}
S_n^{(\mathtt{q})}=\frac{1}{1-n}\frac{1}{n^{\mathtt{q}-2}}\log Z_n^{(\mathtt{q})}=\frac{n^{\mathtt{q}-1}-1}{n-1}\frac{1}{n^{\mathtt{q}-2}}\log 2,
\end{align}
and its $n\to1$ limit is
\begin{align}
S^{(\mathtt{q})}=(\mathtt{q}-1)\log 2.
\end{align}

By using this result, we obtain a simple expression of genuine quadripartite multi-entropy of GHZ state in a four-qubit system as follows
\begin{align}
\GM[4]_{{|\text{GHZ}_4\rangle}}=\left(\frac{1}{3}-a\right)\log2,
\end{align}
which is nonnegative if $a\le1/3$. The signature is completely opposite to the behavior in holographic settings: genuine quadripartite multi-entropy in holographic settings is nonnegative if $a\ge1/3$ as argued in the previous section. However the fact that they are nonzero and thus, both contains $\mathtt{q}=4$ genuine multi-partite entanglement is a common feature. In fact, one can explain this difference from the view point of the tripartite information defined by eq.~\eqref{I3n}, 

The tripartite information of GHZ state is positive:
\begin{align}
I_{{3\,|\text{GHZ}_4\rangle}}=\log 2,
\end{align}
which means that GHZ state does not satisfy the monogamy of holographic mutual information \cite{Hayden:2011ag}.

\subsection{Genuine quadripartite Rényi multi-entropy $(n=2)$ of various four-qubit states}

As further concrete examples, we explicitly evaluate genuine quadripartite Rényi multi-entropy $\GM[4]_{n=2}$ of various four-qubit states surveyed in \cite{Enríquez_2016}. More explicitly, these are, 
\begin{align}
\label{GHZ}
\ket{\text{GHZ}_4}&=\frac{1}{\sqrt{2}}(\ket{0000}+\ket{1111}),\\
\label{W}
\ket{\text{W}_4}&= \frac{1}{2}(\ket{0001}+\ket{0010}+\ket{0100}+\ket{1000}),\\
\label{D4}
\ket{D[4,(2,2)]}&=\frac{1}{\sqrt{6}}\left(|0011\rangle+|1100\rangle+|0101\rangle+|1010\rangle+|0110\rangle+|1001\rangle\right),\\
\ket{\text{HS}}&=\frac{1}{\sqrt{6}}\left(|0011\rangle+|1100\rangle+\omega(|0101\rangle+|1010\rangle)+\omega^2(|0110\rangle+|1001\rangle)\right),\notag\\
\omega&=\exp (2 i \pi / 3),\\
\left|\text{C}_1\right\rangle&=\frac{1}{2}(|0000\rangle+|0011\rangle+|1100\rangle-|1111\rangle),\\
\ket{\Psi_1^{\max} } &=   0.630 |0000\rangle + 0.281 |1100\rangle + 0.202 |1010\rangle + 0.24 |0110\rangle + 0.232 e^{0.494\pi i} |1110\rangle \notag\\
& + 0.059 |1001\rangle + 0.282 |0101\rangle + 0.346 e^{-0.362\pi i} |1101\rangle + 0.218 e^{0.626\pi i} |1011\rangle \notag\\
& + 0.304 |0011\rangle + 0.054 e^{-0.725\pi i} |0111\rangle + 0.164 e^{0.372\pi i} |1111\rangle,\\
\ket{BSSB_4}&=\frac{1}{2 \sqrt{2}}\left(|0110\rangle+|1011\rangle+i(|0010\rangle+|1111\rangle)+(1+i)(|0101\rangle+|1000\rangle)\right),\\
\ket{HD}&=\frac{1}{\sqrt{6}}(|1000\rangle+|0100\rangle+|0010\rangle+|0001\rangle+\sqrt{2}|1111\rangle),\\
\ket{YC}&=\frac{1}{2 \sqrt{2}}(|0000\rangle-|0011\rangle-|0101\rangle+|0110\rangle\notag\\
&\quad \quad \quad +|1001\rangle+|1010\rangle+|1100\rangle+|1111\rangle),\\
\ket{L}&=\frac{1}{2 \sqrt{3}}\big((1+\omega)(|0000\rangle+|1111\rangle)+(1-\omega)(|0011\rangle+|1100\rangle) \notag\\
& \quad \quad \quad +\omega^2(|0101\rangle+|0110\rangle+|1001\rangle+|1010\rangle)\big), \quad \omega=\exp (2 i \pi / 3).
\end{align}

We would like to compute $\GM[4]_{n=1}$, but numerical computations of $\GM[4]_{n}$ for all integer $n\ge2$ and their analytic continuation to $n=1$ are difficult. Hence, we numerically compute $\GM[4]_{n=2}$ of these states and guess their behaviors under the assumption that $\GM[4]_{n=1}$ and $\GM[4]_{n=2}$ are similar.

Our numerical results are summarized in Table \ref{tab:fourqubitstates}. We compute genuine quadripartite Rényi multi-entropy $\GM[4]_{n=2}$  and tripartite information $I_3$ given by eq.~\eqref{I3n=1} to classify various four-qubit states. The tripartite information $I_{3}$ of three highly symmetric states eq.~\eqref{GHZ},\eqref{W},\eqref{D4} is positive, and $I_3$ of the other states is negative. If one state is randomly chosen from all four-qubit states, the symmetry of that state is not high in most cases. Therefore, three highly symmetric states eq.~\eqref{GHZ},\eqref{W},\eqref{D4} are atypical in the sense that $I_3$ is positive, and the other states are typical. Since $I_3$ of the holographic settings is nonpositive, one can argue that the holographic states are typical. Another example of classification by $I_3$ is that thermal states in the (sparse) SYK model are typical, while thermal states in a simple vector model are atypical \cite{Iizuka:2024die}.

\begin{table}[t]
    \centering
    \begin{tabular}{|c|c|c|c|}
        \hline
        \text{ Four-qubit states} & \text{$\GM[4]_{n=2}$}  & \text{$I_{3}$} \\
        \hline
        $\ket{\text{GHZ}_4}$& $\frac{\log 2}{12}-a \log 2$ & $\log2$ \\ 
   \hline
        $\ket{\text{W}_4}$& $-\frac{\log 17}{4}-\frac{4\log 5}{3}+\log
   18+a \log \frac{625}{512}$ & $\log \frac{32}{27}$ \\ 
   \hline
         $\ket{D[4,(2,2)]}$& $\frac{\log2}{12}-\frac{\log 55}{4}+\log 19-\frac{7 \log 3}{4}-a \log 2$ & $\log \frac{32}{27}$ \\
         \hline
         $\ket{\text{HS}}$& $\log \frac{7}{9}+\frac{\log2}{3}+a \log \frac{27}{16}$ &  $-\log \frac{3 \sqrt{3}}{2}$ \\
        \hline
     $\ket{\text{C}_1}$& $-\frac{\log 2}{12}+a \log 2$ & $-\log2$ \\
        \hline
        $\ket{\Psi_1^{\max}}$& $0.0355+0.143a$ & $-0.449$ \\
        \hline
          $\ket{BSSB_4}$& $-\frac{3\log 2}{4}+\frac{2\log 5}{3}-\frac{\log
   3}{2}+a \log \frac{16}{9}$ & $-3\log 2+\frac{\log (3+2 \sqrt{2})}{\sqrt{2}}$ \\
        \hline
          $\ket{HD}$& $\log \frac{7}{9}+\frac{\log2}{3}+a \log \frac{27}{16}$ &  $-\log \frac{27}{16}$\\
        \hline
           $\ket{YC}$&  $-\frac{\log 2}{12}+a \log 2$ & $-\log2$ \\
        \hline
           $\ket{L}$&  $\log \frac{7}{9}+\frac{\log2}{3}+a \log \frac{27}{16}$  & $-\log \frac{27}{16}$\\
        \hline
    \end{tabular}
    \caption{Genuine quadripartite Rényi multi-entropy $\GM[4]_{n=2}$ at $n=2$ and tripartite information $I_3 := I_{3 ,n=1} $ of various four-qubit states.}
    \label{tab:fourqubitstates}
\end{table}

\bibliography{Refs}
\bibliographystyle{JHEP}

\end{document}